\documentclass[12pt]{article}

\usepackage{amssymb}
\usepackage{epsf}
\usepackage{rotating}

\def\ds#1{#1\kern-1ex\hbox{/}}
\def\dsh{h\kern-1.2ex /}

\newcommand{\bea}{\begin{eqnarray}}
\newcommand{\eea}{\end{eqnarray}}

\def\ul{\underline}

\def\beq{\begin{equation}}
\def\eeq{\end{equation}}

\def\beqn{\begin{eqnarray}}
\def\eeqn{\end{eqnarray}}
\def\ba{\begin{eqnarray}}
\def\ea{\end{eqnarray}}

\def\slash#1{#1\hskip-6pt/\hskip6pt}

\newcommand{\beqa}{\begin{eqnarray}}
\newcommand{\eeqa}{\end{eqnarray}}

\newcommand{\la}{\lambda}
\newcommand{\ro}{\rho}
\newcommand{\si}{\sigma}

\begin{document}

%\rightline{LTH 731}
\begin{center}
\vspace{1.5cm}
{\bf\large St\"{u}ckelberg Axions and the Effective Action of Anomalous Abelian Models 1.\\}
{ \bf A unitarity analysis of the Higgs-axion mixing}

\vspace{1cm}
{\bf\large Claudio Corian\`{o} $^{a,b}$  Nikos Irges $^{c}$ and 
Simone Morelli$^{a}$} 

\vspace{.5cm}
{\it  $^a$Dipartimento di Fisica, Universit\`{a} di Lecce, and \\
 INFN Sezione di Lecce,  Via Arnesano 73100 Lecce, Italy\\}
\vspace{.5cm}
{\it  $^b$Department of Mathematical Sciences, University of Liverpool, L69, 3BX, UK \\}

\vspace{.5cm}
{\it
$^c$Department of Physics and Institute of Plasma Physics \\
 University of Crete, 71003 Heraklion, Greece\\}
\vspace{.12in}
~\\
\centerline{\em \bf Dedicated to the Memory of Hidenaga Yamagishi}
\begin{abstract}

We analyze the quantum consistency of anomalous abelian models and of their effective field theories, rendered anomaly-free by a Wess-Zumino term, in the case of multiple abelian symmetries. These models involve the 
combined Higgs-St\"{u}ckelberg mechanism and predict a pseudoscalar axion-like field 
that mixes with the goldstones of the ordinary Higgs sector. We focus our study on the issue of unitarity 
of these models both before and after spontaneous symmetry breaking and detail the set of Ward identities and the organization of the loop expansion in the effective theory. The analysis is performed on simple 
models where we show, in general, the emergence of new effective vertices determined by certain anomalous interactions.

\end{abstract}
\end{center}
\newpage
%\tableofcontents
%%%%%%%%%%%%%%%%%%%%%%%%%%%%%%%%%%%%%%%%%%%%%%%%%%%%%%%%%%%%%%%%%
%\begin{figure}[tbh]
%{\centering \resizebox*{8cm}{!}{\rotatebox{0}
%{\includegraphics{HiggsA.eps}}}\par}
%\caption{}
%\label{g-fusion}
%\end{figure}
\newpage
\section{Introduction}
The search for the identification of possible extensions of the Standard Model (SM) is a challenging area both from the theoretical and the experimental perspectives. It is even more so with the upcoming experiments at the 
LHC, where the hopes are that at least some among the many phenomenological scenarios that have been formulated in the last three decades 
can finally be tested.  The presence of so many wide and diverse possibilities certainly render these studies very 
challenging. Surely, among these, the choice 
of simple abelian extension of the basic gauge structure of the SM is one of 
the simplest to take into consideration. These extensions will probably 
be the easiest to test and be also the 
first to be confirmed or ruled out. Though $U(1)$ extensions are ubiquitous, they are far from being trivial. These theories predict new gauge bosons, the extra Z', with masses that are likely to be detected if they are 
up to 4 or 5 TeV's (see for instance \cite{Langacker,Godfrey,Leike} for an overview and topical studies). 
These extensions are formulated, with a variety of motivations,  
within a well-defined theoretical framework and involve phenomenological studies which are far 
simpler than those required, for instance, in the case of supersymmetry, where a large 
set of parameters and of soft-breaking terms clearly render the theoretical description much more involved.

On the other hand, simple abelian extensions are also quite numerous, since new neutral currents are predicted both by Grand Unified Theories 
(GUT's) and/or by superstring inspired models based on $E_6$ and $SO(10)$ (see \cite{Alon} for instance). 
One of the common features of these models is the absence of an anomalous fermion spectrum,  as for the 
SM, with the anomaly cancelation mechanism playing a key role in fixing the couplings of the fermions to the gauge fields and in guaranteeing their inner consistency. 
In this respect, unitarity and renormalizability, tenets of the effective theory, are preserved. 

When we move to enlarge the gauge symmetry of the SM,
the unitarity has to be preserved, but not necessarily the renormalizability of the model.
In fact, operators of dimension-5 and higher which may appear at higher energies 
have been studied and classified under quite general assumptions \cite{Hagiwara}.

Anomalous abelian models, differently from the non-anomalous ones, show some striking features, which have been 
exploited in various ways, for example in the generation of realistic 
hierarchies among the Yukawa couplings \cite{Binetruy} and to analyze neutrino mixing. There are obvious reasons that justify these studies: the mechanism of anomaly cancelation that Nature selects may not just be based on an anomaly-free spectrum, but may require 
a more complex pattern, similar to the Green-Schwarz (GS) 
anomaly cancelation mechanism 
of string theory, that invokes an axion.  
Interestingly enough, the same pattern appears if, for a completely different and purely dynamical reason, part of the fermion spectrum of an anomaly free theory is integrated out, together with  part of the Higgs 
sector \cite{CorianoIrges}. In both cases, the result is a theory that shows the features 
discussed in this work, though some differences between the two different realizations may remain in the effective theory. For instance, it has been suggested that the PVLAS result can be easily 
explained within this class of models incorporating a single anomalous $U(1)$. The anomaly 
can be real (due to anomaly inflow from extra dimensions, (see \cite{Hill} as an example), 
or effective, due to the partial decoupling of a heavy Higgs, and the St\"uckelberg field is the remnant phase of this partial decoupling. The result is a ``gauging'' of the PQ axion \cite{CorianoIrges}.

\subsection{The quantization of anomalous abelian models and the axion}
The interest on the quantization of anomalous models and their proper field theoretical description has beeen a key topic 
for a long period, in an attempt to clarify under which conditions an anomalous gauge theory may be improved by the introduction of suitable interactions so to become unitary and renormalizable. 
The introduction of the Wess-Zumino term (WZ), a $\theta F\wedge F$ term), which involves a pseudoscalar $\theta$ times the divergence of a topological current, has been proposed as a common cure in order to restore the gauge invariance 
of the theory \cite{Faddeev,Krashnikov}. Issues related to the unitarity of models incorporating Chern-Simons (CS) and anomalous interactions 
in lower dimensions have also been analyzed in the past \cite{Jackiw}.

Along the same lines of thought, also non-local counterterms have been proposed as a way to achieve 
the same objective \cite{Adam}.
The gauge dependence of the WZ term and its introduction into the spectrum so 
to improve the power counting in the loop expansion of the theory has also been a matter of 
debate \cite{Preskill}. Either with or without a WZ term, renormalizability is clearly lost, 
while unitarity, in principle, can be maintained. As we are going to illustrate in specific and realistic 
examples, gauge invariance and anomaly cancelation play a subtle role in guaranteeing the 
gauge independence of matrix elements in the presence of symmetry breaking.

So far, the most interesting application of this line of reasoning in which the Wess-Zumino term acquires a physical meaning 
is in the Peccei-Quinn solution of the 
strong-CP problem of QCD \cite{PQ}, where the SM lagrangean is augmented by a global anomalous U(1) and involves an axion. 

The PQ symmetry, in its original form, is a global symmetry broken only by instanton effects. 
The corresponding axion, which in the absence of non perturbative 
effects would be the massless Nambu-Goldstone boson of the global 
(chiral) symmetry, acquires a tiny mass. In the PQ case 
the mass of the axion and its coupling to the gauge field are correlated,  
since both quantities are defined in terms of the same factor $1/f_a $, 
with $f_a $ being the PQ-breaking scale, which is currently bounded, by 
terrestrial and astrophysical searches, to be very large ($\approx 10^9$ GeV) \cite{ASik}. 

This tight relation between the axion mass and the coupling is a specific feature of models of PQ type where a global symmetry is invoked and, as we are going to see, it can be relaxed if the anomalous interaction is gauged. These issues are briefly 
mentioned here, while more phenomenological details concerning some applications involving the PVLAS experiment 
\cite{PVLAS} will be presented elsewhere.
The axion discussed in this paper and its effective action has some special features that render it an interesting 
physical state, quite distinct from the PQ axion. The term ``gauged axion'' or ``St\"{u}ckelberg axion'' or ``axi-Higgs'' all 
capture some of its main properties. Depending on the size of the PQ-breaking potential, 
the value of the axion mass gets corrected in the form of additional factors which are absent in the standard 
PQ axion. 

 Although some of the motivations to investigate this class of models come from the interest toward special vacua of string theory \cite{cik}, the study of anomalous abelian interactions, in the particular construction that we are going to discuss in this work, are applicable to a wide variety of models which share the typical features of those studied here.

\subsection{The Case of String/Branes Inspired Models}
As we have mentioned, we work under quite general assumptions that apply 
to abelian anomalous models that combine both the Higgs and the St\"{u}ckelberg mechanisms 
\cite{ Stueckelberg} in order to give mass to the extra (anomalous) gauge bosons.  
There are various low-energy effective theories which can be included into this framework, one example being 
low energy orientifold models, but we will try to stress on the generality of the construction rather 
than on its stringy motivations, which, from this perspective, are truly just optional.

These models have been proposed as a possible scenario for physics beyond the Standard Model, with 
motivations that have been presented in \cite{cik}. Certain features of these models have been studied in some generality \cite{Kiritsis}, and their formulation relies on the Green-Schwarz mechanism of anomaly cancelation that incorporates axionic and Chern-Simons interactions. At low energy, the Green-Schwarz term is nothing but the long known Wess-Zumino term. 
In particular, the mechanism of spontaneous symmetry breaking, that now involves 
both the St\"{u}ckelberg field (the axion) and the Standard Model Higgs, has been elucidated \cite{cik}. 
While the general features of the theory have been presented before, the selection of a specific gauge structure (the number of anomalous $U(1)$'s in 
\cite{cik} was generic), in our case of a single additional anomalous $U(1)$, allows us to specify the model in much 
more detail and discuss the structure of the effective action to a larger extent. This study is needed and provides a new 
step toward a phenomenological analysis for the LHC that we will 
present elsewhere.

This requires the choice of a specific 
and simplified gauge structure which can be amenable to experimental testing. While the number of anomalous abelian gauge groups 
is, in the minimal formulation of the models derived from intersecting branes, larger or equal to three, 
the simplest (and the one for which a quantitative phenomenological analysis is possible) case is the one in which a single anomalous 
$U(1)$ is present. This simplified structure appears once the assumption that the masses of the new abelian gauge interactions are widely (but not too widely) 
separated so to guarantee an effective decoupling of the heavier $Z'$, is made. 
Clearly, in this simplified setting, the analysis of 
\cite{cik} can be further specialized and extended and, more interestingly, 
one can try to formulate possible experimental predictions. 
\subsection{The content of this work} 
This work and \cite{CIM2} address the construction of anomalous 
abelian models in the presence of an extra anomalous $U(1)$, called $U(1)_B$. 
This extra gauge boson becomes massive via a combined Higgs-St\"uckelberg 
mechanism and is accompanied by one axion, $b$. We illustrate the physical 
role played by the axion when both the Higgs and the St\"uckelberg mechanisms are present. The physical axion, that emerges in the scalar sector when $b$ is 
rotated into its physical component (the axi-Higgs, denoted by $\chi$) 
interacts with the gauge bosons with dimension-5 operators (the WZ terms). 
The presence of these interactions renders the theory non-renormalizable 
and one needs a serious study of its unitarity in order to make sense of it,  
which is the objective of these two papers. Here the analysis 
is exemplified in the case of two simple models (the A-B and Y-B models) where 
the non-abelian sector is removed. A complete model will be studied in the second part. Beside the WZ term the theory clearly shows that additional 
Chern-Simons interactions become integral part of the effective action.

\subsection{The role of the Chern-Simons interactions} 
There are some very interesting features of these models which deserve a careful study, and which differ from the case of the Standard Model (SM). 
In this last case the cancelation of the anomalies is enforced by charge assignments. As a result of 
this, before electroweak symmetry breaking, all the anomalous trilinear gauge interactions vanish.  This cancelation continues to hold also after symmetry breaking if all the fermions of each generation are mass 
degenerate. Therefore, trilinear gauge interactions containing axial couplings are only sensitive to the mass differences among the 
fermions. In the case of extensions of the SM which include an anomalous $U(1)$ this pattern changes considerably, since the massless contributions in anomalous diagrams do not vanish. In fact, these theories become consistent only if a suitable set of axions and Chern-Simons (CS) interactions are included as counterterms in the defining lagrangean. The role of the CS interactions is to re-distribute the partial anomalies 
among the vertices of a triangle so to restaure the gauge invariance of the 1-loop effective action before symmetry breaking. For instance, a hypercharge current involving a generator $Y$,would be anomalous at 1-loop level in a trilinear interaction of the form $YBB$ or $YYB$, if B is an anomalous gauge boson. In fact, while anomalous diagrams of the form $YYY$ are automatically vanishing by charge assignment, the former ones are not. The theory requires that in these anomalous interactions the CS counterterm moves the partial anomaly from the $Y$ vertex to the $B$ vertex, rendering in these diagrams the hypercharge current effectively vector-like. The B vertex then carries all the anomaly of the trilinear interaction, but $B$ is accompanied by a Green-Schwarz (GS) axion $(b)$ and its anomalous gauge variation is canceled by the 
GS counterterm. It is then obvious that these theories show some new features which have never fully discussed in the past and require a very careful study. In particular, one is naturally forced to develope 
a regularization scheme that allows to keep track correctly of the distributions of the anomalies 
on the various vertices of the theory. This problem is absent in the case of the SM since the vanishing of the anomalous vertices in the massless phase renders any momentum parameterization of the diagrams acceptable. We have described in detail some of these more technical points in several appendices, 
where we illustrate how these theories can be treated consistently in dimensional regularization but with the addition of suitable shifts that take the form of CS counterterms.

%%%%%%%%%%%%%%%%%%%%%%%%%%%%%%%%%%%%%%%%%%%%%%%%%
\section{Massive $U(1)$'s a la St\"{u}ckelberg} 
%%%%%%%%%%%%%%%%%%%%%%%%%%%%%%%%%%%%%%%%%%%%%%%%%
%
One of the ways to render an abelian $U(1)$ gauge theory massive is by the mechanism proposed 
by St\"{u}ckelberg \cite{Stueckelberg}, extensively studied in the past, before that another mechanism, 
the Higgs mechanism, was proposed as a viable and renormalizable method to give mass both to abelian and to non abelian gauge theories. There are various ways in which, nowadays, this 
mechanism is implemented, and St\"{u}ckelberg fields appear quite naturally 
in the form of {\em compensator} fields in many supergravity and string models. On the phenomenological side, one of the first successfull investigations of this mechanism for model building has been presented in \cite{KN}, while, rather recently, supersymmetric extensions of this mechanism have been investigated \cite{Nath}. In other recent work some of its perturbative aspects have also been addressed, 
in the case of non anomalous abelian models. For instance in \cite{MM}, 
among other results, it has been shown that the mass renormalization and the wave function renormalization of the abelian vector field, in this model, are identical.

In the seventies, 
the St\"{u}ckelberg field (also called the ``St\"{u}ckelberg ghost'') re-appared in the analysis of the 
properties of renormalization of abelian massive Yang-Mills theory by Salam and Strathdee \cite{Salam}, 
Delbourgo \cite{Delbourgo} and others \cite{RuizAltaba}, while Gross and Jackiw \cite{GrossJackiw} introduced it in their analysis of the role of the 
anomaly in the same theory. According to these analysis the perturbative properties of a massive Yang-Mills 
theory, which is not renormalizable in its direct formulation, can be ameliorated by the introduction 
of this field. Effective actions in massive Yang-Mills theory 
have been also investigated in the past, and shown to have 
some predictivity also without the use of the St\"{u}ckelberg variables \cite{CY}, but clearly 
the advantages of the Higgs mechanism and its elegance remains a firm result of the current formulation of the Standard Model. We briefly review these points to make our treatment self-contained but also to show 
that the role of this field completely changes in the presence of an anomalous fermion spectrum, when 
the need to render the theory unitary requires the introduction of an $b F\tilde F$ interaction, 
spoiling renormalizability, but leaving the resulting theory, for the rest, well defined as an 
{\em effective} theory. For this to happen one needs to check explicitly the unitarity 
of the theory, which is not obvious, especially if the Higgs and St\"{u}ckelberg mechanisms are combined. 
This study is the main objective of the first part of this investigation, which is focused on the issues of unitarity of simple models which include 
both mechanisms. Various technical aspects of this analysis are important 
for the study of realistic models, as discussed in \cite{CIM2}, where we move toward the study of an extension of the SM with two abelian factors, one of them being the standard 
hypercharge (Y-B). The charge assignments for the anomalous diagrams 
involving a combinations of both gauge bosons are such that additional Ward identities are needed to render the theory unitary, starting from gauge 
invariance. We study most of the features of this model in depth, and show 
how the neutral vertices of the model are affected by the new anomaly cancelation mechanism. We will work out an application of the theory in the process 
$Z\to \gamma \gamma $, which can be tested at forthcoming experiments at the 
LHC.

 \subsection{The St\"{u}ckelberg action from a field-enlarging transformation}
We start with a brief introduction on the derivation of an action 
of St\"{u}ckelberg type to set the stage for further elaborations. 

A massive Yang Mills theory can be viewed as a gauge-fixed version of a more general action involving the 
St\"{u}ckelberg scalar. A way to recognize this is to start from the standard lagrangian 
\beq
\mathcal{L}= -\frac{1}{4} F_{\mu\nu}F^{\mu\nu} + \frac{1}{2} M_1^2 (B_\mu)^2
\label{fixed}
\eeq
with $F_{\mu\nu}=\partial_\mu B_\nu - \partial_\nu B_\mu$ and perform a field-enlarging transformation 
(see the general discussion presented in \cite{Alfaro})
\beq
B_\mu = B_\mu' - \frac{1}{M_1}\partial_\mu b,
\eeq
that brings the original (gauge-fixed) theory (\ref{fixed}) into the new form 
\beq
\mathcal{L}=-\frac{1}{4} F_{\mu\nu}F^{\mu\nu} + \frac{1}{2} M_1^2 (B_\mu)^2 +
\frac{1}{2}(\partial_\mu b)^2 - M_1 B_\mu \partial^\mu b 
\label{enlarged}
\eeq
which now reveals a peculiar gauge symmetry. It is invariant under the transformation 
\beqa
b&\to&b'= b - M \theta \nonumber \\
B_\mu &\to& B_\mu'=B_\mu + \partial_\mu \theta. 
\eeqa    
We can trace back our steps and gauge-fix this lagrangean in order to obtain a new version of the original lagrangean that now contains a scalar. One can choose to remove the mixing between $B_\mu$ and $b$ by the gauge-fixing condition 
\beq
\mathcal{L}_{gf} =- \xi \left( \partial\cdot B + \frac{M_1}{2 \xi}b\right)^2 
\eeq
giving the gauge-fixed lagrangian 
\beq
\mathcal{L}=-\frac{1}{4} F_{\mu\nu}F^{\mu\nu} + \frac{1}{2} M_1^2 (B_\mu)^2 +
\frac{1}{2}(\partial_\mu b)^2 - \xi (\partial B)^2 - \frac{M_1^2}{4 \xi} b^2.
\eeq
It is easy to show that the BRST charge of this model generates exactly the St\"{u}ckelberg condition on the 
physical subspace, decoupling the unphysical Faddeev-Popov ghosts from the physical spectrum. 

Different gauge choices are possible. 
The choice of a unitary gauge $(b=0)$ in the lagrangean (\ref{enlarged}) brings us back to the original 
massive Yang Mills model (\ref{fixed}). In the presence of a chiral fermion, the same field-enlarging transformation trick goes 
through, though this time we have to take into account the contribution of the anomaly 
\beq
\mathcal{L}=-\frac{1}{4}F_B^2 + \frac{M_1^2}{2}(B_\mu + \frac{1}{M_1} \partial_\mu b)^2 + 
i \overline{\psi}_L \gamma^\mu(\partial_\mu + i g_1 B_\mu + i g \partial_\mu b)\psi_L,
\eeq
where $\psi_L=\frac{1}{2}(1- \gamma^5)$ is the left handed anomalous fermion.
The Fujikawa method can be used to derive from the anomalous variation 
of the measure the relation 
\beq
g \overline{\psi}_L \gamma^\mu \partial_\mu b\psi_L = \frac{g^3}{32 \pi^2} 
\epsilon^{\mu\nu\rho\sigma}F_{\mu\nu}F_{\rho\sigma}
\eeq
thereby obtaining the final anomalous action 
\beq
\mathcal{L}=-\frac{1}{4}F_B^2 + \frac{M_1^2}{2}(B_\mu + \frac{1}{M_1}\partial_\mu b)^2 + 
i \overline{\psi}_L \gamma^\mu(\partial_\mu + i g_1 B_\mu)\psi_L
- \frac{g^3}{32 \pi^2}b\,\epsilon^{\mu\nu\rho\sigma}F_{\mu\nu}F_{\rho\sigma}.
\eeq

Notice that the $b$ field can be integrated out \cite{GrossJackiw}.
In this case one obtains an alternative effective action of the form
\beqa
\mathcal{L} &=& -\frac{1}{4}F_B^2 + \frac{M_1^2}{2}(B_\mu)^2 + 
i \overline{\psi}_L \gamma^\mu(\partial_\mu + i g_1 B_\mu)\psi_L \nonumber \\
&& -
\frac{g^3}{96 \pi^2}\int d^4 y F_B^{\alpha\beta}\tilde{F}_{B\alpha \beta}(x)D(x-y|M_1^2 \xi)
F^{\mu\nu}(y)\tilde{F}_{\mu\nu}(y) 
\eeqa
with $(\square + M_1^2 \xi^2)D(x|M_1^2)=-\delta^4(x)$. The locality of the description is clearly 
lost. It is also obvious that the role of the axion, in this case, is to be an unphysical field. However, in the case 
of a model incorporating both spontaneous symmetry breaking and the St\"{u}ckelberg mechanism, the axion plays a physical 
role and can be massless or massive depending whether it is part of the scalar potential or not. 
Our interest, in this work, is to analyze in detail the contribution to the 
1-loop effective action of anomalous abelian models, here defined as the classical lagrangean plus its 
anomalous trilinear fermionic interactions. Anomalous Ward identities in these 
effective actions are eliminated once the divergences from the triangles are removed either by 1) suitable charge assignments for some of generators, or by 2) 
shifting axions 
or 3) by a judicious (and allowed) distribution of the partial anomalies on each vertex. 

Since this approach of anomaly cancelations is more involved than in the SM case, we have decided to analyze it in depth using some simple 
(purely abelian) models as working examples, before considering a realistic extension of the Standard Model. This extension is addressed in \cite{CIM2}. There, all the methodology developed in this work will 
be widely applied to the analysis of a string-inspired model derived from the orientifold construction \cite{cik}.
In fact, this analysis tries to clarify some unobvious issues that naturally appear once an effective anomaly-free gauge theory 
is generated at lower energies from an underlying renormalizable theory at a higher energy. For this purpose we will use a simple approach based on s-channel unitarity, inspired by the classic work of Bouchiat, Iliopoulos and Meyer \cite{BIM}.
% We will 
%try to put into evidence, in these simplified cases, all the interesting features intrinsic to these new cancelations and, in particular, we will focus our attention on the issue of unitarity in the broken phases of these models. There are some additional comments that are in order before we start our technical discussion. 
%%some relevant issues concerning the unitarity of some of these models in the presence of a Higgs-axion mixing. 

%We remind that it is believed by many that the only shortcoming of an anomalous gauge theory lays in its lack of renormalizability. The addition to the theory of a dimension-5 operator, such as the Wess-Zumino term (WZ), a non-renormalizable interaction by itself, only 
%provides a way to ameliorate the description of these theories and does not worsten their already compromised 
%condition in that respect. However, it is acceptable to have a non renormalizable theory on which power 
%counting and a loop expansion can be carried out more easily, and this has to be seen, undoubtedly, as an 
%improvement. In the presence of a spontaneous symmetry breaking the role of the axion is surely more 
%complex since it ceases to be a gauge artifact, as we are going to ilustrate 
%next, and its mixing with the Higgs plays a decisive role in guaranteeing 
%the gauge independence of the S-matrix elements of the theory. 
\subsection{Implications at the LHC}
A second comment concerns the possible prospects for the discovery of a $Z'$ of anomalous origin. Clearly  with $Z'$'s being ubiquitous in GUT's and other SM extensions, discerning an anomalous $Z'$ from a non-anomalous one is subtle, but possible. 
In \cite{CIM2} we propose the Drell-Yan mechanism as a possible way to make this distinction, since some new effects related to the treatment of the 
anomalies are already (at least formally) apparent near the 
$Z$ resonance already in this process. Anomalous vertices involving the $Z$ gauge boson 
appear both in the production mechanism and in its
decay into two gluons or two photons. In the usual Drell-Yan 
process, computed in the SM, these contributions, because of anomaly cancelations, 
are sensitive only to the mass difference between the fermion of a given 
generation and are usually omitted in NNLO computations. 
If these resonances, predicted by theories with extra abelian gauge structures, are very weakly coupled, then a precise determination of the QCD background is necessary to detect them.

%%%%%%%%%%%%%%%%%%%%%%%%%%%%%%%%%%%
\section{The Effective Action in the AB model}
%%%%%%%%%%%%%%%%%%%%%%%%%%%%%%%%%%%

As we have already mentioned, we will focus our analysis on the anomalous effective actions of simple 
abelian theories. We will analize two models: a first one called ``A-B'', with a A vector-like 
(and anomaly-free) and B 
axial-vector like and anomalous; and a second model, called the ``Y-B'' model where B is anomalous and 
Y is anomaly-free but has both vector and axial-vector interactions. Differently from the A-B model, which will be introduced in the next section, the Y-B model will be teated in one of the final sections. 

%%%%%%%%%%%%%%%%%%%%%%%%%%%%%%%%%%%%%%%%%%%%%%%%%%%
%
%%%%%%%%%%%%%%%%%%%%%%%%%%%%%%%%%%%%%%%%%%%%%%%

We start defining a model that we will analyze next. We call it the ``AB'' model, defined by the lagrangean 
\beqa
\mathcal{L}_0 &=& |(\partial_{\mu} + i g^{}_B q^{}_B B_{\mu} ) \phi | ^{2} -\frac{1}{4} F_{A}^{2}
-\frac{1}{4} F_{B}^{2}   + \frac{1}{2}( \partial_{\mu} b + 
M_1\ B_{\mu})^{2} -\lambda( |\phi|^{2} - \frac{v^{2}}{2})^{2}   \nonumber\\
&& + \overline{\psi} i \gamma^{\mu} ( \partial_{\mu} +i e A_{\mu}
+ i g^{}_{B} \gamma^{5} B_{\mu}  ) \psi - \lambda_1 \overline{\psi}_L \phi \psi_R 
- \lambda_1 \overline{\psi}^{}_R \phi^{*} \psi^{}_L
\label{lagrangeBC}
\eeqa
and contains a non anomalous ($A$) and an anomalous ($B$) gauge interaction.

\begin{figure}[t]
{\centering \resizebox*{10cm}{!}{\rotatebox{0}
{\includegraphics{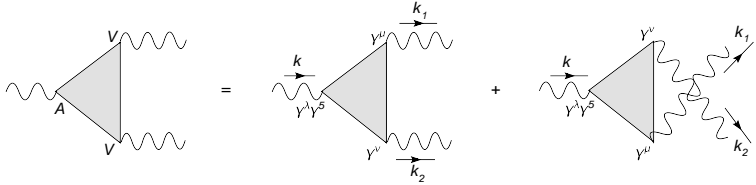}}}\par}
\caption{The ${\bf AVV}$ diagrams}
\label{VAA1}
\end{figure}
%%%%%%%%%%%%%%%%%%%%%%%%%%%%%%%%%%%%%%%%%%%%%%%%
%
%
%%%%%%%%%%%%%%%%%%%%%%%%%%%%%%%%%%%%%%%%%%%%%%
\begin{figure}[t]
{\centering \resizebox*{10cm}{!}{\rotatebox{0}
{\includegraphics{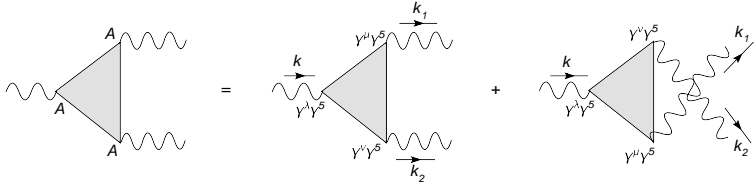}}}\par}
\caption{The ${\bf AAA}$ diagrams}
\label{AAA1}
\end{figure}
%%%%%%%%%%%%%%%%%%%%%%%%%%%%%%%%%%%%%%%%%%%%%%%

Its couplings are summarized in Tables \ref{tab1_AB} and \ref{tab2_AB}, where ``S'' refers to the presence of a St\"{u}ckelberg mass term for the corresponding gauge boson, if present. We have indicated, in this and in the model below, with a small lowercase (i.e. $b$ and $c$) the corresponding axions. The $U(1)_A$ symmetry is unbroken 
while $B$ gets its mass by the combined Higgs-St\"{u}ckelberg mechanism. 
Another feature of the model, as we are going to see, is the presence of 
an Higgs-axion mixing generated not by a scalar potential (such as $V(\phi,b)$), 
as we will show in other examples, but by the fact that both mechanisms communicate their mass 
to the same gauge boson $B$. The axion remains a massless field in this case.

%
%%%%%%%%%%%%%%%%%%%%%%%%%%%%%%%%%%%%%%%%%%%%%%%%%%%%%%%%%%%%%%%%%%%%%%
%%%%%%%%%%%%%%%%%%%% ASSEGNAZIONI PER IL MODELLO AB %%%%%%%%%%%%%%%%%
%%%%%%%%%%%%%%%%%%%%%%%%%%%%%%%%%%%%%%%%%%%%%%%%%%%%%%%%%%%%%%%%%%%%%%
%
\begin{table}[tbh]
\begin{center}
\begin{tabular}{|c|c|c|} \hline  
                         &     A                   &             B                    \\
                 \hline
 $\psi$               & $q^{A}_{L}=q^{A}_{R}=1$   &   $q^{B}_{R}= -q^{B}_{L}=1$       \\
                 \hline
\end{tabular}
\end{center}
\caption{Fermion assignments, A-B Model}
\label{tab1_AB}
%\end{table}
%%%%%%%%%%%%%%%%%%%%
%
%
%%%%%%%%%%%%%%%%%%%%
%\begin{table}[tbh]
\begin{center}
\begin{tabular}{|c|c|c|} \hline  
                   &     $\phi$         &    $S$        \\
                 \hline
 $A$               &   $q^{A}= 0$    &     $0$     \\
                 \hline
  $B$              & $q^{B}= -2 $     &    $b$    \\
                 \hline
\end{tabular}
\end{center}
\caption{Gauge structure, A-B Model}
\label{tab2_AB}
\end{table}

Our discussion relies on the formalism of the 1-loop effective action, which is the generating functional of the one-particle irreducible correlation functions of a given model. The correlators are multiplied by external classical fields and the formalism allows to derive quite directly the anomalous Ward identities of the theory. The reader can find a discussion of the formalism in the appendix, where we 
study the properties of the Chern-Simons and Wess Zumino vertices of the model and their gauge variations.

\begin{figure}[tbh]
{\centering \resizebox*{16cm}{!}{\rotatebox{0}
{\includegraphics{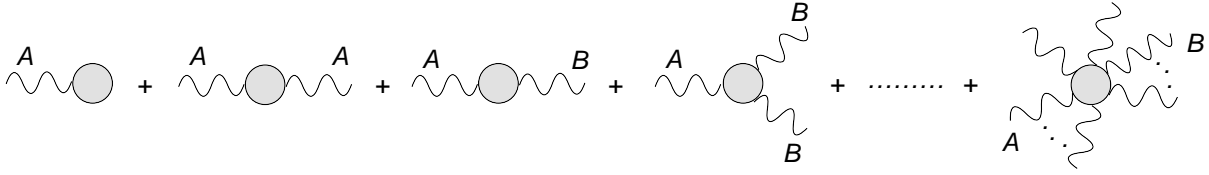}}}\par}
\caption{Expansion of the effective action}
\label{effective11}
\end{figure}

\begin{figure}[tbh]
{\centering \resizebox*{12cm}{!}{\rotatebox{0}
{\includegraphics{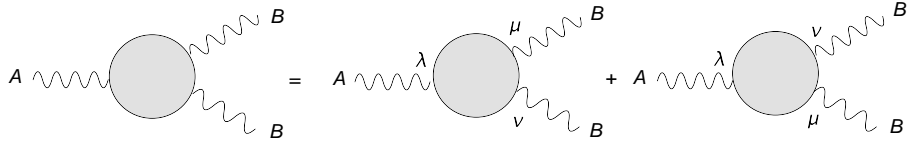}}}\par}
\caption{Triangle diagrams with permutations}
\label{effective22}
\end{figure}

 In the A-B model, this will involve the classical defining action plus the anomaly 
diagrams with fermionic loops and we will require its invariance under gauge transformations.   
The structure of the (total) effective action is summarized, in the case of, say, 
one vector ($A$) and one axial vector ($B$) interaction by an expansion of the form
\beqn
W[A,B] &=& 
 \sum^{\infty}_{n_{1} = 1}  \sum^{\infty}_{n_{2} = 1} \frac{i^{\,n_1 + n_2}}{n_{1}! n_{2}!} \int dx_{1}...dx_{n_{1}}
dy_{1}...dy_{n_{2}} T^{\lambda_{1}...\lambda_{n_{1}} \mu_{1}...\mu_{n_{2}}}(x_{1}...x_{n_1}, y_1...y_{n_2})   \nonumber\\
&& \hspace{3.5cm}  B^{\lambda_1}(x_1)...B^{\lambda_{n_1}}(x_{n_1}) 
A_{\mu_1}(y_1)...A_{\mu_{n_2}}(y_{n_2}),
\eeqn
corresponding to the diagrams in Fig.~\ref{effective11}
%
%%%%%%%%%%%%%%%%%%%%%%%%%%%%%%%%%%%%%%%%%%%%%%%%
%%%%%%%%%%%%%%%%%%%%%%%%%%%%%%%%%%%%%%%%%%%%%%%%
%
where we sum, for each diagram, over the symmetric exchanges of all the indices (including the momentum) of the identical gauge 
bosons (see also Fig.~\ref{effective22}). 
As we are going to discuss next, also higher order diagrams of the form, 
for instance, AVVV will be affected by the presence of an undetermined shift in the triangle 
amplitudes, amounting to Chern-Simons interactions (CS). 
They turn to be well-defined once the distribution of the anomaly on 3-point functions is performed according 
to the correct Bose symmetries of these correlators of lower order. 
% 
%%%%%%%%%%%%%%%%%%%%%%%%%%%%%%%%%%%%%%%%%%%%%%%%
%%%%%%%%%%%%%%%%%%%%%%%%%%%%%%%%%%%%%%%%%%%%%%%%
%
%
%%%%%%%%%%%%%%%%%%%%%%%%%%%%%%%%%%%%%%%%%%%%%%%%
\begin{figure}[tbh]
{\centering \resizebox*{11cm}{!}{\rotatebox{0}
{\includegraphics{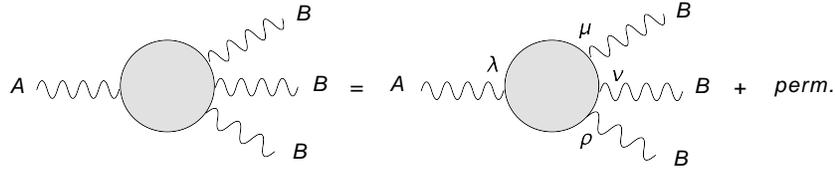}}}\par}
\caption{Symmetric expansion}
\label{effective33}
\end{figure}
%%%%%%%%%%%%%%%%%%%%%%%%%%%%%%%%%%%%%%%%%%%%%%%%
%
%

Computing the variation of the generating functional we obtain 
\beqn
\delta_{B} W^{an} [A,B] &=& \frac{1}{2!} \int dx dy dz \,T^{\lambda \mu \nu}(z,x,y)\, \delta B_{\lambda}(z)
A_{\mu}(x)A_{\nu}(y) \nonumber\\ 
&&+ \frac{1}{3!} \int dx dy dz dw \,T^{\lambda \mu \nu \rho}(z,x,y,w) B_{\lambda}(z)A_{\mu}(x)A_{\nu}(y)A_{\rho}(w),
\eeqn
using $\delta B_{\lambda}(z) = \partial_{\mu} \theta_{B}(z)$ and integrating 
by parts we get 
\beqn
\delta_{B} W^{an} [A,B] &=& -\frac{1}{2!} \int dx dy dz \,\frac{\partial}{\partial z^{\lambda}}T^{\lambda \mu \nu}(z,x,y) 
\,\theta_{B}(z) A_{\mu}(x) A_{\nu}(y) \nonumber\\ 
&&- \frac{1}{3!} \int dx dy dz dw \,\frac{\partial}{\partial z^{\lambda}} T^{\lambda \mu \nu \rho}(z,x,y,w) \,\theta_{B}(z) 
A_{\mu}(x) A_{\nu}(y) A_{\rho}(w). \nonumber \\
\eeqn
Notice that in configuration space the 4- and the 3- point function correlators 
are related by 
\beqn
\frac{\partial}{\partial z^{\lambda}} \overline{T}^{\lambda \mu \nu \rho}(z,x,y,w) = \delta(z-w) \overline{T}^{\rho \mu \nu}(w,x,y,) 
- \delta(z- x) \overline{T}^{\mu \nu \rho}(x,y,w) \mbox{+ perm.}
\eeqn
For $T$ we will be using the same conventions as for $\Delta$, with 
$\overline{T}^{\lambda \mu \nu}$ indicating a single diagram with non-permuted external gauge lines, while 
$T^{\lambda \mu \nu}$ will denote the symmetrized one (in $\mu \nu$).
Clearly, the 1-loop effective theory of this model contains anomalous interactions that need to be cured by the introduction 
of suitable compensator fields. The role of the axions ($b$, for instance) is to remove the anomalies associated to the triangle 
diagrams which are correlators of 1 and 3 chiral currents respectively 

\beq
T^{(\bf AVV)}_{\la\mu\nu}(x,y,z)= \langle 0| T\left( J_\mu(x) J_\nu(y) J^5_\la(z)\right) |0\rangle
\eeq
and 
\beq
T^{(\bf AAA)}_{\la\mu\nu}(x,y,z)= \langle 0| T\left( J^5_\mu(x) J^5_\nu(y) J^5_\la(z)\right) |0\rangle,
\eeq

where
\beq 
J_\mu=-\overline{\psi}\gamma_\mu \psi \qquad J^5_\mu=-\overline{\psi}\gamma_\mu \gamma_5 \psi.
\eeq
We denote by $\Delta(k_1,k_2)$ and $\Delta_3(k_1,k_2)$ their corresponding expressions 
in momentum space 
\beq
(2 \pi)^4 \delta(k - k_1 - k_2) 
\Delta^{\la\mu\nu}(k_1,k_2)=\int dx \,dy \,dz \,e^{ i k_1\cdot x + i k_2\cdot y - i k\cdot z} \, T^{\la\mu\nu}_{(\bf AVV)}(x,y,z)
\label{dd1}
\eeq
\beq
(2 \pi)^4 \delta(k - k_1 - k_2) 
\Delta_3^{\la\mu\nu}(k_1,k_2)=\int dx \,dy \,dz \, e^{ i k_1\cdot x + i k_2\cdot y - i k\cdot z} \, T^{\la\mu\nu}_{(\bf AAA)}(x,y,z).
\label{dd2}
\eeq

Another point to remark is that the invariant amplitudes linear in momenta 
in the definition of the {\bf AVV }trangle diagram correspond, in configuration space, 
to Chern-Simons interactions. In momentum space these are proportional to
\beq
V_{CS}^{\la\mu\nu}(k_1,k_2)=- i\epsilon^{\la\mu\nu\sigma}(k_1^\sigma - k_2^\sigma)
\eeq
and we denote with $T_{CS}^{\la\mu\nu}$ the corresponding contribution to the effective lagrangian 
in Minkowski space (see the appendices)
%\beqn
%\int dx \, dy \, dz\, T_{CS}^{\la\mu\nu}(z,x,y) \, B^{\lambda}(z) A^{\mu}(x) A^{\nu}(y)   
\beq
\mathcal{L}_{CS, ABA}=\int dx A^{\mu}(x)B^{\nu}(x) F_A^{\rho\sigma}(x) \varepsilon_{\mu\nu\rho\sigma}.
\eeq
%with 
%\beq
%T_{CS}^{\la\mu\nu}(z,x,y)=\delta(x-z)\delta(y-z) 
%\eeq

Moving to the anomalous part of the effective action, this takes the form, for generic gauge bosons $A_i $
\beq
\mathcal{S}_{eff}=\mathcal{S}_0 + \mathcal{S}_1,
\eeq
where 
\beq
\mathcal{S}_1= \sum_{i\, j\, k}\frac{1}{n_i! n_j! n_k!} g_{i j k}\int dx dy dz A_i^\la(x)A_j^\mu(y)A_k^\nu(z)  T^{\la\mu\nu}_{A_i A_j A_k}(x,y,z),
\label{Seff}
\eeq
and where $A_i$ indicates an $A$ or a $B$ gauge boson, while 
$g_{ijk}$ is a short-hand notation for the product of the 3 coupling constants $g_{A_i} g_{A_j} g_{A_k}$, with an additional normalization due to a counting of identical external gauge bosons $(n_i!)$. All the anomalous contributions are included in the definition.
In order to derive its explicit structure in our simplified cases, we consider the case of the $BAA$ vertex, the other examples being similar. We have, for instance, the partial contribution 
\beq
\mathcal{S}^{BAA}= g_B\int d x d y d z B_{\la}(z)\langle J_5^\la (z) e^{i g_A \int d^4 x J^\mu (x) A_\mu (x)}\rangle, 
\eeq
where the gauge fields are treated as classical fields and $\langle,\rangle$ indicate the vacuum expectation value. 
Expanding to second order, we keep only the connected contributions obtaining for instance
\beq
\mathcal{S}^{BAA}= \frac{i^2}{2!} g_B g_A^2\int d x d  y d z B_{\la}(z)A_\mu(x) A_\nu(y) \langle J_5^\la (z) J^\mu(x) J^\nu(y)\rangle.
\eeq
This expresssion is our starting point for all the 
further analysis. Most of the manipulations concerning the proof of gauge-invariance of the effective 
action are more easily worked out in this formalism. Moving from momentum space to configuration space and back, may also be quite useful in order to detail 
the Ward identities of a given anomalous effective action. 

In the presence of 
spontaneous symmetry breaking and of St\"{u}ckelberg mass terms one has to 
decide whether the linear mixing between the St\"{u}ckelberg field and the 
gauge boson is kept or not. 
One can keep the mixing and derive ordinary Ward identities for a given model. 
This is a possibility which is clearly at hand and can be useful. The 
disadvantage of this approach is that there is no gauge fixing parameter that can be used to analyze the gauge dependence of a given set of 
amplitudes and their cancelation. When the mixing is removed by going to a 
$R_\xi $ gauge, one can identify the set of gauge-invariant 
contributions to a given amplitude and identify more easily the conditions under which a given model becomes unitary. We follow this second approach. We are then able to combine gauge-dependent contributions in such a way that the unphysical poles of 
a given amplitude cancel. The analysis that we perform is limited to the s-channel, but the results are easily generalizable to the $t$ and $u$ channels as well. From this simple analysis one can easily extract information on the 
perturbative expansion of the effective action.

%%%%%%%%%%%%%%%%%%%%%%%%%%%%%%%%%%%%%%%%%%%%%%%%%%%%%%%%%%%%%%%%%%%%%%%%%%%%%
\subsection{ The anomalous effective Action of the $A-B$ Model }
%%%%%%%%%%%%%%%%%%%%%%%%%%%%%%%%%%%%%%%%%%%%%%%%%%%%%%%%%%%%%%%%%%%%%%%%%%%%%
%
We start from the simplest model. 
In the $A-B$ model, defined in  Eq.~(\ref{lagrangeBC}), the contribution to the anomalous effective action is given by 
\beqa
\mathcal{S}_{an} &=& \mathcal{S}_1 + \mathcal{S}_3 \nonumber \\
\mathcal{S}_{1} &=&  \int d x \, d y \, d z \, \left( \frac{g^{}_{B} \, g^{2}_{A}}{2!} \, 
T_{\bf AVV}^{\la\mu\nu}(x,y,z) B_{\la}(z) A_{\mu}(x)A_{\nu}(y)
 \right)\nonumber \\
\mathcal{S}_3 &=&  \int d x \, d y \, d z \, \left( \frac{ g^{3}_{B}}{3!} \, 
 T_{\bf AAA}^{\la\mu\nu}(x,y,z) B_\la(z) B_\mu(x) B_\nu(y) \right), 
\nonumber \\
\eeqa
where we have collected all the anomalous diagrams of the form ({\bf AVV} and {\bf AAA}). We can easily express the 
gauge transformations of 
$A$ and $B$ in the form
\beqa
 \frac{1}{2!} \delta_B  \langle T_{\bf AVV} BAA \rangle &=&  \frac{i}{2!} a_3(\beta) \frac{1}{4} \langle F_A \wedge F_A  
\theta_B \rangle \nonumber\\  
 \frac{1}{3!} \delta_B\langle T_{\bf AAA} BBB \rangle &=&  \frac{i}{3!} \frac{a_n}{3} \frac{3}{4} \langle  F_B \wedge F_B  
\theta_B \rangle,
\eeqa
where we have left open the choice over the parameterization of the loop momentum, 
denoted by the presence of the arbitrary parameter $\beta$ with
\beq
a_3(\beta)=- \frac{i}{4 \pi^2} + \frac{i}{2 \pi^2} \beta\qquad a_3\equiv \frac{a_n}{3}=-\frac{i}{6 \pi^2},
\eeq
while
\beqa
\frac{1}{2!}\delta_A\langle T_{\bf AVV} BAA\rangle =  \frac{i}{2!}  a_1(\beta) \frac{2}{4} \langle F_B \wedge F_A 
\theta_A \rangle. 
\eeqa
We have the following equations for the anomalous variations of the effective lagrangean 
\beqa
\delta_B \mathcal{L}_{an} &=&  \frac{i g^{}_{B} g^{\,2}_{A}}{2!} \, a_3(\beta) \frac{1}{4}  F_A \wedge F_A  \theta_B 
+  \frac{ i g^{\,3}_{B} }{3!}\, \frac{a_n}{3}  \frac{3}{4} F_B \wedge F_B \theta_B 
\nonumber \\
\delta_A \mathcal{L}_{an} &=&  \frac{i g^{}_{B} g^{\,2}_{A}}{2! } \, a_1(\beta) \frac{2}{4} F_B \wedge F_A \theta_A, 
\eeqa
while $\mathcal{L}_{b,c}$, the axionic contributions (Wess-Zumino terms), needed to 
restore the gauge symmetry violated at 1-loop level, are given by 
\beq
\mathcal{L}_{b} =  \frac{C^{}_{AA}}{M} b \, F_A \wedge F_A  + \frac{C^{}_{BB}}{M} b \, F_B \wedge F_B.  
\eeq
Notice that since the axion shifts only under a gauge variation of the anomalous U(1) gauge field B (and not under A), gauge invariance of the effective 
action under a gauge transformation of the gauge field $A$ requires that
\beq
\delta_A \mathcal{L}_{an}=0.
\eeq
Clearly, this condition fixes $\beta= -1/2\equiv \beta_0$ and is equivalent to the CVC condition on A that we had relaxed at 
the beginning. 
Imposing gauge invariance under B gauge transformations, on the other hand, we obtain 
\beqa
\delta_B \left(  \mathcal{L}_b + \mathcal{L}_{an}  \right) = 0 \nonumber \\
\eeqa
which implies
\beq
C^{}_{AA} =  \frac{i \,g^{}_{B}  g^{\,2}_{A} }{2!} \frac{1}{4} \, a_3(\beta_0) \, \frac{M}{M_1}, 
\qquad C^{}_{BB} =  \frac{i g^{\,3}_{B}}{3!} \frac{1}{4} \, a_n \, \frac{M}{M_1}.
\eeq
These conditions on the coefficients $C$ are sufficient to render gauge-invariant the total lagrangean. We observe that the presence of an abelian symmetry which has 
to remain exact and is not accompanied by a shifting axion has important 
implications on the consistency of the theory. We have brought up this example 
because in more complex situations in which a given gauge symmetry is broken 
and the pattern of breakings is such to preserve a final symmetry (for instance 
$U(1)_{em}$), the structure of the anomalous correlators, in some case, is drastically 
constrained to assume the CVC form. However this is not a general result. 

Under a more general assumption, we could have allowed some Chern-Simons contributions in the counterterm lagrangean. This is an interesting variation 
that can be worked out at a diagrammatic level in order to identify the role 
played by the CS interactions. We will get back to this point once we start our diagrammatic analysis of these simple models.
%

%%%%%%%%%%%%%%%%%%%%%%%%%%%%%%%%%%%%%%%%%%%%%%%%%%%%%%%%%%%%%%%%%%%%%%%%%%%
\section{Higgs-Axion Mixing in $U(1)$ Models: massless  axi-Higgs}
%%%%%%%%%%%%%%%%%%%%%%%%%%%%%%%%%%%%%%%%%%%%%%%%%%%%%%%%%%%%%%%%%%%%%%%%%%%
Having discussed how to render consistent to all orders the effective 
action, we need to discuss the role played by the shifting axions in the spectrum of the theory. We have already pointed out that the axion will mix with the 
remaining scalars of the model. In the presence of a Higgs sector such a mixing can take place at the level of the 
scalar potential, with drastic implications on the mass and the coupling of the axion to the remaining particles of 
the model. Naturally, one would like to 
understand the way the mixing occurs and this is exemplified in the case 
of the $AB$ model.

This model has two scalars: the Higgs and the St\"uckelberg fields. We assume that the Higgs field takes a non-zero vev
and, as usual, the scalar field is expanded around the minimum $v$
\beq
\phi = \frac{1}{\sqrt{2}}\left( v + \phi^{}_{1} + i \phi^{}_{2} \right), 
\eeq
while from the quadratic part of the lagrangean we can easily read out the mass terms and the 
goldstone modes present in the spectrum in the broken phase. 
This is given by 
\beqa
\mathcal{L}_q &=& \frac{1}{2} \left(\partial_\mu \phi_{1}\right)^2 +
\frac{1}{2} \left(\partial_\mu \phi_{2}\right)^2 + \frac{1}{2}\left(\partial_\mu b\right)^2  
+ \frac{1}{2}\left(M_1^2 + (q^{}_{B} g^{}_{B} v)^2\right) B_\mu B^{\mu} 
 - \frac{1}{2} m_{1}^2 \phi_{1}^2   \nonumber\\
&& + B_\mu \partial^\mu \left(M_1  b +  v g^{}_{B} q^{}_{B} \phi^{}_{2} \right),   
\eeqa
from which, after diagonalization of the mass terms we obtain 
\beqa
\mathcal{L}_q &=&\frac{1}{2} \left(\partial_\mu \chi^{}_{B}\right)^2 +
\frac{1}{2} \left(\partial_\mu G^{}_{B}\right)^2  + \frac{1}{2}\left(\partial_\mu h^{}_{1} \right)^2  
 + \frac{1}{2} M_B^2 B_\mu B^{\mu} - \frac{1}{2} m_{1}^2 h_{1}^2     \nonumber\\
&& + M_B B^\mu\partial_\mu G^{}_{B} 
\eeqa
where we have redefined $\phi^{}_{1}(x) = h^{}_{1}(x)$ and $m^{}_{1}=v \sqrt{2 \lambda}$, for the Higgs field and its 
mass. We have identified the linear combinations 
\beqa
\chi^{}_{B} &=& \frac{1}{M_B} \left(- M^{}_1 \, \phi^{}_{2} + q^{}_{B} g^{}_{B} v \, b\right),   \nonumber\\
G^{}_{B} &=& \frac{1}{M_B}\left(q^{}_{B} g^{}_{B} v \, \phi^{}_{2} + M^{}_1 \, b\right), 
\eeqa
corresponding to a massless particle, the axi-Higgs $\chi^{}_{B}$, and a massless goldstone mode $G^{}_{B}$. 
The rotation matrix that allows the change of variables $(\phi^{}_{2},b) \to (\chi^{}_{B},G^{}_{B})$ is given by 
%
%%%%%%%%%%%%%%%%%%%%%%%%
\beq
U=\left(
\begin{array}{ll}
 -\cos \theta^{}_{B} & \sin \theta^{}_{B} \\
 \sin  \theta^{}_{B} & \cos \theta^{}_{B}
\end{array}
\right)
\eeq
%%%%%%%%%%%%%%%%%%%%%%%
%
with $\theta^{}_{B}={\arccos} ({M_1/M_B})={\arcsin}(q^{}_{B} g^{}_{B} v/ M_B)$. 
The axion $b$ can be expressed as linear combination of the rotated fields $\chi^{}_{B}$,$G^{}_{B}$ as 
\beqa
b = \alpha_1 \chi^{}_{B} + \alpha_2 G^{}_{B} = \frac{q^{}_{B} g^{}_{B} v}{M_B} \chi^{}_{B} + \frac{M_1}{M_B} G^{}_{B}, 
\label{projection}
\eeqa
while the gauge fields $B^{}_\mu$ get its mass $M^{}_B$ through the combined Higgs-St\"{u}ckelberg 
mechanism  
\beq
M_B=\sqrt{M_1^2 + (q^{}_{B} g^{}_{B} v)^2}.
\eeq
 
To remove the mixing between the gauge fields and the goldstones we work in the $R_\xi$ gauge. 
The gauge-fixing lagrangean is given by 
\beq
\mathcal{L}_{gf}= -\frac{1}{2}\mathcal{G}_B^2 
\eeq
where
\beq
\mathcal{G}_B=\frac{1}{\sqrt{\xi_B}}\left( \partial\cdot B - \xi_B M_B G^{}_{B}\right), 
\eeq
and the corresponding ghost lagrangeans 
\beqa
\mathcal{L}_{B\,gh}&=& \overline{c^{}_B}\left( - \square - \xi^{}_B  v^{}_{u} (h^{}_{1} + v^{}_{u}) - \xi^{}_B M_1^2\right) c^{}_B. 
\eeqa
For convenience we report the form of the full lagrangean in the physical basis for future reference. 
After diagonalization of the mass matrix this becomes 
\beqa
\mathcal{L}&=& -\frac{1}{4} F_A^2 -\frac{1}{4} F_B^2  + 
\mathcal{L}_{B gh} + \mathcal{L}_{f} + \mathcal{L}_{B}
\eeqa
where 
\beqa
\mathcal{L}_B&=& \frac{1}{2} \left(\partial_\mu \chi\right)^2
 -\frac{1}{2 \xi_B}(\partial \cdot B)^2 
  +\frac{1}{2}
   (\partial_\mu G_B)^2+\frac{1}{2} (\partial_{\mu} h_1)^2 -\frac{1}{2} m_{1}^2
   h_1^2 +\frac{1}{2}
   M_B^2 B_\mu ^2 - 4  \frac{v  g_B^2}{M_B}B_\mu  G_B \partial^{\mu} h_1\nonumber \\
&&
-\frac{4 \lambda v^4  g_B^4}{ M_B^4}G_{B}^4 + \frac{8 v^2  g_B^4}{ M_B^2}(B_\mu) ^2
   G_B^2 + \frac{8 \lambda M_1 v^3
    g_B^3}{M_B^4}\chi^{}_{B} G_B^3 - \frac{8 M_1 v  g_B^3}{M_B^2}(B_\mu) ^2
   \chi^{}_B G_B   
   \nonumber \\
&&
-\frac{ 4 g_B^2 \la  v^3}{M_B^2}G_B^2 h_1 + 4 g_B^2
   (B_\mu)^2 h^{}_{1} v + 2 \frac{g_B^2 M_1^2}{ M_B^2}(B_\mu)^2
   \chi^2 + 2 g_B^2 (B_\mu)^2
   h_1^2 \frac{v  g_B^2}{M_B}B_\mu  h_1 \partial^\mu G_B \nonumber \\
&& +\frac{ 2 \lambda M_1
   v  g^{}_B}{M_B^2} \chi G_B h_1^2 +  \frac{ 2 g^{}_B \lambda M_1^3 v }{M_B^4}G_B
   \chi^3 +  \frac{4  g_B \lambda M_1 v^2
   }{M_B^2}G_B h_1 \chi^{}_B -\frac{2 g_B
   M_1 }{M_B }B^\mu \partial_\mu\chi  h_1 \nonumber \\
&&-\frac{\lambda M_1^4 }{4
   M_B^4}\chi^4 +  \frac{2 g_B M_1 }{M_B}B^\mu  \partial_{\mu} h_1
   \chi^{}_B -    \frac{1}{4} \lambda
   h_1^4 - \lambda v h_1^3 + \frac{3 \lambda  M_1^4}{2
   M_B^4} \chi^2 G_B^2 -\frac{3 \lambda 
   M_1^2}{2 M_B^2}\chi^2 G_B^2 \nonumber \\
&&-\frac{1}{2} \lambda h_1^2 G_B^2-\frac{1}{2}
   M_B^2 \xi_B G_B^2-\frac{\lambda M_1^2 }{2 M_B^2}\chi^2
   h_1^2 + \frac{\lambda M_1^2 }{2
   M_B^2}G_B^2 h_1^2 -\frac{\lambda M_1^2 v }{M_B^2}\chi^2
   h_1 \nonumber \\
\eeqa
where $\mathcal{L}_f$ denotes the fermion contribution. 

At this stage there are some observations to be made. 
In the St\"uckelberg phase the axion $b$ is a goldstone mode, since it can be set to vanish by a gauge transformation on the B gauge boson, while $B$ is massive (with a mass $M_1$) and has 3 degrees of freedom 
(dof). Therefore in this phase the number of physical dof's is  
3 for $B$, 2 for $A$, 2 for the complex scalar Higgs $\phi$, for a total of 7.
After electroweak symmetry breaking we have 3 d.o.f.'s for $B$, 2 for $A$ which remains massless in this model, 1 real Higgs field $h^{}_{1}$ and 1 physical axion $\chi$, for a total of 7.
The axion, in this case, on the contrary of what happens in the case of ordinary 
symmetry breaking is a {\em massless physical } scalar, being not part of the scalar potential. 
Not much surprise so far. Let's now move to the analysis of the case when the axion is part of the 
scalar potential. In this second case the physical axion (the axi-Higgs) gets its mass by the combined 
Higgs-St\"uckelberg mechanisms and shows some interesting features.  

%
%%%%%%%%%%%%%%%%%%%%%%%%%%%%%%%%%%%%%%%%%%%%%%%%%%%%%%%%%%%%%%%%%%%%%%%%%%%
\subsection{Higgs-Axion Mixing in $U(1)$ Models: massive  axi-Higgs}
%%%%%%%%%%%%%%%%%%%%%%%%%%%%%%%%%%%%%%%%%%%%%%%%%%%%%%%%%%%%%%%%%%%%%%%%%%%
We now illustrate the mechanism of mass generation for the physical axion $\chi$. We focus on the breaking of 
the $U(1)^{}_{B}$ gauge symmetry of the $AB$ model. We have a gauge-invariant Higgs potential given by 
\beqn
V_{PQ} = \mu^{2} \phi^{*} \phi + \lambda \left( \phi^{*} \phi \right)^2
\eeqn
plus the new PQ-breaking terms, allowed by the symmetry \cite{cik}
\beqn
V_{\slash P \slash Q} = b_{1} \left( \phi \, e^{- i q^{}_{B} g^{}_{B} \frac{b}{M^{}_{1}}}   \right) 
+ \lambda_1 \left( \phi \, e^{- i q^{}_{B} g^{}_{B} \frac{b}{M^{}_1} }   \right)^{2} 
+2 \lambda_{2} \left(  \phi^{*} \phi  \right)   \left( \phi  \, e^{- i q^{}_B g^{}_{B} \frac{b}{M^{}_1} } \right)    + \mbox{c.c.}
\eeqn
so that the complete potential considered is given by
\beqn
V(H, b) = V_{PQ} +  V_{\slash P \slash Q} + V^{*}_{\slash P \slash Q}.
\label{ppqq}
\eeqn 
We require that the minima of the potential are located at 
\beq
\langle b \rangle=0 \qquad \langle \phi \rangle =v,
\eeq
which imply that the mass parameter satisfies
\beqn
\mu^2 =   - \frac{b_1}{v}  -  2 v^2  \lambda - 2  \lambda_1 - 6 v \lambda_2.
\eeqn
We are interested in the matrix describing the mixing of the CP-odd Higgs sector with the axion field $b$, given by
\beqn
\left( \,\,\, \phi_2, \,\,\,  b \,\,\,  \right)  {\mathcal M_{2} }
\left(   \begin{array}{c}
 \phi_2\\
b
 \end{array}   \right)
\eeqn
where $\mathcal M_{2}$ is a symmetric matrix
 \beqn
 \mathcal M_{2}  = -\frac{1}{2} c_{\chi} v^2 \pmatrix{ 1    &  - v \frac{q^{}_B g^{}_{B}}{M^{}_1}  \cr
                                                     - v \frac{q^{}_B g^{}_{B}}{M^{}_1}  &  v^2 \frac{q^2_B g^{2}_{B}}{M^2_1}  \cr}
 \eeqn
and where the dimensionless coefficient multiplied in front is given by
\beqn
c_{\chi} = 4 \left( \frac{b_1}{v^3} + \frac{4 \lambda_1}{v^2} + \frac{2 \lambda_2}{v} \right).
\label{chimass}
\eeqn
Notice that this parameter plays an important role in establishing the size of the mass of the physical axion, after diagonalization. It encloses all the dependence of the mass from the 
PQ corrections to the standard Higgs potential. They can be regarded as corrections 
of order $p/v$, with $p$ being any parameter of the PQ potential. If $p$ is very small, which is the case if 
the $V_{\slash{P} \slash{Q}}$ term of the potential is generated non-perturbatively
(for instance by instanton effects in the case of QCD), the mass of the axi-Higgs 
can be pushed far below the typical mass of the electroweak 
breaking scenario (the Higgs mass), as discussed in \cite{CorianoIrges}. 

The mass matrix has 1 zero eigenvalue corresponding to the goldstone boson G and 1 non-zero eigenvalue corresponding to a physical 
axion field $- \chi -$ with mass
\beqn
m^2_{\chi} =  - \frac{1}{2} c_{\chi} v^2 \left[  1 + \frac{q^2_B g^{2}_{B} v^2}{M^2_1} \right] = - \frac{1}{2} c_{\chi} \, v^2 \, 
\frac{M^{2}_{B}}{M^{2}_{1}}. 
\eeqn 
The mass of this state is positive if $c_{\chi} < 0$.
The rotation matrix that takes from the interaction eigenstates to the mass eigenstates is denoted by 
$O^{\chi}$
\beqn
\left(   \begin{array}{c}
    \chi     \\
   G
 \end{array}   \right) =   O^{\chi}  \left(   \begin{array}{c}
    \phi_{2}     \\
      b
 \end{array}   \right) 
\eeqn
so that we obtain the rotations
\beqn
\phi_2  &=& \frac{1}{M_B}  (- M^{}_1 \, \chi + q^{}_B g^{}_{B} \, v \, G   )   \\
b &=& \frac{1}{M_B} (  q^{}_B g^{}_{B} \, v \,\chi  +  M^{}_1 \, G   ).   
\eeqn
The mass squared matrix can be diagonalized as
\beqn
(\,\, \chi , \,G \,\,)  O^{\chi} \, {\mathcal M_2} (O^{\chi})^T  \left(   \begin{array}{c}
    \chi     \\
   G
 \end{array}   \right)   = (\,\, \chi,\, G\,\,)       \pmatrix{   m^2_\chi   &   0     \cr
                                                       0   &   0  \cr}   \left(   \begin{array}{c}
    \chi     \\
   G
 \end{array}   \right) 
\eeqn
so that G is a massless goldstone mode and $m_\chi$ is the mass of the physical axion. In 
\cite{CorianoIrges} one can find a discussion of some physical implications of this field when its mass 
is driven to be small in the instanton vacuum, similarly to the Peccei-Quinn axion 
of a global symmetry. However, given the presence of both mechanisms, the St\"uckelberg and the Higgs, 
 it is not possible to decide whether this axion can be a valid dark-matter candidate. In the same work 
it is shown that the entire St\"uckelberg mechanism can be the result of a partial decoupling of 
a chiral fermion.

%%%%%%%%%%%%%%%%%%%%%%%%%%%%%%%%%%%%%%%%%%%%%%%%%%%%%%%%%%%%%%%%%%%%%%%
\section{Unitarity  issues in the A-B model in  the exact phase} 
%%%%%%%%%%%%%%%%%%%%%%%%%%%%%%%%%%%%%%%%%%%%%%%%%%%%%%%%%%%%%%%%%%%%%%%
In this section we start discussing the issue of unitarity of the model that we have introduced. This is a rather involved 
topic that can be addressed by a diagrammatic analysis of those Feynman amplitudes with s-channel exchanges of gauge particles, 
the axi-Higgs and the 
NG modes, generated in the various phases of the theory 
(before and after symmetry breaking, with/without Yukawa couplings). The analysis could, of course, be repeated in the other channels (t,u) as well, but no further condition would be 
obtained. 
We will gather all the information coming from the study of the S-matrix amplitudes  
to set constraints on the parameters of the model. We have organized our analysis 
as a case-by-case study verifying the cancelation of the unphysical singularities in the amplitude 
in all the phases of the theory, establishing also their gauge independence. This is worked out in the 
$R_\xi$ gauge so to make evident the disappearance of the gauge-fixing parameter in each amplitude. 
The scattering amplitudes are built out of two anomalous diagrams with s-channel exchanges of gauge dependent propagators, and in all the cases we are brought back to the analysis of a set of anomalous 
Ward identities to establish our results.
%
%%%%%%%%%%%%%%%%%%%%%%%%%%%%%%%%%%%%%%%%%%%%%%%%%%%%%%%%%%%%%%%%%
\subsection{Unitarity and CS interactions in the $A-B$ model}
%%%%%%%%%%%%%%%%%%%%%%%%%%%%%%%%%%%%%%%%%%%%%%%%%%%%%%%%%%%%%%%%%
%
The first point that we address in this section concerns the role played by the CS interactions 
in the unitarity analysis of simple s-channel amplitudes. This analysis clarifies that CS interactions can be 
included or kept separately from the anomalous vertices with no consequence. 
To show this,  we consider the following modification on the $AB$ model, where the CS interactions are generically introduced as possible counterterms in the 1-loop effective action, which is given by 

\beqn
{\mathcal L} = {\mathcal L_0} + {\mathcal L_{GS}} + {\mathcal L_{CS}},
\eeqn
where ${\mathcal L_0}$ is already known from previous sections, but in particular we focus on the components
\beqn
{\mathcal L_{GS}} =  \frac{C^{}_{AA}}{M} b F^{}_{A} \wedge F^{}_{A} +\frac{C^{}_{BB}}{M} b F^{}_{B} \wedge F^{}_{B} 
\eeqn
and
\beqn
{\mathcal L_{CS}} =  d^{}_{1}B^{\mu} A^{\nu} F^{\rho \sigma}_{A} \epsilon_{\mu \nu \rho \sigma}
\equiv d^{}_{1}BA\wedge F^{A}.
\eeqn
Under an $A$-gauge transformation we have \footnote{In the language of the effective action the multiplicity factors are proportional to the number $(n!)$ of external gauge lines of a given type. We keep these factors explicitly to backtrack their origin.}
\beqn
\delta^{}_{A} {\mathcal L} = \frac{d^{}_{1}}{2}\, \theta^{}_{A}  F^{}_{B} \wedge F^{}_{A} 
+ \frac{i}{2!} a^{}_{1}(\beta) \frac{2}{4} \, \theta^{}_{A}  F^{}_{B} \wedge F^{}_{A}, 
\eeqn
so that we obtain
\beqn
\frac{d^{}_{1}}{2} + \frac{i a^{}_{1}(\beta)}{4} = 0 \,\,\,\,\,\leftrightarrow \,\,\,\,\, d^{}_{1}= -\frac{i}{2} a^{}_{1}(\beta).
\eeqn
Analogously, under a $B$-gauge transformation we have 
\beqn
&&\delta^{}_{B} {\mathcal L} = -\frac{d^{}_{1}}{2} \theta^{}_{B}  F^{}_{A} \wedge F^{}_{A} 
+ \frac{i}{2!} a^{}_{3}(\beta) \frac{1}{4}  F^{}_{A} \wedge F^{}_{A} \theta^{}_{B} 
- C^{}_{AA} \frac{ M^{}_{1} \theta^{}_{B} }{M}\,  F^{}_{A} \wedge F^{}_{A}  \nonumber\\
&&\,\,\,\,\,\,\,\,\,\,\,\,\,\,\,\,\,\,- C^{}_{BB} \frac{ M^{}_{1} \theta^{}_{B} }{M} F^{}_{B} \wedge F^{}_{B} 
+ \frac{i}{3!} \frac{a^{}_{n}}{3} \frac{3}{4} \theta^{}_{B} F^{}_{B} \wedge F^{}_{B}, 
\eeqn
to obtain
%
%\beqn
%- C^{}_{BB} \frac{ M^{}_{1} \theta^{}_{B} }{M} F^{}_{B} \wedge F^{}_{B} +  
% \frac{i}{3!} \frac{a^{}_{n}}{3} \frac{3}{4} \theta^{}_{B} F^{}_{B} \wedge F^{}_{B} = 0,
%\eeqn
%
\beqn
-\frac{d^{}_{1}}{2} - C^{}_{AA} \frac{M^{}_{1}}{M} + \frac{i}{2!} a^{}_{3}(\beta) \frac{1}{4} = 0 \,\,\,\,\,
\leftrightarrow  \,\,\,\,\,
C^{}_{AA}= \left( - \frac{d^{}_{1}}{2} + \frac{i}{2!}a^{}_{3}(\beta)\frac{1}{4}  \right)\frac{M}{M^{}_{1}}.
\label{GSexplicit}
\eeqn
%
%
%%%%%%%%%%%%%%%%%%%%%%%%%%%%%%%%%%%%%%
\begin{figure}[t]
{\centering \resizebox*{14cm}{!}{\rotatebox{0}
{\includegraphics{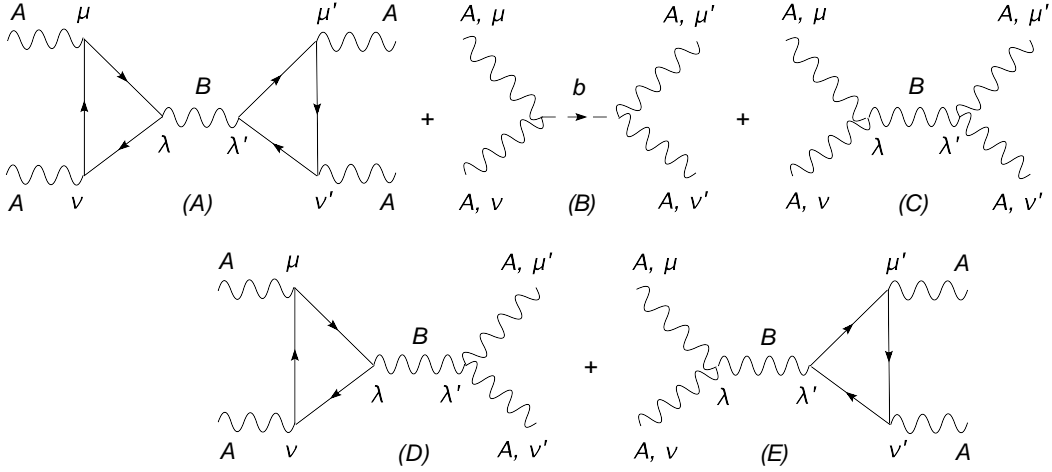}}}\par}
\caption{Diagrams involved in the unitarity analysis with external CS interactions.}
\label{chern}
\end{figure}
%%%%%%%%%%%%%%%%%%%%%%%%%%%%%%%%%%%%%%
%
We refer the reader to one of the appendices where the computation is performed in detail.

We have shown that the presence of external Ward identities forcing the 
invariant amplitudes of a given anomalous triangle diagram to assume a specific form, allow to 
re-absorb the CS coefficients inside the triangle, thereby simplifyig the computations. In this specific case the CVC condition for $A$ is a property of the theory. In other cases this does 
not take place. For instance, instead of the condition $a_1(\beta)=0$, a less familiar condition such as $a_3(\beta)=0$ (conserved axial current, CAC) may be needed. In this sense, 
if we define the CVC condition to be the ``standard case'', the CAC condition points toward a {\em new} anomalous interaction. We remark once more that $\beta$ remains ``free'' in the 
SM, since the anomaly traces cancel for all the generators, differently 
from this new situation. The theory allows new CS interactions, with the understanding 
that, at least in these cases, these interactions can be absorbed into a redefinition 
of the vertex. However, the presence of a Ward identity, that allows us to re-express $a_1$ and $a_2$ in terms of $a_3\dots a_6$ in {\em different} 
ways, at the same time allows us to come up with different gauge invariant 
expressions of the same vertex function (fixed by a CVC or a CAC condition, depending on the case). These different versions of these 
{\bf AVV} 3-point functions are characterized by different (gauge-variant) contact 
interactions since $a_1$ and $a_2$ in Minkowski space contain, indeed, CS 
interactions. We will elaborate on this point in a following section where we discuss the structure of the effective action in Minkowski space.
 
\begin{figure}[t]
 {\centering \resizebox*{15cm}{!}{\rotatebox{0}
 {\includegraphics{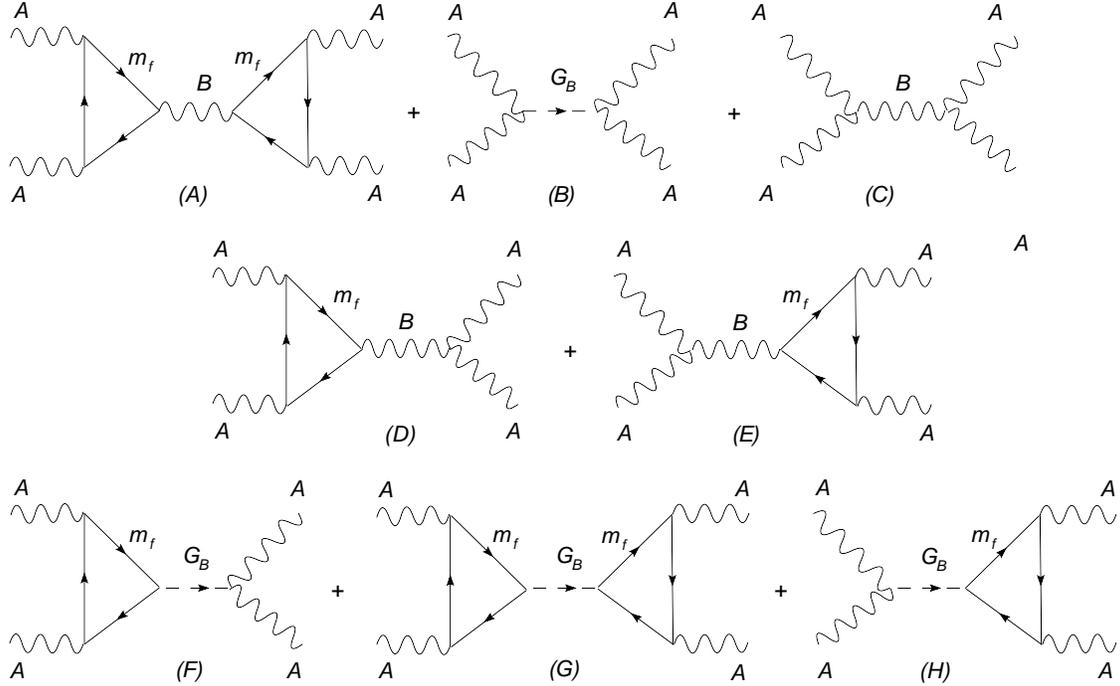}}}\par}
 \caption{Unitarity issue for the AB model in the broken phase.}
 \label{chernmassive}
 \end{figure}

The extension of this pattern to the broken Higgs phase can be understood 
from Fig.~\ref{chernmassive} where the additional contributions have been 
explicitly included. We have 
depicted the CS terms as separate contributions and shown perfect cancelation also in this case. 

%
%%%%%%%%%%%%%%%%%%%%%%%%%%%%%
%%%%%%%%%%%%%%%%%%%%%%%%%%%%%
%
The complete set of diagrams is shown in Fig.~(\ref{chernmassive})
\beq
S^{}_{\xi} = A^{}_{\xi} + B^{}_{\xi} + C^{}_{\xi} + D^{}_{\xi} + E^{}_{\xi} + F^{}_{\xi} + 
G^{}_{\xi} + H^{}_{\xi} \eeq
and the total gauge dependence vanishes.
Details can be found in the appendix.

\section{Gauge independence of the A self-energy to higher orders and the loop counting}
In this section we move to an analysis performed to higher orders that illustrates how the loop expansion and the counterterms get organized so to have a consistent gauge independent theory.

\begin{figure}[t]
{\centering \resizebox*{13cm}{!}{\rotatebox{0}
{\includegraphics{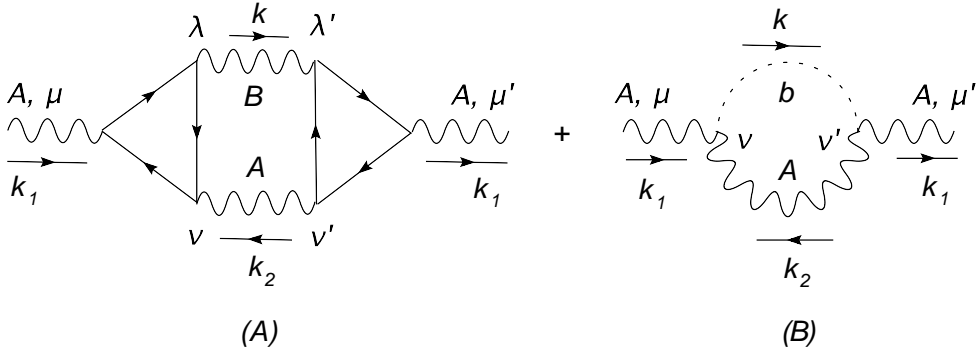}}}\par}
\caption{3 loop cancelations of the gauge dependence.}
\label{box}
\end{figure}

For this purpose, let's consider the diagrams in Fig.~\ref{box}, which are relevant in order to verify this 
cancelation in the massless fermion case. It shows the self-energy of the A gauge boson. From now on we are dropping all the coupling constants to simplify the notation, which can be 
re-inserted at the end. We have omitted diagrams which are 
symmetric with respect to the 
two  intermediate lines of the B and A gauge bosons, for simplicity. This symmetrization is responsible for the 
cancelation of the gauge dependence of the propagator of $A$ and the vector interaction of B, while the gauge 
dependence of the axial-vector contribution of B is canceled by the corresponding goldstone (shown).
Diagram 
(A) involves 3 loops and therefore we need to look for cancelations induced by a diagram involving 
the s-channel exchange both of an $A$ and of a $B$ gauge boson plus the 1-loop 
interactions involving the relevant counterterms. In this case one easily identifies diagram (B) as the 
only possible additional contribution. 

To proceed with the demonstration we first isolate the gauge dependence in the propagator for the gauge boson exchanged in the s-channel 
\beqa
\frac{- i}{k^2 - M_1^2}\left[  g^{\, \lambda\, \lambda^{\prime}} - \frac{k^\lambda \, k^{\lambda^\prime}}{k^2 - \xi^{}_B M_1^2}
(1 - \xi_B) \right]  &=& 
\frac{- \,i}{ k^2 - M_1^2} \left( g^{\,\lambda\, \lambda^\prime} - \frac{k^\lambda \, 
k^{\lambda^\prime}}{M_1^2} \right) 
+  \frac{- \,i}{ k^2 - \xi_B \,M_1^2}  \left( \frac{ k^{\lambda} k^{\lambda^\prime}}{M_1^2}    \right)  \nonumber\\
&=& P_0^{\lambda \, \lambda^\prime} +  P_{\xi}^{\lambda \, \lambda^\prime}.
\eeqa

%
%%%%%%%%%%%%%%%%%%%%%%%%%
\begin{figure}[t]
{\centering \resizebox*{16cm}{!}{\rotatebox{0}
{\includegraphics{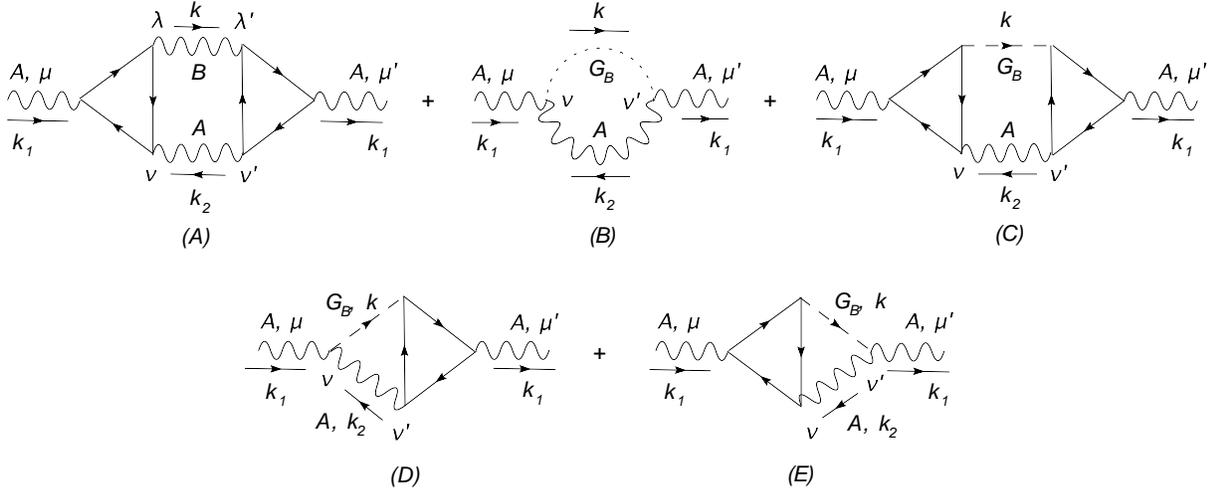}}}\par}
\caption{The complete set of diagrams in the broken phase.}
\label{boxmass}
\end{figure}

%%%%%%%%%%%%%%%%%%%%%%%%%
% 
and using this separation, the sum involving the two diagrams gives
\beqn
S = \int \frac{d^{4} k^{}_{2}}{(2 \pi)^{4}} ( {\mathcal A} + {\mathcal B} ),
\eeqn
with the gauge-dependent contributions being given by
\beqn
{\mathcal A}^{}_{ \xi 0} &=&  \, \Delta^{\lambda \mu \nu} (-k^{}_{1}, -k^{}_{2})  P^{\lambda \lambda'}_{\xi} 
\Delta^{\lambda' \mu' \nu'} (k^{}_{1}, k^{}_{2})    P^{\nu \nu'}_{o}    \nonumber\\
&=&  \, \Delta^{\lambda \mu \nu} (-k^{}_{1}, -k^{}_{2})
\left[ \frac{- i }{k^{2} - \xi^{}_{B} M^{2}_{1}} \left( \frac{k^{\lambda} k^{\lambda'}}{ M^{2}_{1}} \right)  \right] 
\Delta^{\lambda' \mu' \nu'} (k^{}_{1}, k^{}_{2}) \left[  \frac{-i g^{\nu \nu'}}{k^{2}_{2}}  \right]   \nonumber\\
{\mathcal B}_{\xi 0} &=& 4 \times  \left( \frac{4}{M} C^{}_{AA} \right)^{2} \epsilon^{\, \mu \nu \rho \sigma} 
k^{}_{1\rho} k^{}_{2\sigma} \frac{i}{k^{2} - \xi^{}_{B} M^{2}_{1}} \epsilon^{\, \mu' \nu' \rho' \sigma'} 
k^{}_{1\rho'} k^{}_{2\sigma'}   P^{\nu \nu'}_{o}. 
\eeqn
Using the anomaly equations and substituting the appropriate value already obtained for the WZ-coefficient, we obtain a vanishing expression
\beqn
{\mathcal A}^{}_{ \xi 0} + {\mathcal B}^{}_{ \xi 0}  
&=&   ( - a^{}_{3}(\beta) \epsilon^{\, \mu \nu \rho \sigma} 
k^{}_{1\rho} k^{}_{2\sigma})
\left[ \frac{- i }{k^{2} - \xi^{}_{B} M^{2}_{1}}  \frac{1}{ M^{2}_{1} }  \right] 
 ( a^{}_{3}(\beta) \epsilon^{\, \mu' \nu' \rho' \sigma'} k^{}_{1\rho'} k^{}_{2\sigma'} )  P^{\nu \nu'}_{o}  \nonumber\\
&+& 4 \, \frac{16}{ M^{2} } \left( \frac{i}{2!} \frac{1}{4} a^{}_{3}(\beta)  \frac{M}{M^{}_{1}} \right)^{2}
 \epsilon^{\, \mu \nu \rho \sigma} 
k^{}_{1\rho} k^{}_{2\sigma} \frac{i}{k^{2} - \xi^{}_{B} M^{2}_{1}} \epsilon^{\, \mu' \nu' \rho' \sigma'} 
k^{}_{1\rho'} k^{}_{2\sigma'}P^{\nu \nu'}_{o}= 0. \nonumber \\
\eeqn
After symmetry breaking, with massive fermions, the pattern gets far more involved and is described in 
Fig.~\ref{boxmass}. Also in this case we have 
%
%%%%%%%%%%%%%%%%%%%%%%
%%%%%%%%%%%%%%%%%%%%%%%
%
\beqn
S = \int \frac{d^{4} k^{}_{2}}{(2 \pi)^{4}} ( {\mathcal A} + {\mathcal B}+ {\mathcal C} + {\mathcal D} + {\mathcal E} )
\eeqn
and it can be shown by direct computation that the gauge dependences cancel in this combination. 
The interested reader can find the discussion in the appendix (``Gauge cancelations in the self-energy 
diagrams''). 
Notice that the pattern to follow in order to identify the relevant diagrams is quite clear, 
and it is not difficult to identify at each order of the loop expansion
contributions with the appropriate s-channel exchanges that combine into gauge invariant amplitudes. These are built identifying direct contributions and counterterms in an appropriate fashion, counting the counterterms at their appropriate order in Planck's constant $(h)$. The identification of similar patterns in the broken phase is more cumbersome and is addressed below. 

%%%%%%%%%%%%%%%%%%%%%%%%%%%%%%%%%%%%%%%%%%%%%%%%%%%%%%%%%%%%%%%%%%%%%%%%%%
\section{ Unitarity Analysis of the AB model in the broken phase}
%%%%%%%%%%%%%%%%%%%%%%%%%%%%%%%%%%%%%%%%%%%%%%%%%%%%%%%%%%%%%%%%%%%%%%%%%%
%
In the broken phase and in the presence of Yukawa couplings there are some modifications that take place, since the s-channel 
exchange of the $b$ axion is rotated on the two components (the goldstone and physical axion $\chi$) as shown later in Fig.~\ref{gs}. 
The introduction of the Yukawa interaction and the presence of a symmetry breaking phase determines an interaction of the axion with the 
fermions. Therefore, let's consider the Yukawa lagrangean    
\beqa
{\cal{L}}_Y =- \lambda^{}_1 \, \overline{\psi}_{L} \phi \, \psi^{}_{R} 
- \lambda^{}_1 \, \overline{\psi}_{R} \phi^{*} \psi^{}_{L},
\eeqa
which is needed to extract the coupling between the axi-Higgs and the 
fermions. We focus on the term
\beqa
{\cal{L}}_Y(\phi^{}_{2}) =- \frac{ \lambda^{}_1 }{2} \left[ \overline{\psi} ( 1 + \gamma^5) \psi 
\, \frac{ i \phi_{2} }{\sqrt 2} 
-  \overline{\psi} (1 - \gamma^5) \psi\, \frac{i \phi_{2} }{\sqrt 2}   \right],
\eeqa
having expanded around the Higgs vacuum. Performing a rotation to express the pseudoscalar 
Higgs phase $\phi^{}_{2}$ in  terms of the physical axion and the NG boson 
$$\phi^{}_{2} = - \frac{M_1}{M_B} \chi^{}_{B} +  \frac{ q^{}_{B} g^{}_{B} v}{M_B} G^{}_{B},$$ one extracts a 
$\chi\overline{\psi} \psi$ coupling of the kind
\beqa 
{\cal{L}}_Y(\chi^{}_{B}) =  \frac{\lambda^{}_1}{\sqrt 2} \frac{M_1}{M_B} i \, \overline{\psi}  \chi^{}_{B} \gamma^5 \psi,
\eeqa
and a coupling $G\overline{\psi} \psi$ for the goldstone mode
\beqa
{\cal{L}}_Y(G^{}_{B}) = -
\frac{\lambda^{}_1}{\sqrt 2} \frac{q^{}_{B} g^{}_{B} v}{M_B} i  \,  \overline{\psi} \, G^{}_{B}  \gamma^5 
\psi =  2 g^{}_{B} \frac{ m^{}_{f} }{M_B} 
i \overline{\psi} \gamma^5 \psi G^{}_{B}.  
\eeqa
Having fixed the Yukawa couplings of the model, we move to the 
analysis of the same diagrams of the previous section in the broken phase. Preliminarily, we need to identify the 
structure of the anomaly equation for the fermionic 3-point functions with their complete mass dependence.  
In the case of massive fermions the anomalous Ward identities for an {\bf AVV} triangle are of the form 
\beqn
k_{1\mu}\Delta^{\lambda\mu\nu}(\beta,k_1,k_2)&=& a_1(\beta) \varepsilon^{\lambda\nu\alpha\beta}
k_1^\alpha k_2^\beta,\nonumber\\
k_{2\nu}\Delta^{\lambda\mu\nu}(\beta,k_1,k_2)&=& a_1(\beta) \varepsilon^{\lambda\mu\alpha\beta}
k_2^\alpha k_1^\beta,\nonumber\\
k_\lambda\Delta^{\lambda\mu\nu}(\beta,k_1,k_2)&=& a_3(\beta) \varepsilon^{\mu\nu\alpha\beta}
k_1^\alpha k_2^\beta +  2 m_{f} \Delta^{\mu \nu},
\label{bbshift}
\eeqn
and in the case of {\bf AAA} triangle $\Delta_{3}^{\lambda \mu \nu}(\beta, k_1, k_2)= \Delta_{3}^{\lambda \mu \nu}( k_1, k_2)$, with  Bose symmetry providing a factor 1/3 for the distribution of the anomalies among the 3 vertices
\beqn
k_{1\mu}\Delta_{3}^{\lambda\mu\nu}(k_1,k_2)&=& \frac{a_n}{3} \varepsilon^{\lambda\nu\alpha\beta}
k_1^\alpha k_2^\beta +   2 m_{f} \Delta^{\lambda \nu}       ,\nonumber\\
k_{2\nu}\Delta_{3}^{\lambda\mu\nu}(k_1,k_2)&=& \frac{a_n}{3} \varepsilon^{\lambda\mu\alpha\beta}
k_2^\alpha k_1^\beta +   2 m_{f} \Delta^{\lambda \mu}          ,\nonumber\\
k_\lambda \Delta_{3}^{\lambda\mu\nu}(k_1,k_2)&=& \frac{a_n}{3} \varepsilon^{\mu\nu\alpha\beta}
k_1^\alpha k_2^\beta +   2 m_{f} \Delta^{\mu \nu},
\label{bbshift}
\eeqn
where we have dropped the appropriate coupling constants common to both sides. The amplitude $\Delta^{\mu\nu}$ is given by 
\beqn 
\Delta^{\mu \nu} =  \int \frac{ d^4 q }{ ( 2 \pi )^4 } \frac{ Tr \left[ \gamma^{5} ( \slash{q} - \slash{k} + m_{f}) 
\gamma^{\nu} \gamma^{5} ( \slash{q} - \ds{k}_1 + m_{f})  \gamma^{\mu} \gamma^5 (\slash{q} + m_{f}) \right] }{
[ q^2 - m_{f}^2] [ ( q - k )^2 - m_{f}^2 ] [ (q - k_1)^2 - m_{f}^2 ] } \mbox{+ exch.}
\label{chidec}
\eeqn
and can be expressed as a two-dimensional integral using the Feynman parameterization. We find
%
%%%%%%%%%%%%%%%%%%%%%%%
%\begin{figure}[t]
%{\centering \resizebox*{5cm}{!}{\rotatebox{0}
%{\includegraphics{chiBB.eps}}}\par}
%\caption{Amplitude for $\chi$ decay into B B. The WZ contributions have not been included }
%\label{axiontriangle}
%\end{figure}
%%%%%%%%%%%%%%%%%%%%%%%
%
%
\beqn 
\Delta^{\mu\nu}&=&  \epsilon^{\alpha \beta \mu \nu }  k_1^{\alpha} 
k_2^{\beta} m_f \left( \frac{1}{ 2 \pi^2}  \right) I, \nonumber \\
\label{chigg}
\eeqn
with $I$ denoting the formal expression of the integral

\beq
 I\equiv \int^{1}_{0} dx \int^{1-x}_{0} dy  \frac{(1 - 2x - 2y)}{\Delta(x,y,m_f,m_\chi,M_B)}. 
\eeq

We have dropped the charge dependence since we have normalized the charges to unity and we have defined 

\beq
\Delta(x,y,m_f,m_\chi,M_B) = \Sigma^2 - D =  m_f^2 - x\,y \,m_{\chi}^2 + M_B^2 (x+y)^2 - x M_B^2 - y M_B^2 \equiv \Delta(x,y)
\label{delta}.
\eeq

\begin{figure}[t]
{\centering \resizebox*{10cm}{!}{\rotatebox{0}
{\includegraphics{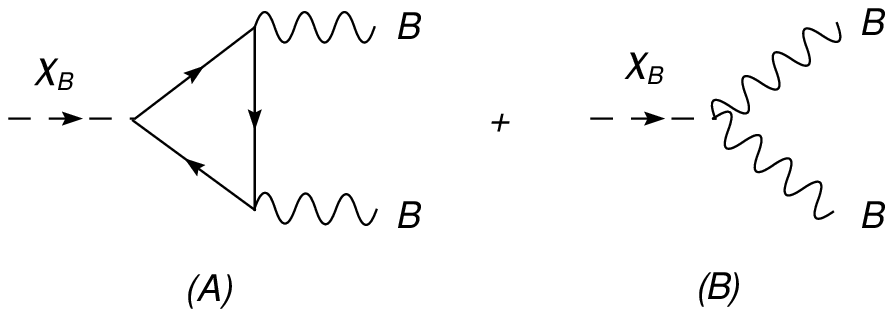}}}\par}
\caption{Decay amplitude for $\chi\to B B $.}
\label{chi-decay}
\end{figure}

\begin{figure}[t]
{\centering \resizebox*{12cm}{!}{\rotatebox{0}
{\includegraphics{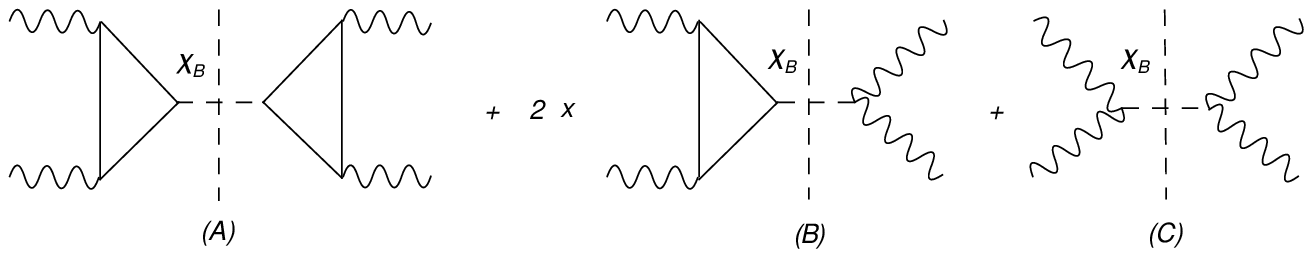}}}\par}
\caption{Decay cross section for $\chi \to B B $.}
\label{chi-generic}
\end{figure}

We can use this amplitude to compute the 1-loop decay of the axi-Higgs in this simple model, 
shown in Fig.~\ref{chi-decay}, which is given by
\beqn
{\mathcal M}_{\chi \rightarrow BB} &=&  {\mathcal A} + {\mathcal B}   \nonumber\\
&=& i \frac{\lambda^{}_{1}}{\sqrt 2} \frac{M^{}_{1}}{M^{}_{B}} \Delta^{\mu \nu} (k^{}_{1}, k^{}_{2})
+ \alpha^{}_{1} \frac{4}{M} C^{}_{BB}    \nonumber\\
&=& i \frac{\lambda^{}_{1}}{\sqrt 2} \frac{M^{}_{1}}{M^{}_{B}} \Delta^{\mu \nu} (k^{}_{1}, k^{}_{2})
 -    \frac{2 g^{}_{B} v}{M^{}_{B}} \left(  \frac{4}{M} \frac{i}{3!} \frac{1}{4} a^{}_{n} \frac{M}{M^{}_{1}}  \right),
\eeqn 
where $\alpha^{}_{1}$ is the coefficient that rotates the axion $b$ on the axi-higgs particle $\chi$. The related cross section is shown in Fig.~\ref{chi-generic}. 

%%%%%%%%%%%%%%%%%%%%%%%%%%%%%%%%%%%%%%%%

%%%%%%%%%%%%%%%%%%%%%%%%%%%%%%%%%%%%%%%%
%%%%%%%%%%%%%%%%%%%%%%%%%%%%%%%%%%%%%%%%

%

%
%%%%%%%%%%%%%%%%%%%%%%%%%%%%%%%%%%%%%%%%%%%%%%%%%%%%%%%%%%%%%%%%%%%%%%%%%%%%%%%%%%%%%%
\subsection{A-B model: BB $\rightarrow$ BB mediated by a B gauge boson in the broken phase}
%%%%%%%%%%%%%%%%%%%%%%%%%%%%%%%%%%%%%%%%%%%%%%%%%%%%%%%%%%%%%%%%%%%%%%%%%%%%%%%%%%%%%%
%
%%%%%%%%%%%%%%%%%%%%%%%%%%%%%%%%%%%%%%%%%%%%%%%%%%%%%%%%%%%%%%%%%%%%%%%%%%%%%%
\begin{figure}[t]
{\centering \resizebox*{15cm}{!}{\rotatebox{0}
{\includegraphics{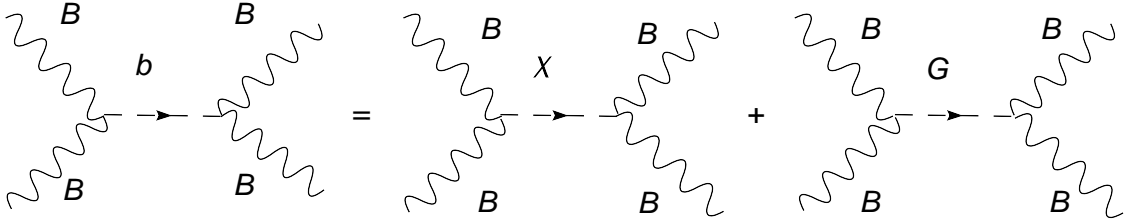}}}\par}
\caption{Diagrams from the Green-Schwarz coupling after symmetry breaking.}
\label{gs}
\end{figure}
\begin{figure}[tbh]
{\centering \resizebox*{13cm}{!}{\rotatebox{0}
{\includegraphics{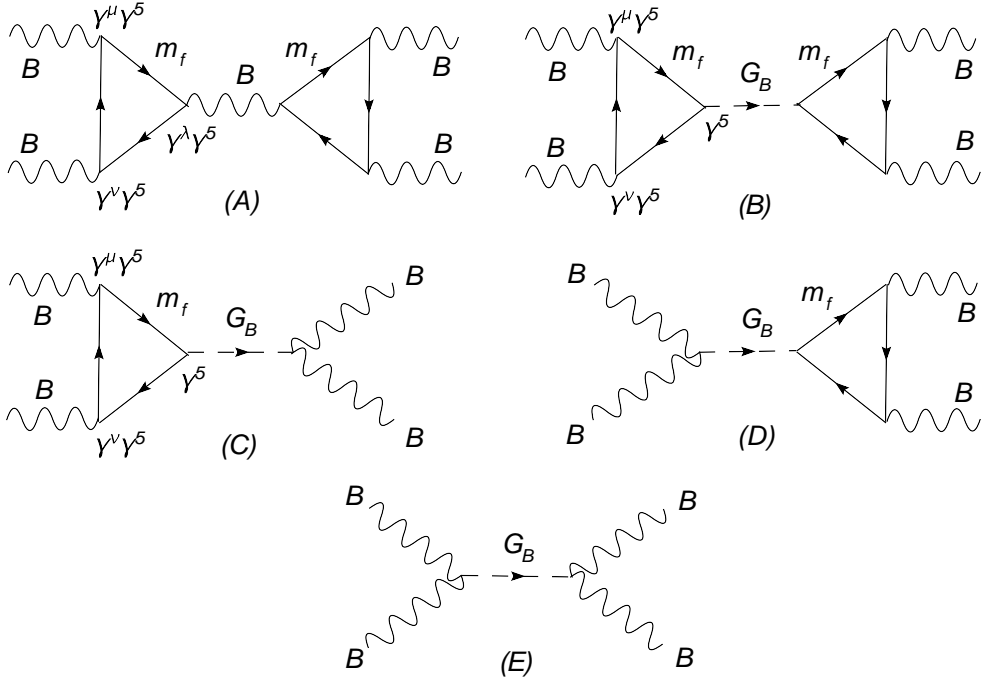}}}\par}
\caption{Cancelation of the gauge dependence after spontaneous symmetry breaking.}
\label{brokenphase}
\end{figure}

%%%%%%%%%%%%%%%%%%%%%%%%%%%%%%%%%%%%%%%%%%%%%%%%%%%%%%%%%%%%%%%%%%%%%%%%%%%%%%
%
A second class of contributions that require a different distribution of the partial anomalies are those involving $BBB$ diagrams. They appear in the $BB\to BB$ amplitude, mediated by the exchange of a $B$ gauge boson of mass 
$M_B = \sqrt{M_1^2 + (2 g^{}_{B} v)^2}$. Notice that $M_B$ gets its mass both from the Higgs and the St\"{u}ckelberg sectors. 
 The relevant diagrams for this check are shown in Fig.~\ref{brokenphase}. 
We have not included the exchange of the physical axion, since this is not gauge-dependent. We are only 
after the gauge-dependent contributions. 
Notice that the expansion is valid at 
2-loop level and involves 2-loop diagrams built as combinations of 
the original diagrams and of the 1-loop counterterms. There 
are some comments that are due in order to appreciate the way the 
cancelations take place. If we neglect the Yukawa couplings the diagrams 
(B), (C) and (D) are absent, since the goldstone does not couple to a massless fermion. In this case, the axion $b$ is rotated, as in the previous sections, into a goldstone mode $G_B$ and a physical axion $\chi$ (see Fig.~\ref{gs}). On the other hand, if we include the Yukawa couplings then the entire set of diagrams is needed. 
%
%%%%%%%%%%%%%%%%%%%%%%%%%%%%%%%%%%
%%%%%%%%%%%%%%%%%%%%%%%%%%%%%%%%%%
%
From diagram (E) we obtain the partial contribution 
\beqn
\mathcal{E}^{}_{\xi} = 
4 \times \left( \frac{4}{M} \alpha^{}_2 C^{}_{BB} \right)^2 \varepsilon^{\mu\nu\rho\sigma} k_{1}^\rho k_{2}^\sigma  
 \left( \frac{i}{k^2 - \xi_B M_{B}^2} \right)  \varepsilon^{\mu^\prime  \nu^\prime   \rho^\prime \sigma^\prime} 
k_{1}^{\rho^\prime} k_{2}^{\sigma^\prime},
\eeqn
where the overall factor of 4 in front is a symmetry factor, 
the coefficient $\alpha^{}_2$ comes from the rotation of the b axion 
over the goldstone boson ($\alpha_2 = \frac{M_1}{M_B}$), and the coefficient $C^{}_{BB}$ has already been determined from the condition of 
gauge invariance of the anomalous effective action before symmetry breaking.
Similarly, from diagram (B) we get the term 
\beqn
\mathcal{B}_{\xi} = 
( g^{}_{B} )^3 \Delta^{\mu \nu}(-k^{}_1,- k^{}_2) \left( 2i \, \frac{ m_{f} }{M_B} \right)  \frac{i}{k^2 - \xi_B M_B^2} 
\left( 2 i \frac{ m_{f} }{M_B} \right)
( g^{}_{B})^3 \Delta^{\mu^\prime \nu^\prime}(k^{}_1, k^{}_2),
\eeqn 
while diagram (C) gives
\beqn
\mathcal{C}_{\xi} = 
2 \times ( g^{}_{B} )^3 \Delta^{\mu \nu}(-k^{}_1,- k^{}_2) \left( 2\,i\frac{ m_{f} }{M_B} \right)  \frac{i}{k^2 - \xi_B M_B^2}  
\left( \frac{4}{M} \alpha^{}_2 C^{}_{BB} \, 
\varepsilon^{\mu^\prime \nu^\prime \rho^\prime \sigma^\prime}  k_1^{\rho^\prime}  k_2^{\sigma^\prime} \right),
\eeqn
having introduced also in this case a symmetry factor. 
Finally (D) gives 
\beqn
\mathcal{D}^{}_{\xi} = 
2\times \left( \frac{4}{M} \alpha_2 C^{}_{BB} \varepsilon^{\mu \nu \rho \sigma}  k_1^{\rho}  k_2^{\sigma} \right)
\frac{i}{k^2 - \xi_B M_B^2} \left(2\,i \frac{ m_{f} }{M_B} \right) \Delta^{\mu^\prime \nu^\prime}(k^{}_{1}, k^{}_{2}) ( g^{}_{B} )^3.
\eeqn
We will be using the anomaly equation in the massive fermion case, having distributed the anomaly equally among the three B vertices 

\beqn
k^\lambda \Delta^{\lambda \mu \nu}=  \frac{a_n}{3} \varepsilon^{\mu \nu \alpha \beta} k_1^{\alpha} k_2^{\beta} 
+ 2 m_{f} \Delta^{\mu \nu}. 
\eeqn 
Separating in diagram (A) the gauge dependent part of the propagator of the 
boson B  from the rest we obtain
\beqa
\mathcal{A}^{}_{\xi}&=& \Delta^{\lambda \mu \nu } \frac{- i }{k^2 - \xi_B M_B^2} \left( \frac{k^\lambda k^{\lambda^\prime}}{M^{2}_{B}} \right) 
\Delta^{\lambda^\prime \mu^\prime \nu^\prime}  \nonumber \\ 
&=&  \left( \frac{a_n}{3} \varepsilon^{\mu \nu \alpha \beta} k_1^{\alpha} k_2^{\beta} 
+ 2 m^{}_{ f} \Delta^{\mu \nu} \right) ( g^{}_{B})^{3} \frac{ i }{k^2 - \xi^{}_B M_B^2} \frac{1}{M_B^2} \left( \frac{a_n}{3} 
\varepsilon^{\mu^\prime \nu^\prime \alpha^\prime \beta^\prime} 
k_1^{\alpha^\prime} k_2^{\beta^\prime} +  2 m^{}_{ f} \Delta^{\mu^\prime \nu^\prime}   \right)  ( g^{}_{B})^{3}  \nonumber\\ 
&=& \frac{ i }{k^2 - \xi_B M_B^2} \frac{g^{\,6}_{B}}{M_B^2}  \left[  \left( \frac{a_n}{3} \right)^2 \varepsilon^{\mu \nu \alpha \beta } 
\varepsilon^{\mu^\prime \nu^\prime \alpha^\prime \beta^\prime } k_1^\alpha k_2^\beta k_1^{\alpha^\prime} 
k_2^{\beta^\prime} \right. \nonumber\\ 
&&\left. +\frac{a_n}{3} \varepsilon^{\mu \nu \alpha \beta}  k_1^\alpha k_2^\beta \, 2 m^{}_{ f} \Delta^{\mu^\prime \nu^\prime} 
 + 2 m^{}_{ f} \Delta^{\mu \nu} \frac{a_n}{3} \varepsilon^{\mu^\prime \nu^\prime \alpha^\prime \beta^\prime } k_1^{\alpha^\prime}
k_2^{ \beta^\prime} \right.  \nonumber\\
&&\left.+ \left( 2 m^{}_{ f} \Delta^{\mu \nu} \right) ( 2 m^{}_{ f} \Delta^{\mu^\prime \nu^\prime}) \right].  
\label{brokenABC}
\eeqa
The first term in (\ref{brokenABC}) is exactly canceled by the contribution 
from diagram (E).
The last contribution cancels by the contribution from diagram (B). Finally diagrams (C) 
and (D) cancel against the second and third 
contributions from diagram (A).

%
%%%%%%%%%%%%%%%%%%%%%%%%%%%%%%%%%%%%%%%%%%%%%%%%%%%%%
\section{The effective action in the Y-B model}
%%%%%%%%%%%%%%%%%%%%%%%%%%%%%%%%%%%%%%%%%%%%%%%%%%%%%
%
%
We anticipate in this section some of the methods that will be used 
in \cite{CIM2} in the analysis 
of a realistic model. In the previous model, in order to simplify the 
analysis, we have assumed that the coupling of the B gauge boson was 
purely axial while $A$ was purely vector-like. Here we discuss 
a gauge structure which allows both gauge bosons to have combined vector and axial-vector 
couplings. We will show that the external Ward identities of the model involve a specific 
definition of the shift parameter in one of the triangle diagrams that forces the axial-vector current to be conserved ($a_3(\beta)=0$). This result, new compared to the case of the SM, shows the presence of an effective CS term in some amplitudes.

The lagrangean that we choose to exemplify this new situation is given by 
\beqn
{\mathcal L}_{0} &=& 
| (\partial^{}_{\mu} + i g^{}_{Y} q^{Y}_{u} Y^{}_{\mu} + i g^{}_{B} q^{B}_{u} B^{}_{\mu}) \phi_{u} |^{2} + 
| (\partial^{}_{\mu} + i g^{}_{Y} q^{Y}_{d} Y^{}_{\mu} + i g^{}_{B} q^{B}_{d} B^{}_{\mu}) \phi_{d} |^{2}
- \frac{1}{4} F^{2}_{Y} - \frac{1}{4} F^{2}_{B}   \nonumber\\
&+& \frac{1}{2} (\partial^{}_{\mu} b + M^{}_{1}B^{}_{\mu})^{2} - \lambda^{}_{u} \left( |\phi^{}_{u}|^{2} 
- \frac{v^{}_{u}}{2} \right)^{2} - \lambda^{}_{d} \left( |\phi^{}_{d}|^{2} 
- \frac{v^{}_{d}}{2} \right)^{2} + {\mathcal L}_{f} + {\mathcal L}_{Yuk},
\eeqn
where the Yukawa couplings are given by
\beqn
{\mathcal L}_{Yuk} =- \lambda^{}_{1} \overline{\psi}_{1L} \phi^{}_{u} \psi^{}_{1R} - \lambda^{}_{1} \overline{\psi}_{1R}
 \phi^{*}_{u} \psi^{}_{1L}
-  \lambda^{}_{2} \overline{\psi}_{2L} \phi^{}_{d} \psi^{}_{2R} - \lambda^{}_{2} \overline{\psi}_{2R} \phi^{*}_{d} \psi^{}_{2L},
\eeqn

with $L$ and $R$ denoting left- and right- handed fermions. 

The fermion currents are
\beqn
{\mathcal L}_{f} &=& \overline{\psi}_{1L} i \gamma^{\mu} \left[ \partial^{}_{\mu} + i g^{}_{Y} q^{Y}_{1L} Y^{}_{\mu} 
+ i g^{}_{B} q^{B}_{1L}B^{}_{\mu}  \right] \psi^{}_{1L} + 
 \overline{\psi}_{1R} i \gamma^{\mu} \left[ \partial^{}_{\mu} + i g^{}_{Y} q^{Y}_{1R} Y^{}_{\mu} 
+ i g^{}_{B} q^{B}_{1R}B^{}_{\mu}  \right] \psi^{}_{1R} \nonumber\\ 
&+& \overline{\psi}_{2L} i \gamma^{\mu} \left[ \partial^{}_{\mu} + i g^{}_{Y} q^{Y}_{2L} Y^{}_{\mu} 
+ i g^{}_{B} q^{B}_{2L}B^{}_{\mu}  \right] \psi^{}_{2L}  + 
 \overline{\psi}_{2R} i \gamma^{\mu} \left[ \partial^{}_{\mu} + i g^{}_{Y} q^{Y}_{2R} Y^{}_{\mu} 
+ i g^{}_{B} q^{B}_{2R}B^{}_{\mu}  \right] \psi^{}_{2R} \nonumber\\ 
\eeqn
so that, in general, without any particular charge assignment, both gauge bosons show vector and axial-vector couplings.
In this case we realize an anomaly-free charge assignment for the 
hypercharge by requiring that $q^{Y}_{2L} = - q^{Y}_{1L}, q^{Y}_{2R} = - q^{Y}_{1R}$,
which cancels the anomaly for a YYY triangle since 
\beqn
\sum^{}_{f=1,2} (q^{Y}_{f})^{3} = (q^{Y}_{1R})^{3} -  (q^{Y}_{1L})^{3} + (q^{Y}_{2R})^{3} - (q^{Y}_{2L})^{3} 
=  (q^{Y}_{1R})^{3} -  (q^{Y}_{1L})^{3} - (q^{Y}_{1R})^{3} 
+ (q^{Y}_{1L})^{3} = 0. \nonumber \\
\eeqn
This condition is similar to the vanishing of the (YYY) anomaly for the hypercharge in the 
SM, and for this reason we will assume that it holds also in our simplified model.

\begin{figure}[t]
{\centering \resizebox*{15cm}{!}{\rotatebox{0}
{\includegraphics{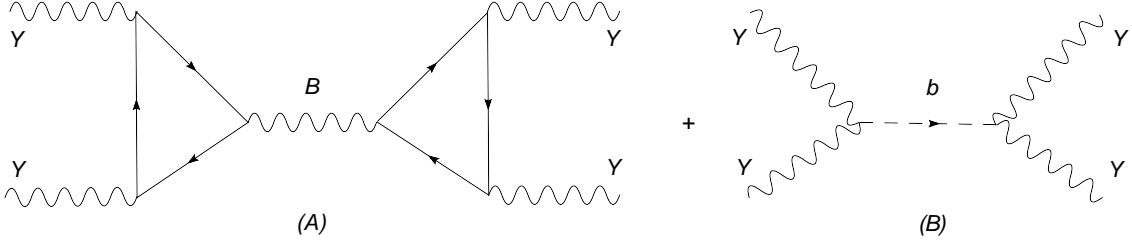}}}\par}
\caption{Unitarity diagrams in the Y-B model}
\label{yb-unitarity}
\end{figure}

Before symmetry breaking the B gauge boson has a 
goldstone coupling coming from the St\"{u}ckelberg mass term due to the presence of a Higgs field.
The effective action for this model is given by 
\beqn
{\mathcal S}_{eff} =  {\mathcal S}_{an} + {\mathcal S}_{WZ},  
\eeqn
which reads
\beqn
{\mathcal S}_{an} 
&=&  \frac{1}{3!}\langle T^{\lambda \mu \nu}_{BBB}(z,x,y) B^{\lambda} B^{\mu} B^{\nu}\rangle + 
\frac{1}{2!}\langle  T^{\lambda \mu \nu}_{BYY}(z,x,y) B^{\lambda} Y^{\mu} Y^{\nu} \rangle
+ \frac{1}{2!} \langle T^{\lambda \mu \nu}_{YBB}(z,x,y) Y^{\lambda} B^{\mu} B^{\nu}  \rangle \nonumber \\ 
\eeqn
with
\beqn
{\mathcal S}_{WZ} =  \frac{C^{}_{YY}}{M}\,\langle b\, F^{Y} \wedge F^{Y}\rangle + 
 \frac{C^{}_{BB}}{M}\, \langle b\, F^{B} \wedge F^{B}\rangle + \frac{C^{}_{BY}}{M}\,\langle b\, F^{B} \wedge F^{Y} \rangle
\eeqn
denoting the WZ counterterms.
Only for the triangle BBB we have assumed an anomaly symmetrically distributed, all the other anomalous diagrams having an {\bf AVV} anomalous 
structure, given in momentum space by
\beqn
\Delta^{\lambda \mu \nu}_{ijk} = \frac{i^{3}}{2} \sum_{f}\left[ q^{i}_{f} q^{j}_{f} q^{k}_{f} \right] 
\int \frac{d^{4} p}{(2 \pi)^{4}} 
\frac{ Tr[ \gamma^{\lambda} \gamma^{5} \slash{p} \gamma^{\mu} 
(\slash{p} - \ds{k}_{1}) \gamma^{\nu} (\slash{p} - \ds{k}) ]}{p^{2} 
(p-k^{}_{1})^{2} (p-k)^{2} } 
+ \{ \mu, k^{}_{1} \,\, \leftrightarrow \,\, \nu, k^{}_{2} \}, \nonumber \\
\eeqn
with indices running over i, j, k =Y, B. The sum over the fermionic spectrum involves the charge operators in the chiral basis
\beqn
D^{}_{ijk} = \frac{1}{2} \sum_{f=1,2} \left[ q^{i}_{f} q^{j}_{f} q^{k}_{f} \right] \equiv
\frac{1}{2} \sum_f\left( q^{i}_{fR} q^{j}_{fR} q^{k}_{fR} - q^{i}_{fL} q^{j}_{fL} q^{k}_{fL}\right).  
\eeqn
Computing the Y-gauge variation for the effective one loop anomalous action under the trasformations 
$Y^{}_{\mu} \rightarrow Y^{}_{\mu} + \partial^{}_{\mu} \theta^{}_{Y}$ we obtain
\beqn
\delta_{Y} {\mathcal S}_{an} = \frac{i}{2!} a^{}_{1}(\beta^{}_{1}) \frac{2}{4} \theta^{}_{Y} F^{}_{B} \wedge F^{}_{Y} D_{BYY} +  
\frac{i}{2!} a^{}_{3}(\beta^{}_{2}) \frac{1}{4} \theta^{}_{Y} F^{}_{B} \wedge F^{}_{B} D^{}_{YBB},
\eeqn
and, similarly,  for B-gauge transformations $B^{}_{\mu} \rightarrow B^{}_{\mu} + \partial^{}_{\mu} \theta^{}_{B}$ we obtain
\beqn
\delta_{B} {\mathcal S}_{an} &=&  \frac{i}{3!} \frac{a^{}_{n}}{3} \frac{3}{4}  \theta^{}_{B} F^{}_{B} \wedge F^{}_{B} D^{}_{BBB}  + 
 \frac{i}{2!} a^{}_{3}(\beta^{}_{1}) \frac{1}{4} \theta^{}_{B} F^{}_{Y} \wedge F^{}_{Y} D^{}_{BYY}   \nonumber\\
&&+ \frac{i}{2!} a^{}_{1}(\beta^{}_{2}) \frac{2}{4} \theta^{}_{B} F^{}_{B} \wedge F^{}_{Y} D^{}_{YBB},  
\eeqn
so that to get rid of the anomalous contributions due to gauge variance we have to fix the parameterization of the loop momenta with parameters
\beqn
\beta^{}_{1} = \overline{\beta^{}}_{1}  = - \frac{1}{2},  \qquad  
\beta^{}_{2} = \overline{\beta^{}}_{2} = + \frac{1}{2}.
\eeqn
Notice that while $\beta^{}_{1}$ corresponds to a canonical choice (CVC condition), the second amounts to a 
condition for a conserved axial-vector current, which can be interpreted as a condition that forces a CS counterterm in the parameterization of the triangle amplitude. 
Having imposed these conditions to cancel the anomalous variations for the Y gauge boson, we can determine the WZ coefficients as
\beqn
C^{}_{BB}=  \frac{M}{M^{}_{1}} \frac{i}{3!} a^{}_{n} \frac{1}{4} D^{}_{BBB}, \,\,
C^{}_{YY}=  \frac{M}{M^{}_{1}} \frac{i}{2!} a^{}_{3}(\overline{\beta}_{1}) \frac{1}{4} D^{}_{BYY},   
\,\,  C^{}_{BY}=  \frac{M}{M^{}_{1}} \frac{i}{2!} a^{}_{1}(\overline{\beta}_{2}) \frac{2}{4} D^{}_{YBB}. \nonumber \\
\eeqn 
Having determined all the parameters in front of the counterterms we can test the unitarity of the model. Consider the process $YY \rightarrow YY$ 
mediated by an B gauge boson depicted in Fig.~(\ref{yb-unitarity}), one can easily check that the gauge dependence vanishes. 
%
%%%%%%%%%%%%%%%%%%%%%%%%%%%%%
%%%%%%%%%%%%%%%%%%%%%%%%%%%%%
%
In fact we obtain
\beqn
{\mathcal S}^{}_{\xi} &=& 
{\mathcal A}_{\xi} + {\mathcal B}_{\xi}  \nonumber\\ 
&=&\Delta^{\lambda \mu \nu}(-k^{}_{1}, -k^{}_{2}) 
\left[  \frac{- i}{ k^2 - \xi_{B} M^{2}_{1}} \left( \frac{k^{\lambda} k^{\lambda'}}{M^{2}_{1}}
\right) \right]  \Delta^{\lambda' \mu' \nu'}(k^{}_{1}, k^{}_{2})  (D^{}_{BYY})^{2}  \nonumber\\
&&+ 4 \left(  \frac{4}{M} C_{YY}  \right)^{2}  \varepsilon^{\mu \nu \rho \sigma} 
k^{\rho}_{1} k^{\sigma}_{2}  \left( \frac{i}{k^2 - \xi_{B} M^{2}_{1}} \right)  
 \varepsilon^{\mu' \nu' \rho' \sigma'} k^{\rho'}_{1} k^{\sigma'}_{2}  
  \nonumber\\
&=&   \frac{- i}{ k^2 - \xi_{B} M^{2}_{1}} \frac{1}{M^{2}_{1}} 
 \left(  - a^{}_{3}( \overline{\beta^{}}_{1} ) \varepsilon^{\mu \nu \rho \sigma} k^{\rho}_{1} k^{\sigma}_{2} \right) 
  \left( a^{}_{3}(\overline{\beta^{}}_{1} )    \varepsilon^{\mu' \nu' \rho' \sigma'} k^{\rho'}_{1} k^{\sigma'}_{2} \right) 
 (D^{}_{BYY})^{2}    \nonumber\\
 &&+ 4 \frac{16}{M^{2}}\left( \frac{M}{M^{}_{1}} \frac{i}{2!} a^{}_{3}( \overline{\beta}_{1}) \frac{1}{4} D^{}_{BYY}\right)^{2}  
 \varepsilon^{\mu \nu \rho \sigma} 
k^{\rho}_{1} k^{\sigma}_{2}  \left( \frac{i}{k^2 - \xi_{B} M^{2}_{1}} \right) 
  \varepsilon^{\mu' \nu' \rho' \sigma'} k^{\rho '}_{1} k^{\sigma '}_{2}    = 0,
\nonumber \\
\eeqn 
where we have included the corresponding symmetry factors. There are some comments that are in order. 
In the basis of the interaction eigenstates, characterized Y and B before symmetry breaking, the CS counterterms can be absorbed into the diagrams, thereby obtaining a re-distribution of the partial anomalies on each anomalous gauge interaction. As we have already mentioned, the role of the CS terms is to render vector-like an axial vector current at 1-loop level in an anomalous trilinear coupling. The anomaly is moved 
from the Y vertex to the B vertex, and then canceled by a WZ counterterm. However, after symmetry 
breaking, in which Y and B undergo mixing, the best way to treat these anomalous interactions is to 
keep the CS term, rotated into the physical basis, separate from the triangular contribution. This separation is scheme dependent, being the CS term gauge variant. These theories are clearly characterized 
by direct interactions which are absent in the SM which can be eventually 
tested in suitable processes at the LHC \cite{CIM2}

%
%%%%%%%%%%%%%%%%%%%%%%%%%%%%%%%%%%%%%%%%%%%%%%%%%%%%%%%%%%%
%%%%%%%%%%%%%%%%%%%%%%%%%%%%%%%%%%%%%%%%%%%%%%%%%%%%%%%%%%%
%
%%
\section{The fermion sector} 
Moving to the analyze the gauge consistencly of the fermion sector, 
we summarize some of the features of the organization of some typical fermionic 
amplitudes. These considerations, naturally, can also be 
generalized to more complex cases. Our discussion is brief and we omit 
details and work directly in the A-B model for simplicity. Applications of 
this analysis can be found in \cite{CIM2}.

\begin{figure}[tbh]
{\centering \resizebox*{9cm}{!}{\rotatebox{0}
{\includegraphics{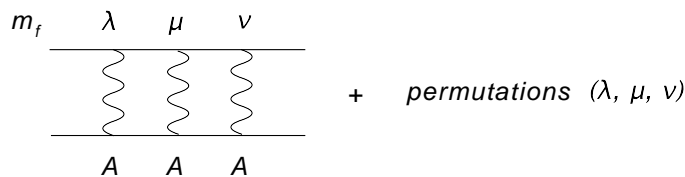}}}\par}
\caption{ The massive fermion sector with massive vector exchanges 
in the t-channel }
\label{fermion1}
\end{figure}

\begin{figure}[tbh]
{\centering \resizebox*{9cm}{!}{\rotatebox{0}
{\includegraphics{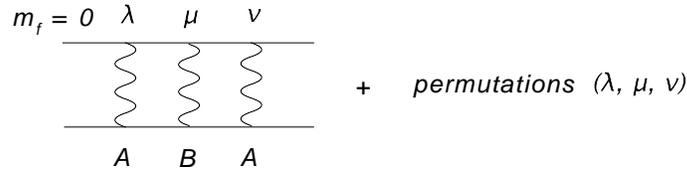}}}\par}
\caption{t-channel exchanges with vector and axial-vector interactions of 
massive gauge bosons. Being the fermion massless, permutation of the exchanges is sufficient to generate a gauge invariant result.}
\label{fermion2}
\end{figure}

\begin{figure}[tbh]
{\centering \resizebox*{13cm}{!}{\rotatebox{0}
{\includegraphics{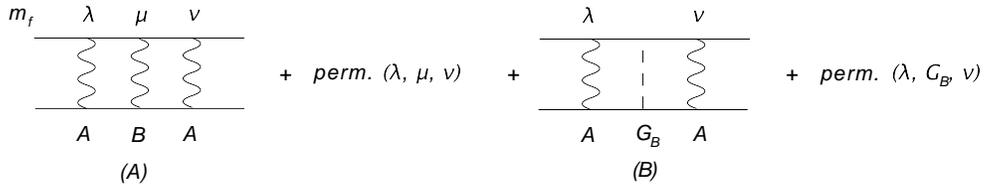}}}\par}
\caption{As in Fig.~\ref{fermion2} but in the massive case and with a goldstone }
\label{fermion3}
\end{figure}

 We start from Fig.~\ref{fermion1} that describes the t-channel exchange of A-gauge bosons. We have 
explicitly shown the indices $(\lambda\mu\nu)$ over which we perform permutations. In the absence of axial-vector interactions the gauge independence of diagrams of these types is obtained just 
with the symmetrization of the A-lines, both in the massive and in the massless fermion ($m_f$) case. When, 
instead, we allow for a B exchange in diagrams of the same topology, 
the cases $m_f=0$ and $m_f\neq 0$ involve a different (see Fig.~\ref{fermion2}) organization of the expansion. In the first case, the derivation of the gauge independence in this class of diagrams 
is obtained again just by a permutation of the attachments of the gauge boson lines. In the massive fermion case, instead, we need to add to this class of diagrams also the corresponding goldstone 
exchanges together with their similar symmetrizations (Fig.~\ref{fermion3}). 
%
%%%%%%%%%%%%%%%%%%%%%%%%%
%%%%%%%%%%%%%%%%%%%%%%%%%
%
%%%%%%%%%%%%%%%%%%%%%%%%%
%%%%%%%%%%%%%%%%%%%%%%%%%
%
%%%%%%%%%%%%%%%%%%%%%%%%%
%%%%%%%%%%%%%%%%%%%%%%%%%
%
%%%%%%%%%%%%%%%%%%%%%%%%%
\begin{figure}[tbh]
{\centering \resizebox*{6cm}{!}{\rotatebox{0}
{\includegraphics{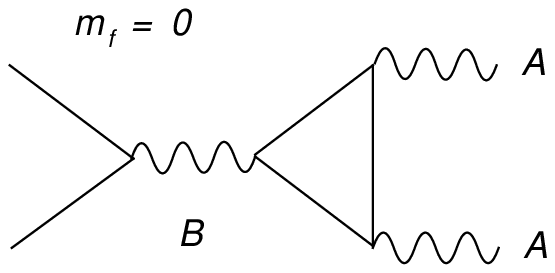}}}\par}
\caption{  $f$$\overline{f}$ annihilation in the massless case}
\label{fermion4}
\end{figure}
%%%%%%%%%%%%%%%%%%%%%%%%%
%
%%%%%%%%%%%%%%%%%%%%%%%%%
\begin{figure}[tbh]
{\centering \resizebox*{9cm}{!}{\rotatebox{0}
{\includegraphics{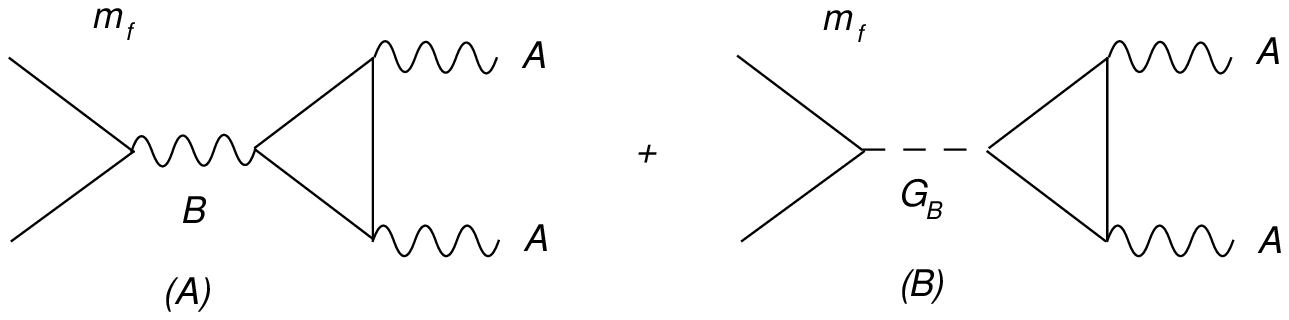}}}\par}
\caption{ $f$$\overline{f}$ in the massive case}
\label{fermion5}
\end{figure}
%%%%%%%%%%%%%%%%%%%%%%%%%
%
%%%%%%%%%%%%%%%%%%%%%%%%%
%\begin{figure}[tbh]
%{\centering \resizebox*{13cm}{!}{\rotatebox{0}
%{\includegraphics{fermion6.eps}}}\par}
%\caption{ Axi-higgs production cross section}
%\label{fermion6}
%\end{figure}
%%%%%%%%%%%%%%%%%%%%%%%%%
%
%%%%%%%%%%%%%%%%%%%%%%%%%
\begin{figure}[tbh]
{\centering \resizebox*{13cm}{!}{\rotatebox{0}
{\includegraphics{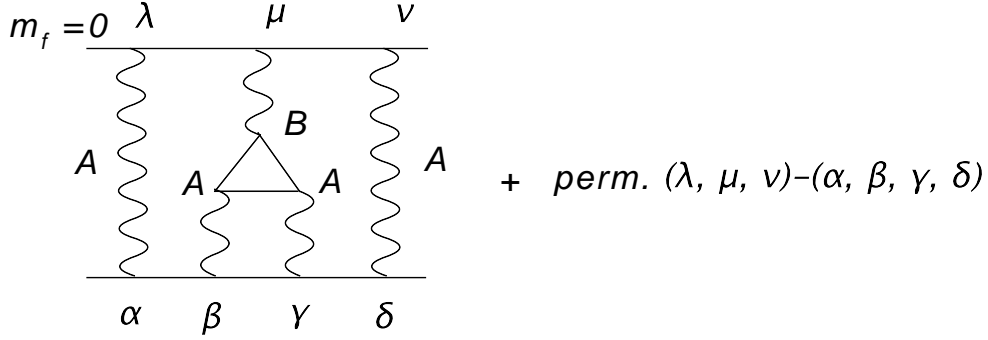}}}\par}
\caption{Anomaly in the t-channel}
\label{fermion7}
\end{figure}
%%%%%%%%%%%%%%%%%%%%%%%%%
%
%%%%%%%%%%%%%%%%%%%%%%%%%
\begin{figure}[tbh]
{\centering \resizebox*{13cm}{!}{\rotatebox{0}
{\includegraphics{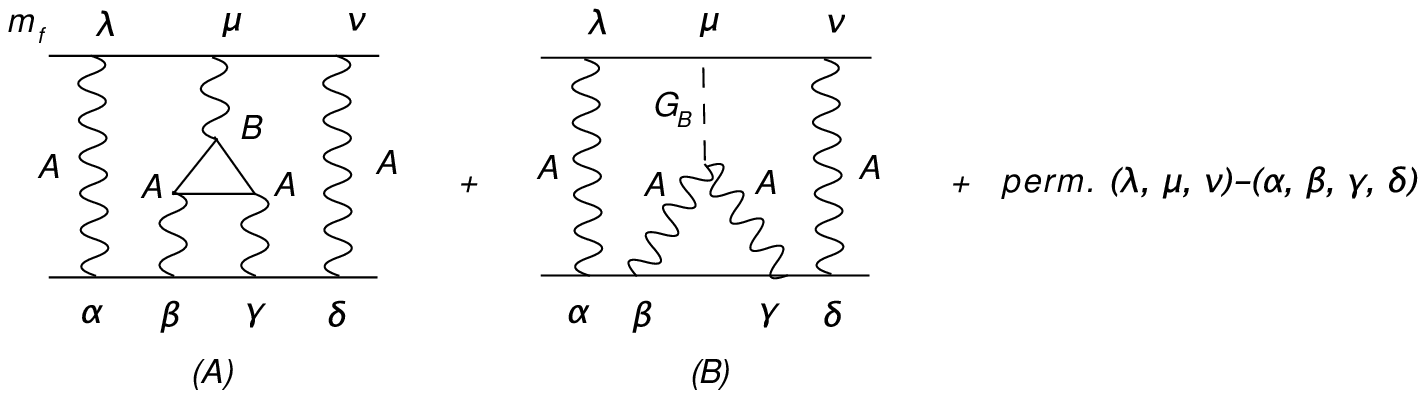}}}\par}
\caption{The WZ counterterm for the restauration of gauge invariance}
\label{fermion8}
\end{figure}

As for the annihilation channel of two fermions (f), 
we illustrate in Figs.~\ref{fermion4} and \ref{fermion5} 
the organization of the expansion to lowest orders for a process of the type $f\bar{f} \to A A$ which is the analogous of $q\bar{q} \to \gamma \gamma$ in this simple model. The presence of a 
goldstone exchange takes place, obviously, only in the massive case. 
Finally, we have included 
the set of gauge-invariant diagrams describing the exchange of A and B gauge bosons in the 
t-channel and with an intermediate triangle anomaly diagram (BAA) (Figs. \ref{fermion7} and \ref{fermion8}). 
In the massless fermion case gauge invariance is 
obtained simply by adding to the basic diagram all the similar ones obtained by permuting the attachments of the gauge 
lines (this involves both the lines at the top 
and at the bottom) and summing only over the topologically independent configurations. 
In the massive case one needs to add to this set of diagrams 2 additionals sets: those containing a goldstone exchange and those involving a WZ interaction. The contributions of these additional diagrams have 
to be symmetrized as well, by moving the attachments of the gauge 
boson/scalar lines.

\section{ The effective action in configuration space
}
Hidden inside the anomalous 3-point functions are some Chern-Simons interactions. 
Their ``extraction'' can be done quite easily if we try to integrate out 
completely a given diagram and look at the structure of the effective action that is so generated directly in configuration space. The resulting action is non-local but contains a contact term that is present independently from the type of external Ward identities that need to be imposed on the external vertices. 
This contact term is a dimension-4 contribution that is identified with a CS 
interaction, while the higher dimensional contributions have 
a non-trivial structure. The variation of both the local and the non-local effective vertex is still a local operator, proportional to $F\wedge F$. The coefficient in front of the local CS interaction changes, 
depending on the external conditions imposed on the diagram (the external Ward identities). In this sense, different vertices may carry different 
CS terms. 

To illustrate this issue in a simple way, we proceed as follows. 
Consider the special case in which the two lines $\mu \nu  $ are on shell, so that $k_1^2=k_2^2=0$. This simplifies our derivation, though 
a more general analysis can also be considered. We work in the specific parameterization in which the vertex satisfies the vector Ward identity on the 
$\mu \nu$ lines, with the anomaly brought entirely on the $\lambda$ line. 

In this case we have 
\beqa
A_1(k_1,k_2)&=&k_1\cdot k_2 A_3(k_1,k_2) \nonumber \\ 
A_2(k_1,k_2) &=& -A_1(k_2,k_1) \nonumber \\
A_5(k_1,k_2) &=& -A_4(k_2,k_1) \nonumber \\
A_6(k_1,k_2) &=& -A_3(k_2,k_1), \nonumber \\
\eeqa 
and defining 
$s=2 k_1\cdot k_2= k^2$, the explicit expressions of $A_1$ and $A_2$ are summarized in the form 
\beq
A_1(k^2)= -i \frac{1}{4 \pi^2} + i C_0(m_f^2,s)
\eeq
with $C_0$ a given function of the ratio $m_f^2/s $ that we redefine as $R(m_f^2/s)$. The typical expression of these functions can be found in the appendix. Here we assume that $s > 4 m_f^2  $, but other regions can be reached by suitable analytic continuations. The important point to be appreciated is the presence of a constant term in this 
invariant amplitude. Notice that the remaining amplitudes do not share this property. If we denote as $T_{c}$ the vertex in configuration space, the contribution to the effective action becomes 

\beqa
\langle T_c^{\lambda\mu\nu}(z,x,y)B_{\lambda}(z) A_\mu(x)A_\nu(y) \rangle 
&=& \frac{1}{4 \pi^2} \langle \epsilon_{\mu\nu\rho\sigma}B^\lambda A^\mu 
F_A^{\rho\sigma} \rangle \nonumber \\
&&
+ \langle R\left(-m_f^2/\square_z)\right)
\left[ \delta(x-z)\delta(y-z)\right]B^{\lambda}(z)A^\mu(x) A^\nu(y) \rangle\nonumber \\
\eeqa
where the $1/\square_z$ operator acts only on the distributions inside the 
squared brackets ($\left[\,\, \right]$). It is not difficult to show that if we perform a gauge variations, say under $B$, of the vertex written in this form, 
then the first term trivially gives the $F_A\wedge F_A$ contribution, while 
the second (non-local) expression summarized in $R$, vanishes identically after an integration by parts. For this one needs to use the Bianchi identities 
of the A, B gauge bosons. 

A similar computation on $A_3, A_4$ etc, can be carried out, but this time 
these contributions do not have a contact interaction as $A_1$ and $A_2$, 
but they are, exactly as the $R()$ term, purely non-local. Their gauge variations are also vanishing. When we impose a parameterization of the triangle 
diagram that redistributes the anomaly in such a way that some axial 
interactions are conserved or, for that reason, any other distribution of the 
partial anomalies on the single vertices, than we are actually introducing 
into the theory some specific CS interactions. We should think of 
these vertices as new effective vertices, fixed by the Ward identities 
imposed on them. Their form is dictated by the conditions 
of gauge invariance. These conditions may appear with an axion term if 
the   corresponding gauge boson, such as $B$ in this case, is anomalous. 
Instead, if the gauge boson is not paired to a shifting axion, such as for 
Y, or the hypercharge in a more general model, 
then gauge invariance under Y is restaured by suitable CS terms.  
The discussion of the phenomenological relevance of these vertices will be 
addressed in related work.

\section{Summary and Conclusions} 
We have analyzed in some detail unitarity issues that emerge in the context of anomalous abelian models 
when the anomaly cancelation mechanism involves a Wess Zumino term, CS interactions and traceless conditions on some 
of the generators. We have investigated the 
features of these types of theories both in their exact and in their broken phases, and we have used s-channel 
unitarity as a simple strategy to achieve this. 
We have illustrated in a simple model (the ``A-B model'') how the axion ($b$) is decomposed into a physical 
field ($\chi$) and a goldstone field ($G_B$) (Eq. \ref{projection}), and how the cancelation of the gauge dependences 
in the S-matrix involves either $b$ or $G_B$, in the St\"uckelberg or the Higgs-St\"uckelberg phase respectively. In 
the St\"uckelberg phase the axion is a Goldstone mode. The physical component of the field, $\chi$, 
appears after spontaneous symmetry breaking, and becomes massive via a combination of both the St\"uckelberg and the Higgs mechanism. Its mass can be driven 
to be light if the Peccei-Quinn breaking contributions in the scalar potential (Eq.\ref{ppqq}) appear with small parameters $(b_1,\lambda_1, \lambda_2)$ compared to the Higgs vev (Eq. \ref{chimass}). Then we have performed a unitarity analysis of this model first in the St\"uckelberg phase and then in the Higgs-St\"uckelberg phase, summarized 
in the set of diagrams collected in Fig. \ref{chern} and Fig. \ref{chernmassive} respectively. A similar analysis has been presented in sections 6 and 7, and is summarized in Figs. \ref{box} and \ref{boxmass} respectively. In the broken phase, the most demanding pattern of cancelation is the 
one involving several anomalous interactions (BBB), and the analysis is summarized 
in Fig. (\ref{brokenphase}). The amplitude for the decay of the axi-Higgs in this model has 
been given in (\ref{chidec}). We have also shown (section 5.1) that in the simple models discussed in this work, Chern-Simons interactions can be absorbed into the triangle diagrams 
by a re-definition of the momentum parameterization, if one rewrites a given amplitude in the basis of the interaction eigenstates. Isolation of the Chern-Simons terms may 
however help in the computation of 3-linear gauge interactions in realistic extensions of the SM 
and can be kept separate from the fermionic triangles. 
Their presence is the indication that the theory requires external Ward identities to be correctly defined at 1-loop. 
Our results will be generalized 
and applied to the analysis of effective string models derived from the orientifold construction which are discussed in related work.

\centerline{\bf Acknowledgements} 
We thank Marco Guzzi, Theodore Tomaras and Marco Roncadelli for discussions. 
The work of C.C. was supported (in part) 
by the European Union through the Marie Curie Research and Training Network ``Universenet'' (MRTN-CT-2006-035863) and by a grant from the UK Royal Society. 
He thanks the Theory Group at the Department 
of Mathematics of the University of Liverpool and in particular Alon Faraggi for discussions and for the kind hospitality.
S.M. and C.C. thank the Physics Department at the University of Crete and in particular Theodore Tomaras for the kind hospitality. 
\newpage

\section{Appendix. The triangle diagrams and their ambiguities}
We have collected in this and in the following appendices some of the more technical material 
which is summarized in the main sections. We present also a 
rather general analysis of the main features of anomalous diagrams, some of 
which are not available in the similar literature on the Standard Model, for instance
due to the different pattern of cancelations of the anomalies required in our case study.

The consistency of these models, in fact,
requires specific realizations of the vector Ward identity for gauge trasformations 
involving the vector currents, which implies a specific parameterization of the fermionic triangle diagrams. 
While the analysis of these triangles is well known in the massless fermion 
case, for massive fermions it is slightly more involved. We have gathered here some 
results concerning these diagrams. 

The typical ${\bf AVV}$ diagram with two vectors and one 
axial-vector current (see Fig.~\ref{VAA1}) is  described in this work using a 
specific parameterization of the loop momenta given by
\beq 
\Delta_{\bf AVV}^{\lambda\mu\nu} = \Delta^{\lambda\mu\nu} =  i^3\int \frac{d^4 q}{(2\pi)^4} 
\frac{Tr\left[\gamma^\mu(\ds{q}+m) \gamma^\lambda\gamma^5(\ds{q}-\ds{k}+m)\gamma^\nu(\ds{q}-\ds{k}_1+m)\right]}
{(q^2-m^2)[(q-k_1)^2-m^2][(q-k)^2-m^2]} + \makebox{exch}.
\eeq
Similarly, for the ${\bf AAA}$ diagram we will use the parameterization
\beq 
\Delta_{\bf AAA}^{\lambda\mu\nu}=\Delta_3^{\lambda\mu\nu}= i^3 \int \frac{d^4 q}{(2\pi)^4} 
\frac{Tr\left[\gamma^\mu\gamma^5 (\ds{q}+m) \gamma^\lambda\gamma^5(\ds{q}-\ds{k}+m)\gamma^\nu\gamma^5(\ds{q}-\ds{k}_1+m)\right]}
{(q^2-m^2)[(q-k_1)^2-m^2][(q-k)^2-m^2]} + \makebox{exch}.
\eeq
In both cases we have included both the direct and the exchanged contributions 
\footnote{Our conventions differ from \cite{Zee} by an overall (-1) since our currents 
are defined as $j_\mu^B=-q_B g_B \overline{\psi}\gamma_\mu \psi$}.
 
In our notation $\overline{\Delta}^{\lambda \mu \nu}$ denotes a single diagram while we will use the 
symbol $\Delta$ to denote the Bose symmetric expression
\beqn
\Delta^{\lambda \mu \nu} = \overline{\Delta}^{\lambda \mu \nu}(k_1, k_2) + \mbox{exchange of}\,\, \{(k_1, \mu),(k_2, \nu)\}.
\eeqn
To be noticed that the exchanged diagram is equally described by a diagram equal to the first diagram but with a reversed fermion flow. Reversing the fermion flow is sufficient to guarantee Bose symmetry of the two {\bf V} lines. Similarly, for an ${\bf AAA}$ diagram, the exchange of any two {\bf A} lines is sufficient to render the entire diagram completely symmetric under cyclic permutations of the three ${\bf AAA}$ lines. 

Let's now consider the ${\bf AVV}$ contribution and work out some preliminaries.
It is a simple exercise to show that the parameterization that we have used 
above indeed violates the 
vector Ward identity (WI) on the $\mu\nu $ vector lines giving 
\beqa
k_{1\mu} \Delta^{\la\mu\nu}(k_1,k_2) = a_1 \epsilon^{\lambda\nu\alpha\beta} 
k_1^\alpha k_2^\beta \nonumber \\
k_{2\nu} \Delta^{\la\mu\nu}(k_1,k_2) = a_2 \epsilon^{\lambda\mu\alpha\beta} 
k_2^\alpha k_1^\beta \nonumber \\
k_{\la} \Delta^{\la\mu\nu}(k_1,k_2) = a_3 \epsilon^{\mu\nu\alpha\beta} 
k_1^\alpha k_2^\beta, \nonumber \\
\eeqa
where
\beq
 a_1=-\frac{i}{8 \pi^2} \qquad a_2=-\frac{i}{8 \pi^2} \qquad a_3=-\frac{i}{4 \pi^2}.
\label{basic}
\eeq
Notice that $a_1=a_2 $, as expected from the Bose symmetry of the two {\bf V} lines. It is also well known that the total anomaly 
$a_1+a_2 + a_3 \equiv  a_n$ is regularization scheme independent 
($a_n=-\frac{i}{2 \pi^2}$). We do not impose any WI on the {\bf V} lines, 
conditions which would bring the anomaly only to the axial vertex, as done for the SM case,   
but we will determine consistently the value of the three anomalies at a later stage from the requirement of gauge invariance of the effective action, with the inclusion of 
the axion terms. 
%
%It is obvious that a shift 
%of the internal momentum will always redefine the values of $a_1$ and $a_2$ - which will stay equal for any parameterization  and can even be 
%set to vanish, thereby obtaining CVC (conserved vector current) on the two V vertices and the total anomaly only on the A vertex 
%($a_3=a_n$), as is the case in the SM. 
To render our discussion self-contained, and define our notations, we briefly review the issue of the shift dependence of these diagrams. 

We recall that a shift of the momentum in the integrand $(p\rightarrow p + a)$ where $a$ is the most general momentum written in terms of the two independent external momenta of the triangle diagram $(a=\alpha (k_1 + k_2) + \beta(k_1 - k_2))$ induces on  $\Delta$ changes that 
appear only through a dependence on one of the two parameters characterizing $a$, that is 

\beq
\Delta^{\la\mu\nu}(\beta,k_1,k_2)= \Delta^{\la\mu\nu}(k_1,k_2) - \frac{i}{4 \pi^2}\beta \epsilon^{\lambda\mu\nu\sigma}\left( k_{1\sigma} - 
k_{2\sigma}\right).
\eeq

We have introduced the notation $\Delta^{\la\mu\nu} (\beta,k_1,k_2)$ to denote the shifted 3-point function, while 
$\Delta^{\la\mu\nu}(k_1,k_2)$ denotes the original one, with a 
vanishing shift.
In our parameterization, the choice $\beta=-\frac{1}{2}$ corresponds to conservation of the vector current 
and brings the anomaly to the axial vertex
\beqn
k_{1\mu}\Delta^{\lambda\mu\nu}(a,k_1,k_2)&=& 0,\nonumber\\
k_{2\nu}\Delta^{\lambda\mu\nu}(a,k_1,k_2)&=&0,\nonumber\\
k_\lambda\Delta^{\lambda\mu\nu}(a,k_1,k_2)&=&-\frac{i}{2 \pi^2}\varepsilon^{\mu\nu\alpha\beta}k_1^\alpha k_2^\beta
\label{bshift}
\eeqn
with $a_n=a_1+a_2+a_3=-\frac{i}{2 \pi^2}$ still equal to the total anomaly. Therefore, starting from generic values of $(a_1=a_2,a_3)$, for instance from the values deduced from the 
basic parameterization (\ref{basic}), an additional shift with parameter $\beta' $ gives

\beq
\Delta^{\lambda\mu\nu}(\beta',k_1,k_2)= \Delta^{\lambda\mu\nu}(\beta,k_1,k_2)-\frac{i \beta'}{4 \pi^2}
\varepsilon^{\lambda\mu\nu\sigma}(k_1-k_2)_\sigma
\eeq
and will change the Ward identities into the form 
\beqn
k_{1\mu}\Delta^{\lambda\mu\nu}(\beta',k_1,k_2)&=& (a_1 -\frac{i \beta'}{4 \pi^2})
\varepsilon^{\lambda\nu\alpha\beta}k_1^\alpha k_2^\beta,\nonumber\\
k_{2\nu}\Delta^{\lambda\mu\nu}(\beta',k_1,k_2)&=&(a_2-\frac{i \beta'}{4 \pi^2})
\varepsilon^{\lambda\mu\alpha\beta}k_2^\alpha k_1^\beta,\nonumber\\
k_\lambda\Delta^{\lambda\mu\nu}(\beta',k_1,k_2)&=&(a_3+\frac{i \beta'}{2 \pi^2})
\varepsilon^{\mu\nu\alpha\beta}k_1^\alpha k_2^\beta,
\label{bbshift}
\eeqn
where $a_2=a_1$. There is an intrinsic ambiguity in the definition of the amplitude, which 
can be removed by imposing CVC on the vector vertices, as done, for instance, in Rosenberg's 
original paper \cite{Rosenberg} and discussed in an appendix. We remark 
once more that, in our case, this condition is not automatically required. 
The distribution of the anomaly may, in general, be different and we are 
defining, 
in this way, new effective parameterizations of the 3-point anomalous vertices.

It is 
therefore convenient to introduce a notation that makes explicit this dependence and for this reason we define 
\beqa
a_1(\beta)&=&a_2(\beta)=-\frac{i}{8\pi^2} - \frac{i}{4 \pi^2}\beta \nonumber \\ 
a_3(\beta) &=& -\frac{i}{4\pi^2} + \frac{i}{2 \pi^2}\beta, \nonumber \\ 
\label{a12}
\eeqa
with
\beq
a_1(\beta) + a_2(\beta) + a_3(\beta)=a_n=-\frac{i}{2 \pi^2}.
\eeq
Notice that the additional $\beta$-dependent contribution amounts 
to a Chern-Simons 
interaction (see the appendices). Clearly, this contribution can be moved around at will and is related to the presence of two divergent terms in the 
general triangle diagram that need to be fixed appropriately using the underlying Bose symmetries of the 3-point functions. 

%%%%%%%%%%%%%%%%%%%%%%%%%%%%%%%%%%%%%%%%%%%%%%%%%%%%%%%%%%%%%%%%%%
\begin{figure}[t]
{\centering \resizebox*{12cm}{!}{\rotatebox{0}
{\includegraphics{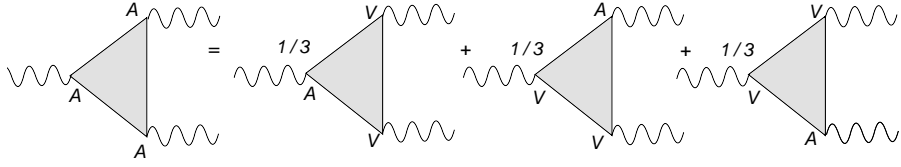}}}\par}
\caption{Distribution of the axial anomaly for the ${\bf AAA}$ diagram}
\label{distribution}
\end{figure}
%%%%%%%%%%%%%%%%%%%%%%%%%%%%%%%%%%%%%%%%%%%%%%%%%%%%%%%%%%%%%%%%%%

The regularization of the $\bf AAA$ vertex, instead, has to respect the complete Bose symmetry of the diagram and this can be achieved with the symmetric expression 
\beq
\Delta_3^{\lambda\mu\nu}(k_1,k_2)=\frac{1}{3}[\Delta^{\lambda\mu\nu}(k_1,k_2)+\Delta^{\mu\nu\lambda}(k_2,-k)+\Delta^{\nu\lambda\mu}
(-k,k_1)].
\label{delta3}
\eeq
%
%\begin{figure}[tbh]
%{\centering \resizebox*{12cm}{!}{\rotatebox{0}
%{\includegraphics{delta3.eps}}}\par}
%\caption{}
%\label{}
%\end{figure}
%
It is an easy exercise to show that this symmetric choice is independent from the momentum shift 
\beq
\Delta_3^{\lambda\mu\nu}(\beta,k_1,k_2)=
\Delta_3^{\lambda\mu\nu}(k_1,k_2)
\eeq
and that the anomaly is equally distributed among the 3 vertices, $a_1=a_2=a_3=a_n/3$, as shown in 
Fig.~ \ref{distribution}.
We conclude this section with some comments regarding the kind of invariant amplitudes appearing in the definition 
of $\Delta$ which help to clarify the role of the CS terms in the parameterization of these diagrams in momentum space. 
For $\overline{\Delta}_{{\la}\mu\nu}^{({\bf AVV})}$, expressed using Rosenberg's parametrization, one obtains 
\beqa
\overline{\Delta}_{{\la}\mu\nu} &=& \hat{a}_1 \epsilon[k_1,\mu,\nu,\la] 
+ \hat{a}_2 \epsilon[k_2,\mu,\nu,\la] +
\hat{a}_3 \epsilon[k_1,k_2,\mu,\la]{k_1}^{\nu}\nonumber \\
&& + \hat{a}_4\epsilon[k_1,k_2,\mu,\la]k_2^{\nu}
 + \hat{a}_5 \epsilon[k_1,k_2,\nu,\la]k_1^\mu 
+ \hat{a}_6 \epsilon[k_1,k_2,\nu,\la]k_2^\mu,
\label{Ros}
\eeqa
originally given in \cite{Rosenberg}, with $\lambda$ being the axial-vector vertex. 
By power-counting, 2 invariant amplitudes are divergent, $a_1$ and $a_2$, while the $a_i$ with $i\geq 3$ are finite\footnote{
We will be using the notation $\epsilon[a,b,\mu,\nu]\equiv \epsilon_{\alpha\beta\mu\nu}
a^\alpha b^\beta$ to denote the structures in the expansion of the anomalous triangle diagrams}. 

Instead, for the direct plus the exchanged diagrams we use the expression
 
\beqa
\Delta_{\underline{\la}\mu\nu} + \Delta_{\underline{\la}\nu\mu} &=& 
(\hat{a}_1 - \hat{a}_2) \epsilon[k_1,\mu,\nu,\la] + (a_2- a_1)\epsilon[k_2,\mu,\nu,\la]\nonumber \\
&& +(\hat{a}_3- \hat{a}_6) \epsilon[k_1,k_2,\mu,\la]{k_1}^{\nu}
 + (\hat{a}_4 - \hat{a}_5)\epsilon[k_1,k_2,\mu,\la]k_2^{\nu}\nonumber \\
&& +(\hat{a}_5 - \hat{a}_4 ) \epsilon[k_1,k_2,\nu,\la]k_1^\mu 
+ (\hat{a}_6 - \hat{a}_3)\epsilon[k_1,k_2,\nu,\la]k_2^\mu \nonumber \\
&=& \underline{a}_1 \epsilon[k_1,\mu,\nu,\la] + 
\underline{a}_2\epsilon[k_2,\mu,\nu,\la] 
+\underline{a}_3 \epsilon[k_1,k_2,\mu,\la]{k_1}^{\nu}\nonumber \\
&& +\underline{a}_4 \epsilon[k_1,k_2,\mu,\la]{k_2}^{\nu} +
\underline{a}_5 \epsilon[k_1,k_2,\nu,\la]k_1^\mu +
\underline{a}_6\epsilon[k_1,k_2,\nu,\la]k_2^\mu
\label{Ros1}
\eeqa

where clearly $\ul{a}_2=-\ul{a}_1, \ul{a}_3=-\ul{a}_6$ and $\ul{a}_4=-\ul{a}_5$. 
The CS contributions are those proportional to the two terms linear in the external momenta. 
We recall that in Rosenberg, these linear terms are re-expressed in terms of the remaining ones by imposing the vector Ward identities on the {\bf V}-lines. 
As already explained, we will instead assume, in our case, that the distribution of the anomaly among the 3 vertices of all the anomalous diagrams of the theory respects 
the requirement of Bose symmetry, with no additional constraint. 
A discussion of some 
technical points concerning the regularization of this and other diagrams both in 4 dimensions and in other schemes, such as Dimensional Regularization (DR) can be found below. For instance, one can find there the proof of the identical vanishing 
of $\Delta_{\bf VVV}$ worked out in both schemes. 
In this last case 
this result is obtained after removing the so called 
hat-momenta of the t'Hooft-Veltman scheme on the external lines. 
In this scheme this is possible since one can choose the external momenta to lay on a four-dimensional subspace (see \cite{CorianoGordon} for a discussion of these methods). 
We remark also that it is also quite useful to be able to switch from momentum space to configuration 
space with ease, and for this purpose we introduce the 
Fourier transforms of (\ref{dd1}) and (\ref{dd2}) in the anomaly equations, obtaining their 
expressions in configuration space
\beqa
\frac{\partial}{\partial x^\mu}T^{\la\mu\nu}_{\bf AVV}(x,y,z)&=& 
i a_1(\beta) \epsilon^{\la\nu\alpha\beta}\frac{\partial}{\partial x^\alpha}
\frac{\partial}{\partial y^\beta}\left(\delta^4(x-z)\delta^4(y-z)\right),
\nonumber \\
\frac{\partial}{\partial y^\nu}T^{\la\mu\nu}_{\bf AVV}(x,y,z)&=& 
i a_2(\beta) \epsilon^{\la\mu\alpha\beta} \frac{\partial}{\partial y^\alpha}
\frac{\partial}{\partial x^\beta}\left(\delta^4(x-z)\delta^4(y-z)\right),
\nonumber \\
\frac{\partial}{\partial z^\la}T^{\la\mu\nu}_{\bf AVV}(x,y,z)&=& 
i a_3(\beta) \epsilon^{\mu\nu\alpha\beta}\frac{\partial}{\partial x^\alpha}
\frac{\partial}{\partial y^\beta}\left(\delta^4(x-z)\delta^4(y-z)\right),
\eeqa
with $a_1, a_2$ and $a_3$ as in (\ref{a12}), for the $\bf AVV$ case
and 
\beqa
\frac{\partial}{\partial x^\mu}T^{\la\mu\nu}_{\bf AAA}(x,y,z)&=& 
i \frac{a_n}{3}\epsilon^{\la\nu\alpha\beta}\frac{\partial}{\partial x^\alpha}
\frac{\partial}{\partial y^\beta}\left(\delta^4(x-z)\delta^4(y-z)\right),
\nonumber \\
\frac{\partial}{\partial y^\nu}T^{\la\mu\nu}_{\bf AAA}(x,y,z)&=& 
i \frac{a_n}{3} \epsilon^{\la\mu\alpha\beta} \frac{\partial}{\partial y^\alpha}\frac{\partial}{\partial x^\beta}\left(\delta^4(x-z)\delta^4(y-z)\right),
\nonumber \\
\frac{\partial}{\partial z^\la}T^{\la\mu\nu}_{\bf AAA}(x,y,z)&=& 
i \frac{a_n}{3} \epsilon^{\mu\nu\alpha\beta}\frac{\partial}{\partial x^\alpha}\frac{\partial}{\partial y^\beta}\left(\delta^4(x-z)\delta^4(y-z)\right),
\eeqa
for the $\bf AAA$ case. Notice that in this last case we have distributed 
the anomaly equally among the three vertices. These relations will be needed when we derive the anomalous variation of the effective action directly in configuration space.

\section{Appendix. Chern Simons cancelations}

Having isolated the CS contributions, as shown in 
Fig.~\ref{chern}, the cancelation of the gauge dependence can be obtained 
combining all these terms so to obtain\footnote{The symmetry factor of each configuration 
is easily identified ias the first factor in each separate contribution.} 

\beqn
S_{\xi} &=& \Delta^{\lambda \mu \nu}(-k^{}_{1}, -k^{}_{2}) 
\left[  \frac{-i}{k^{2} - \xi^{}_{B} M^{2}_{1}} \frac{k^{\lambda} k^{\lambda'}}{ M^{2}_{1} } \right]
 \Delta^{\lambda' \mu' \nu'}(k^{}_{1}, k^{}_{2})   \nonumber\\
&+& 4 \times \left( \frac{4}{M} C^{}_{AA} \right)^{2} 
\epsilon^{\mu\nu\rho\sigma} k^{}_{1\rho} k^{}_{2\sigma} \left[ \frac{i}{k^{2} - \xi^{}_{B} M^{2}_{1}} \right] 
\epsilon^{\mu'\nu'\rho'\sigma'} k^{}_{1\rho'} k^{}_{2\sigma'} \nonumber\\
&+&  4 \times ( i d^{}_{1} \epsilon^{\mu \nu \lambda \sigma} (k^{}_{1}-k^{}_{2})_{\sigma} ) 
\left[ \frac{-i}{k^{2} - \xi^{}_{B} M^{2}_{1}} \frac{k^{\lambda} k^{\lambda'}}{ M^{2}_{1} } \right] 
(- i d^{}_{1} \epsilon^{\mu' \nu' \lambda' \sigma'} (k^{}_{1}-k^{}_{2})_{\sigma'} )  \nonumber\\
&+&2 \times \Delta^{\lambda \mu \nu}(-k^{}_{1}, -k^{}_{2}) 
\left[  \frac{-i}{k^{2} - \xi^{}_{B} M^{2}_{1}} \frac{k^{\lambda} k^{\lambda'}}{ M^{2}_{1} } \right] 
(- i d^{}_{1} \epsilon^{\mu' \nu' \lambda' \sigma'} (k^{}_{1}-k^{}_{2})_{\sigma'} )   \nonumber\\
&+&2 \times ( i d^{}_{1} \epsilon^{\mu \nu \lambda \sigma } (k^{}_{1}-k^{}_{2})_{\sigma} ) 
 \left[  \frac{-i}{k^{2} - \xi^{}_{B} M^{2}_{1}} \frac{k^{\lambda} k^{\lambda'}}{ M^{2}_{1} } \right] 
 \Delta^{\lambda' \mu' \nu'}(k^{}_{1}, k^{}_{2}),
\eeqn
and using the relevant Ward identities these simply so to obtain 
\beqn
S_{\xi}&=&(- a^{}_{3}(\beta) \epsilon^{\mu \nu \rho \sigma} k^{}_{1\rho} k^{}_{2\sigma}  )  
\left[  \frac{-i}{k^{2} - \xi^{}_{B} M^{2}_{1}} \frac{1}{ M^{2}_{1} } \right] 
 ( a^{}_{3}(\beta) \epsilon^{\mu' \nu' \rho' \sigma'} k^{}_{1\rho'} k^{}_{2\sigma'}  )  \nonumber\\
&+&  4 \times \frac{16}{M^{2}} \left[ \left( -\frac{ d^{}_{1} }{2} + \frac{i}{2} a^{}_{3}(\beta)\frac{1}{4} \right)^{2} 
\frac{M^{2}}{M^{2}_{1}} \right] \epsilon^{\mu \nu \rho \sigma} k^{}_{1\rho} k^{}_{2\sigma} 
\left[ \frac{i}{k^{2} - \xi^{}_{B} M^{2}_{1}} \right] \epsilon^{\mu' \nu' \rho' \sigma'} k^{}_{1\rho'} k^{}_{2\sigma'} \nonumber\\
&+&  4 \times d^{2}_{1} \left[  \frac{-i}{k^{2} - \xi^{}_{B} M^{2}_{1}} \frac{1}{ M^{2}_{1} } \right] 
4 \epsilon^{\mu \nu \rho \sigma} k^{}_{1\rho} k^{}_{2\sigma} 
\epsilon^{\mu' \nu' \rho' \sigma'} k^{}_{1\rho'} k^{}_{2\sigma'}   \nonumber\\
&+&  2 \times (- a^{}_{3}(\beta) \epsilon^{\mu \nu \rho \sigma} k^{}_{1\rho} k^{}_{2\sigma}  )  
\left[  \frac{-i}{k^{2} - \xi^{}_{B} M^{2}_{1}} \frac{1}{ M^{2}_{1} } \right] 
(+ i d^{}_{1} 2 \epsilon^{\mu' \nu' \lambda' \sigma'} k^{\lambda'}_{1} k^{\sigma'}_{2} ) \nonumber\\
&+& 2 \times (- i d^{}_{1} 2 \epsilon^{\mu \nu \lambda \sigma } k^{\lambda}_{1} k^{\sigma}_{2} ) 
\left[  \frac{-i}{k^{2} - \xi^{}_{B} M^{2}_{1}} \frac{1}{ M^{2}_{1} } \right]  
( a^{}_{3}(\beta) \epsilon^{\mu' \nu' \rho' \sigma'} k^{}_{1\rho'} k^{}_{2\sigma'}  ) = 0.
\eeqn
Having shown the cancelation of the gauge-dependent terms, the gauge independent contribution becomes 
%
%$P^{\lambda\lambda'} 
%= P^{\lambda\lambda'}_{0} +P^{\lambda\lambda'}_{\xi} $ we obtain 
%
\beqn
S_{0} &=& \Delta^{\lambda \mu \nu}(-k^{}_{1}, -k^{}_{2}) 
\left[  \frac{-i}{k^{2} - M^{2}_{1}} \left( g^{\lambda \lambda'} - \frac{k^{\lambda} k^{\lambda'}}{ M^{2}_{1} }  \right)\right]
 \Delta^{\lambda' \mu' \nu'}(k^{}_{1}, k^{}_{2})   \nonumber\\
&+& 4 \times ( i d^{}_{1} \epsilon^{\,\mu \nu \lambda \sigma} (k^{}_{1}-k^{}_{2})_{\sigma} ) 
\left[    \frac{-i}{k^{2} - M^{2}_{1}} \left( g^{\lambda \lambda'} - \frac{k^{\lambda} k^{\lambda'}}{ M^{2}_{1} } \right) \right] 
(- i d^{}_{1} \epsilon^{\,\mu' \nu' \lambda' \sigma'} (k^{}_{1}-k^{}_{2})_{\sigma'} )  \nonumber\\
&+& 2 \times \Delta^{\lambda \mu \nu}(-k^{}_{1}, -k^{}_{2}) 
\left[  \frac{-i}{k^{2} - M^{2}_{1}} \left( g^{\lambda \lambda'} - \frac{k^{\lambda} k^{\lambda'}}{ M^{2}_{1} } \right) \right] 
(- i d^{}_{1} \epsilon^{\,\mu' \nu' \lambda' \sigma'} (k^{}_{1}-k^{}_{2})_{\sigma'} )   \nonumber\\
&+&2 \times ( i d^{}_{1} \epsilon^{\,\mu \nu \lambda \sigma } (k^{}_{1}-k^{}_{2})_{\sigma} ) 
 \left[  \frac{-i}{k^{2} - M^{2}_{1}} \left( g^{\lambda \lambda'} - \frac{k^{\lambda} k^{\lambda'}}{ M^{2}_{1}} \right)  \right] 
 \Delta^{\lambda' \mu' \nu'}(k^{}_{1}, k^{}_{2}). 
\eeqn  
At this point we need to express the triangle diagrams in terms of their shifting parameter $\beta$ using the shift-relations
\beqn
&&\Delta^{\lambda \mu \nu}(\beta, k^{}_{1}, k^{}_{2}) = \Delta^{\lambda \mu \nu}(k^{}_{1}, k^{}_{2}) 
- \frac{i}{4 \pi^{2}} \beta \epsilon^{\lambda \mu \nu \sigma} (k^{}_{1}- k^{}_{2})_{\sigma},  \\
&& \Delta^{\lambda \mu \nu}(\beta,- k^{}_{1},- k^{}_{2}) = \Delta^{\lambda \mu \nu}(- k^{}_{1},- k^{}_{2}) 
+ \frac{i}{4 \pi^{2}} \beta \epsilon^{\lambda \mu \nu \sigma} (k^{}_{1}- k^{}_{2})_{\sigma}, 
\eeqn
and with the substitution $d^{}_{1} = -i a^{}_{1}(\beta)/2$ we obtain 
\beqn
S^{}_{0} &=& \left( \Delta^{\lambda \mu \nu}(- k^{}_{1},- k^{}_{2}) 
+ \frac{i}{4 \pi^{2}} \beta \epsilon^{\,\lambda \mu \nu \sigma} (k^{}_{1}- k^{}_{2})_{\sigma}  \right) P^{\lambda \lambda'}_{0}
\left( \Delta^{\lambda' \mu' \nu'}( k^{}_{1}, k^{}_{2}) 
- \frac{i}{4 \pi^{2}} \beta \epsilon^{\,\lambda' \mu' \nu' \sigma'} (k^{}_{1}- k^{}_{2})_{\sigma'} \right) \nonumber\\
&+&4 \times \left( \frac{1}{2} a^{}_{1}(\beta)
\epsilon^{\,\mu \nu \lambda \sigma} (k^{}_{1}- k^{}_{2})_{\sigma} \right) P^{\lambda \lambda'}_{0} 
\left( - \frac{1}{2} a^{}_{1}(\beta)
\epsilon^{\,\mu' \nu' \lambda' \sigma'} (k^{}_{1}- k^{}_{2})_{\sigma'} \right)  \nonumber\\
&+&2 \times  \left( \Delta^{\lambda \mu \nu}(- k^{}_{1},- k^{}_{2}) 
+ \frac{i}{4 \pi^{2}} \beta \epsilon^{\,\lambda \mu \nu \sigma} (k^{}_{1}- k^{}_{2})_{\sigma}  \right) P^{\lambda \lambda'}_{0}
\left( - \frac{1}{2} a^{}_{1}(\beta)
\epsilon^{\, \mu' \nu' \lambda' \sigma'} (k^{}_{1}- k^{}_{2})_{\sigma'} \right) \nonumber\\
&+& 2 \times  \left( \frac{1}{2} a^{}_{1}(\beta)
\epsilon^{\, \mu \nu \lambda \sigma} (k^{}_{1}- k^{}_{2})_{\sigma} \right) P^{\lambda \lambda'}_{0} 
\left( \Delta^{\lambda' \mu' \nu'}(k^{}_{1}, k^{}_{2}) 
- \frac{i}{4 \pi^{2}} \beta \epsilon^{\,\lambda' \mu' \nu' \sigma'} (k^{}_{1}- k^{}_{2})_{\sigma'}  \right).
\eeqn
Introducing the explicit expression for $a^{}_{1}(\beta)$, it is an easy 
exercise to show the equivalence between $S^{}_{0}$ and diagram A of Fig.~\ref{chern}, with a choice of the shifting parameter  that corresponds 
to the CVC condition ($\beta=-1/2$) 

\beqn
S^{}_{0} \equiv \left( \Delta^{\lambda \mu \nu}(- k^{}_{1},- k^{}_{2}) 
- \frac{i}{8 \pi^{2}} \epsilon^{\,\lambda \mu \nu \sigma} (k^{}_{1}- k^{}_{2})_{\sigma}  \right) P^{\lambda \lambda'}_{0}
\left( \Delta^{\lambda' \mu' \nu'}( k^{}_{1}, k^{}_{2}) 
+ \frac{i}{8 \pi^{2}} \beta \epsilon^{\,\lambda' \mu' \nu' \sigma'} (k^{}_{1}- k^{}_{2})_{\sigma'} \right). \nonumber \\
\eeqn 
%
%
% \begin{figure}[tbh]
% {\centering \resizebox*{16cm}{!}{\rotatebox{0}
% {\includegraphics{trepuntiVVA.eps}}}\par}
% \caption{The AVV triangle}
% \label{}
% \end{figure}
%
\subsection{Cancelation of gauge dependences in the broken Higgs phase}
In this case we have (see Fig.~\ref{chernmassive})
\beqn
S^{}_{\xi} &=& A^{}_{\xi} + B^{}_{\xi} + C^{}_{\xi} + D^{}_{\xi} + E^{}_{\xi} + F^{}_{\xi} + G^{}_{\xi} + H^{}_{\xi} \nonumber \\
&=& \Delta^{\lambda \mu \nu}(-k^{}_{1}, -k^{}_{2}) 
\left[  \frac{-i}{k^{2} - \xi^{}_{B} M^{2}_{B}} \frac{k^{\lambda} k^{\lambda'}}{ M^{2}_{B} } \right]
 \Delta^{\lambda' \mu' \nu'}(k^{}_{1}, k^{}_{2})   \nonumber\\
&+& 4 \times \left( \frac{4}{M} \alpha^{}_{2} C^{}_{AA} \right)^{2} 
\epsilon^{\mu\nu\rho\sigma} k^{}_{1\rho} k^{}_{2\sigma} \left[ \frac{i}{k^{2} - \xi^{}_{B} M^{2}_{B}} \right] 
\epsilon^{\mu'\nu'\rho'\sigma'} k^{}_{1\rho'} k^{}_{2\sigma'} \nonumber\\
&+&  4 \times ( i d^{}_{1} \epsilon^{\mu \nu \lambda \sigma} (k^{}_{1}-k^{}_{2})_{\sigma} ) 
\left[ \frac{-i}{k^{2} - \xi^{}_{B} M^{2}_{B}} \frac{k^{\lambda} k^{\lambda'}}{ M^{2}_{B} } \right] 
(- i d^{}_{1} \epsilon^{\mu' \nu' \lambda' \sigma'} (k^{}_{1}-k^{}_{2})_{\sigma'} )  \nonumber\\
&+&2 \times \Delta^{\lambda \mu \nu}(-k^{}_{1}, -k^{}_{2}) 
\left[  \frac{-i}{k^{2} - \xi^{}_{B} M^{2}_{B}} \frac{k^{\lambda} k^{\lambda'}}{ M^{2}_{B} } \right] 
(- i d^{}_{1} \epsilon^{\mu' \nu' \lambda' \sigma'} (k^{}_{1}-k^{}_{2})_{\sigma'} )   \nonumber\\
&+&2 \times ( i d^{}_{1} \epsilon^{\mu \nu \lambda \sigma } (k^{}_{1}-k^{}_{2})_{\sigma} ) 
 \left[  \frac{-i}{k^{2} - \xi^{}_{B} M^{2}_{B}} \frac{k^{\lambda} k^{\lambda'}}{ M^{2}_{B} } \right] 
 \Delta^{\lambda' \mu' \nu'}(k^{}_{1}, k^{}_{2})   \nonumber\\
&+&  2 \times \Delta^{\mu \nu}(-k^{}_{1},- k^{}_{2}) \left( 2i \frac{m^{}_{f}}{M^{}_{B}} \right) 
\left[ \frac{i}{k^{2} - \xi^{}_{B} M^{2}_{B} } \right] \left( \frac{4}{M} \alpha^{}_{2} C^{}_{AA} 
\epsilon^{\mu' \nu' \rho' \sigma'} k^{}_{1\rho'}   k^{}_{2\sigma'}\right)  \nonumber\\
&+& \Delta^{\mu \nu}(-k^{}_{1},- k^{}_{2}) \left( 2i \frac{m^{}_{f}}{M^{}_{B}} \right) 
\left[ \frac{i}{k^{2} - \xi^{}_{B} M^{2}_{B} } \right] \left( 2i \frac{m^{}_{f}}{M^{}_{B}} \right) 
\Delta^{\mu' \nu'}(k^{}_{1}, k^{}_{2})    \nonumber\\
&+&  2 \times \left( \frac{4}{M} \alpha^{}_{2} C^{}_{AA} 
\epsilon^{\mu \nu \rho \sigma} k^{}_{1\rho}   k^{}_{2\sigma}\right)  
\left[ \frac{i}{k^{2} - \xi^{}_{B} M^{2}_{B} } \right]   \left( 2i \frac{m^{}_{f}}{M^{}_{B}} \right) 
  \Delta^{\mu \nu}(k^{}_{1}, k^{}_{2})\nonumber \\
\eeqn
The vanishing of this expression cna be checked as in the previous case, using the massive version of 
the anomalous Ward identities in the triangular graphs involving $\Delta$.
%
%%%%%%%%%%%%%%%%%%%%%%%%%%%%%%%%%%%%%%%%%%%%%%%%%%%%%%%%%%%%%%%%
\subsection{Cancelations in the A-B Model: BB $\rightarrow$ BB mediated by a B gauge boson}
\begin{figure}[t]
{\centering \resizebox*{15cm}{!}{\rotatebox{0}
{\includegraphics{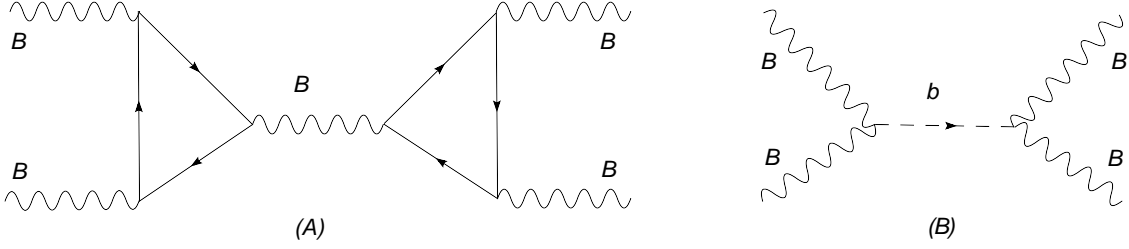}}}\par}
\caption{ Relevant diagrams for the unitarity check before symmetry breaking.}
\label{unitaritycheck}
\end{figure}

%%%%%%%%%%%%%%%%%%%%%%%%%%%%%%%%%%%%%%%%%%%%%%%%%%%%%%%%%%%%%%%%%
%
Let's now discuss the exchange of a B gauge boson in the s-channel before spontaneous symmetry breaking. 
The relevant diagrams are shown in Fig.~\ref{unitaritycheck}. We remark, obviously, that each diagram has to be inserted with the correct multiplicity factor in order 
to obtain the cancelation of the unphysical poles. 

%
%%%%%%%%%%%%%%%%%%%%
%%%%%%%%%%%%%%%%%%%%
%
In this case, from Bose-symmetry, the anomaly is equally distributed among the 3 vertices, $a_1=a_2=a_3=a_n/3$, as we have discussed 
above. We recall that from the variations $\delta_{B} \mathcal{L}_{an}$ and $\delta_{B} \mathcal{L}_{b}$ the relevant terms are
\beqa
\frac{1}{3!} \delta_{B} \langle T_{BBB}^{\lambda \mu \nu} B^{\lambda}(z) 
B^{\mu}(x) B^{\nu}(y) \rangle &=&  \,  \frac{i g^{\,3}_{B} }{ 3! } \, a_n \langle   \theta_{B} \, \, \frac{F^{B}\wedge F^{B}}{4}  \rangle      \nonumber\\
\delta_{B} \langle \frac{ C^{}_{BB} }{M} \, b F^{B}\wedge F^{B}  \rangle &=& - \, C^{}_{BB} \frac{M_1}{M} 
\langle \theta^{}_B F^{B}\wedge F^{B}  \rangle
\; \;\; \mbox{from}  \;\; \;\delta_{B} b = - M_1 \theta_{B}, \nonumber \\
\eeqa
so that from the condition of anomaly cancelation we obtain 
\beqa
- C^{}_{BB} \frac{M_1}{M} + \frac{ i g^{\,3}_{B} }{ 3! }\, \frac{1}{4} \, a^{}_{n} = 0 \,\, \Longleftrightarrow \,\,
  C^{}_{BB} =  \,\frac{i g^{\,3}_{B} }{ 3! }   \, \frac{1}{4}  \,a^{}_n \frac{M}{M_1},
\eeqa
which fixes the appropriate value of the coefficient of the WZ term. 
One can easily show the correspondence between a Green-Schwarz term $\frac{C^{}_{BB}}{M}\, b \, F^B \wedge F^B$ and a vertex 
$4 \frac{ C^{}_{BB}}{M} \, \varepsilon^{\mu \nu \rho \sigma} k_1^{\rho} k_2^{\sigma}$ in momentum representation, a derivation of which can be found in an appendix.
%%%%%%%%%%%%%%%%%%%%%%%%%%%%%%%%%%%%%%%%%%%%%%%%
%The rotation of the axion b into

%\beqa
%b = \frac{gv}{M_B} \chi + \frac{M_1}{M_B} G  \equiv \alpha_1 \chi + \alpha_2 G.
%\eeqa 

%Since the propagator of the physical axion $\chi$ is independent of the parameter $\xi_B$, the cancellation of the $\xi_B$ dependence 
%must take place between the gauge-dependent part of the boson exchange diagram and the diagram with the goldstone exchange. 
%The corresponding vertex is given by

%\beqa
% \frac{4}{M} \, \alpha_2\,  l_2 \, \varepsilon^{\mu \nu \rho \sigma} k_1^{\rho} k_2^{\sigma}.
%\eeqa

%where we have projected b over the goldstone component. 

Taking into account only the gauge-dependent parts of the two diagrams, we have that the diagram with the exchange of the gauge boson 
B can be written as 
\beqa
\mathcal{A}^{}_{\xi} =
\Delta^{\lambda \, \mu \, \nu} (- k^{}_{1}, -k^{}_{2}) \left[  \frac{- \, i}{  k^2  - \xi_{B} M_1^2} \left( 
\frac{k^\lambda \, k^{\lambda^\prime}}{M_1^2} \right)  \right] \Delta^{\lambda^\prime \, \mu^\prime \, \nu^\prime}(k^{}_{1}, k^{}_{2})
\eeqa
and the diagram with the exchange of the axion b is
\beqa
\mathcal{B}^{}_{\xi} = 
4 \times \left( \frac{4}{M} C^{}_{BB}  \right)^2  \varepsilon^{\mu \nu \rho \sigma}
 k_1^\rho k_2^\sigma \left(  \frac{i}{k^2 - \xi_B M_1^2}  \right) 
\varepsilon^{\mu^\prime \nu^\prime \rho^\prime \sigma^\prime} k_1^{\rho^\prime} k_2^{\sigma^\prime}.
\eeqa
Using the anomaly equations for the AAA vertex we can evaluate the first diagram
\beqa
\mathcal{A}^{}_{\xi}&=&\frac{- \, i}{k^2 - \xi_B M_1^2} \frac{1}{M_1^2} \left(k^\lambda \Delta^{\lambda \mu \nu} \right)
\left(k^{\lambda^\prime} \Delta^{\lambda^\prime \mu^\prime \nu^\prime} \right)   \nonumber\\
 &=& \frac{- \, i}{k^2 - \xi_B M_1^2} \frac{1}{M_1^2} \left( - ( \,g^{}_{B})^3 \,  \frac{a_n}{3}  \varepsilon^{\mu \nu \alpha \beta} 
k_1^\alpha k_2^\beta \right)   \left( ( \,g^{}_{B})^3 \, \frac{a_n}{3}  \varepsilon^{\mu^\prime \nu^\prime \alpha^\prime \beta^\prime}
k_1^{\alpha^\prime} k_2^{\beta^\prime} \right)      \nonumber\\
&=& \frac{ \, i}{k^2 - \xi_B M_1^2} \frac{1}{M_1^2}  \left(\frac{a_n}{3}\,g^{\,3}_{B} \right)^2  \,  \varepsilon^{\mu \nu \alpha \beta} \,
\varepsilon^{\mu^\prime \nu^\prime \alpha^\prime \beta^\prime} \, k_1^\alpha k_2^\beta \, k_1^{\alpha^\prime} k_2^{\beta^\prime},
\eeqa 
while the axion exchange diagram gives
\beqa
\mathcal{B}^{}_{\xi}&=&4 \times \left(  \frac{4 C^{}_{BB}}{M} \right)^2   \left(   \frac{i}{k^2 - \xi_B M_1^2} \right) 
 \varepsilon^{\mu \nu \alpha \beta} \,
\varepsilon^{\mu^\prime \nu^\prime \alpha^\prime \beta^\prime} \, k_1^\alpha k_2^\beta \, 
k_1^{\alpha^\prime} k_2^{\beta^\prime}    \nonumber\\
&=&  \frac{64 \,  \,  {C^{\,2}_{BB}} }{M^2}   \frac{i}{k^2 - \xi_B M_1^2}  \,  \varepsilon^{\mu \nu \alpha \beta} \,
\varepsilon^{\mu^\prime \nu^\prime \alpha^\prime \beta^\prime} \, k_1^\alpha k_2^\beta \, 
k_1^{\alpha^\prime} k_2^{\beta^\prime}. 
\eeqa 
Adding the contributions from the two diagrams we obtain
\beqa
\mathcal{A}^{}_{\xi} + \mathcal{B}^{}_{\xi} = 0  \,\,\, \Longleftrightarrow \,\,\,
\frac{1}{M_1^2}  \left(\frac{a_n}{3}\,g^{\,3}_{B} \right)^2 + \frac{64 \,  \,  {C^{\,2}_{BB}}   }{M^2} = 0,
\eeqa
in fact substituting the proper value for the coefficient $C^{}_{BB}$ we obtain an identity 
\beqa
\frac{1}{M_1^2}  \frac{a_n^2}{9} \, g^{\,6}_{B} + \frac{64}{M^2}\left[  \,\frac{i g^{\,3}_{B} }{ 3! }   \, \frac{1}{4}  
\,a^{}_n \frac{M}{M_1}   \right]^2  = \frac{1}{M_1^2} \frac{a_n^2}{9} g^{\,6}_{B} - \frac{64}{M_1^2}
\frac{1}{64} \frac{a_n^2}{9}  g^{\,6}_{B} =0.
\eeqa

This pattern of cancelations holds for a massless fermion ($m_f=0$). 

\subsection{Gauge cancelations in the self-energy diagrams}
In this case, following Fig. \ref{boxmass}, we isolate the following gauge-dependent amplitudes  
\beqn
{\mathcal A}^{}_{ \xi 0} &=&   \, \Delta^{\lambda \mu \nu} (-k^{}_{1}, -k^{}_{2})
\left[ \frac{- i }{k^{2} - \xi^{}_{B} M^{2}_{B}} \left( \frac{k^{\lambda} k^{\lambda'}}{ M^{2}_{B}} \right)  \right] 
\Delta^{\lambda' \mu' \nu'} (k^{}_{1}, k^{}_{2}) \, P^{\nu \nu'}_{o},    \nonumber\\
{\mathcal B}_{\xi 0} &=& 4 \times  \left( \frac{4}{M} \alpha^{}_{2} C^{}_{AA} \right)^{2} \epsilon^{\, \mu \nu \rho \sigma} 
k^{}_{1\rho} k^{}_{2\sigma} \frac{i}{k^{2} - \xi^{}_{B} M^{2}_{B}} \epsilon^{\, \mu' \nu' \rho' \sigma'} 
k^{}_{1\rho'} k^{}_{2\sigma'} \,  P^{\nu \nu'}_{o},   \nonumber\\
{\mathcal C}_{\xi 0} &=& \Delta^{\mu \nu}(-k^{}_{1}, -k^{}_{2}) \left(  2 i \frac{m^{}_{f}  }{ M^{}_{B} }\right) 
\frac{i}{k^{2} - \xi^{}_{B} M^{2}_{B} } \left(  2 i \frac{m^{}_{f}  }{ M^{}_{B} }\right) 
\Delta^{\mu' \nu'}(k^{}_{1}, k^{}_{2}) P^{\nu \nu'}_{o},     \nonumber\\
{\mathcal D}_{\xi 0} &=&  2 \times  \left( \frac{4}{M} \alpha^{}_{2} C^{}_{AA} \epsilon^{\, \mu \nu \rho \sigma} 
k^{}_{1\rho} k^{}_{2\sigma}   \right)  \frac{i}{k^{2} - \xi^{}_{B} M^{2}_{B} } 
\left(  2i \frac{m^{}_{f}  }{ M^{}_{B} }\right) 
\Delta^{\mu' \nu'}(k^{}_{1}, k^{}_{2}) P^{\nu \nu'}_{o},   \nonumber\\
{\mathcal E}_{\xi 0} &=& 
 2 \times  \Delta^{\mu \nu}(-k^{}_{1}, -k^{}_{2}) \left(  2i \frac{m^{}_{f} }{ M^{}_{B} }\right) 
\frac{i}{k^{2} - \xi^{}_{B} M^{2}_{B} }  \left( \frac{4}{M} \alpha^{}_{2} C^{}_{AA}  \epsilon^{\, \mu' \nu' \rho' \sigma'} 
k^{}_{1\rho'} k^{}_{2\sigma'} \right)  P^{\nu \nu'}_{o}, \nonumber\\
\eeqn
so that using the anomaly equations for the triangles
\beqn
k^{\lambda'} \Delta^{\lambda' \mu' \nu'}(k^{}_{1},k^{}_{2}) &=& a^{}_{3}(\beta) \epsilon^{\, \mu' \nu' \rho' \sigma'} 
k^{}_{1\rho'} k^{}_{2\sigma'}  + 2 m^{}_{f} \Delta^{\mu' \nu'},  \nonumber\\
k^{\lambda} \Delta^{\lambda \mu \nu}(-k^{}_{1},-k^{}_{2}) &=& - a^{}_{3}(\beta) \epsilon^{\, \mu \nu \rho \sigma} 
k^{}_{1\rho} k^{}_{2\sigma}  - 2 m^{}_{f} \Delta^{\mu \nu}, \nonumber
\eeqn
and substituting the appropriate value for the WZ-coefficient, with the rotation coefficient of the axion $b$ 
to the goldstone boson given by $\alpha^{}_{2} = M^{}_{1}/M^{}_{B}$, 
one obtains quite straightforwardly that the condition of gauge independence is satisfied
\beqn
{\mathcal A}^{}_{ \xi 0} + {\mathcal B}^{}_{ \xi 0} + {\mathcal C}^{}_{ \xi 0} + {\mathcal D}^{}_{ \xi 0} 
+ {\mathcal E}^{}_{ \xi 0}  = 0.   
\eeqn

%%%%%%%%%%%%%%%%%%%
%
%%%%%%%%%%%%%%%%%%%%%%%%%%%%%%%%%%%%%%%%%%%%%
\section{ Ward identities on the tetragon} 
%%%%%%%%%%%%%%%%%%%%%%%%%%%%%%%%%%%%%%%%%%%%%
%
%%%%%%%%%%%%%%%%%%%%%%%%%%%%%%%%%%%%%%%%%%%%%%%%%%%

%%%%%%%%%%%%%%%%%%%%%%%%%%%%%%%%%%%%%%%%%%%%%%%%%%%
%
As we have seen in the previous sections, the shift dependence from the anomaly on each vertex, parameterized 
by  $\beta, \beta_1, \beta_2$, drops in the actual computation of the unitarity conditions on the s-channel amplitudes, which clearly signals the irrelevance of these shifts in the actual computation, as far as the Bose 
symmetry of the corresponding amplitudes that assign the anomaly on each vertex consistently, are respected.  
It is well known that all the contribution of the anomaly in correlators with more external legs 
is taken care of by the correct anomaly cancelation in 3-point function. It is instructive to illustrate, for generic shifts, 
chosen so to respect the symmetries of the higher point functions, how a similar patterns holds. This takes place 
since anomalous Ward identities for higher order correlators are expressed in terms of standard triangle 
anomalies. This analysis and a similar analysis of other diagrams of 
this type, which we have included in an appendix, is useful for the 
investigation of some rare Z decays (such as Z to 3 photons) which takes place 
with an on-shell Z boson.  

\begin{figure}[t]
{\centering \resizebox*{8cm}{!}{\rotatebox{0}
{\includegraphics{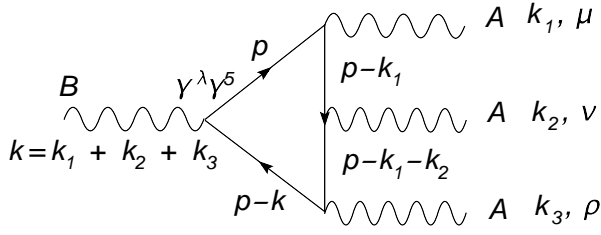}}}\par}
\caption{The tetragon contribution.}
\label{tetragon2}
\end{figure}

\begin{figure}[t]
{\centering \resizebox*{15cm}{!}{\rotatebox{0}
{\includegraphics{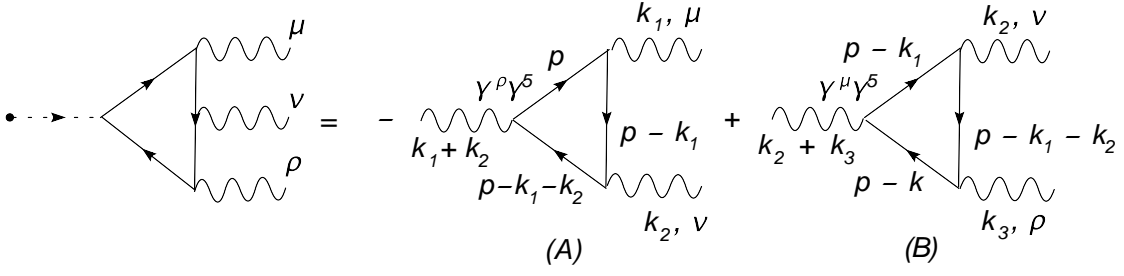}}}\par}
\caption{Distribution of the moments in the external lines in a Ward identity.}
\label{ward}
\end{figure}

Then let's consider the tetragon diagram BAAA shown in fig.(\ref{tetragon2}), where B, being characterized 
by an axial-vector coupling, generates an anomaly in the related Ward identity.  
%  
%%%%%%%%%%%%%%%%%%%%%%%%%%%%%%%%%%%%%%%%
%%%%%%%%%%%%%%%%%%%%%%%%%%%%%%%%%%%%%%%%
%
%
We have the fermionic trace
\beqn
\Delta^{\lambda \mu \nu \rho}(k_{1}, k_{2}, k_{3}) = \overline{\Delta}^{\lambda \mu \nu \rho}(k_{1}, k_{2}, k_{3}) + \mbox{perm.}
\eeqn
where perm. means permutations of $\{(k_{1}, \mu),(k_{2}, \nu),(k_{3}, \rho) \}$.
One contribution to the axial Ward identity comes for instance from 
\beqn
&&k^{\lambda} Tr \left[ \gamma^{\lambda} \gamma^{5} \frac{1}{\slash p - \slash k} \gamma^{\rho} 
\frac{1}{\slash p - \ds{ k}_{1} - \ds{k}_{2} } \gamma^{\nu} 
\frac{1}{\slash p -\ds{k}_{1} } \gamma^{\mu} \frac{1}{\slash p} \right]     \nonumber\\ 
&=& Tr \left[ \slash k \gamma^{5} \frac{1}{\slash p - \slash k} \gamma^{\rho} 
\frac{1}{\slash p - \ds{ k}_{1} - \ds{ k}_{2} } \gamma^{\nu} \frac{1}{\slash p -\ds{ k}_{1} } \gamma^{\mu} 
\frac{1}{\slash p} \right]     \nonumber\\
&=&  -  Tr \left[ \gamma^{5} \frac{1}{\slash p} \gamma^{\rho} 
\frac{1}{\slash p - \ds{k}_{1} - \ds{k}_{2}} \gamma^{\nu} \frac{1}{\slash p -\ds{ k}_{1} } \gamma^{\mu} \right]   \nonumber\\
&&+ Tr \left[ \gamma^{5} \frac{1}{\slash p - \slash k} \gamma^{\rho} 
\frac{1}{\slash p - \ds{ k}_{1} - \ds{ k}_{2} } \gamma^{\nu} \frac{1}{\slash p -\ds{ k}_{1} } \gamma^{\mu} \right],
\label{tetrag}
\eeqn 
which has been rearranged in terms of triangle anomalies 
using
\beqn
\frac{1}{\slash p} \slash k \gamma^{5}  \frac{1}{\slash p -\slash k} = \gamma^{5} \frac{1}{\slash p - \slash k} 
- \gamma^{5} \frac{1}{\slash p}.
\eeqn
Relation (\ref{tetrag}) is diagrammatically shown in Fig.~\ref{ward}.
%  
%%%%%%%%%%%%%%%%%%%%%%%%%%%%%%%%%%%%%%%%
%%%%%%%%%%%%%%%%%%%%%%%%%%%%%%%%%%%%%%%%
%
%
Explicitly these diagrammatic equations become 
\beqn
k^{\lambda} \overline{\Delta}^{\lambda \mu \nu \rho} &=& -  \overline{\Delta}^{\rho \mu \nu}(k_{1}, k_{2}) 
+ \overline{\Delta}^{ \mu \nu \rho}(\beta_1, k_{2}, k_{3}),   \nonumber\\
k^{\lambda} \overline{\Delta}^{\lambda \mu \rho \nu} &=& -  \overline{\Delta}^{\nu \mu \rho}(k_{1}, k_{3}) 
+ \overline{\Delta}^{ \mu \rho \nu}(\beta_2, k_{3}, k_{2}),   \nonumber\\
k^{\lambda} \overline{\Delta}^{\lambda \nu \rho \mu} &=& -  \overline{\Delta}^{ \mu \nu \rho}(k_{2}, k_{3}) 
+ \overline{\Delta}^{ \nu \rho \mu}(\beta_3, k_{3}, k_{1}),   \nonumber\\
k^{\lambda} \overline{\Delta}^{\lambda \nu \mu \rho} &=& -  \overline{\Delta}^{\rho \nu \mu}(k_{2}, k_{1}) 
+ \overline{\Delta}^{ \nu \mu \rho}(\beta_4, k_{1}, k_{3}),   \nonumber\\
k^{\lambda} \overline{\Delta}^{\lambda \rho \mu \nu} &=& -  \overline{\Delta}^{\nu \rho \mu}(k_{3}, k_{1}) 
+ \overline{\Delta}^{ \rho \mu \nu}(\beta_5, k_{1}, k_{2}),   \nonumber\\
k^{\lambda} \overline{\Delta}^{\lambda \rho \nu \mu} &=& -  \overline{\Delta}^{\mu \rho \nu}(k_{3}, k_{2}) 
+ \overline{\Delta}^{ \rho \nu \mu}(\beta_6, k_{2}, k_{1}),   \nonumber\\
\eeqn
where the usual (direct) triangle diagram is given for instance by
\beqn
\overline{\Delta}^{\mu \nu \rho} = \int \frac{d^4 p}{(2 \pi)^4} Tr \left[  \gamma^{\mu} \gamma^{5} \frac{1}{\slash p - \slash k} 
\gamma^{\rho} \frac{1}{\slash p - \ds{ k}_{1} - \ds{ k}_{2}} \gamma^{\nu} \frac{1}{ \slash p - \ds{ k}_{1}} \right].
\eeqn
Adding all the contributions we have 
\beqn
k^{\lambda} \Delta^{\lambda \mu \nu \rho}(k_1, k_2, k_3) &=& - \left[  \Delta^{\rho \mu \nu}(k_1, k_2) 
+ \Delta^{\nu \mu \rho}(k_1, k_3) + \Delta^{\mu \nu \rho}(k_2, k_3) \right]  \nonumber\\
&&+  \left[ \Delta^{\rho \mu \nu}(\beta_5, \beta_6,k_1, k_2) 
+ \Delta^{\nu \mu \rho}(\beta_3, \beta_4,k_1, k_3) 
+ \Delta^{\mu \nu \rho}(\beta_1, \beta_2, k_2, k_3) \right].  \nonumber\\
\label{WARD}
\eeqn
At this point, to show the validity of the Ward identity independently of the chosen value of the CS shifts, 
we recall that under some shifts
\beqn
\Delta^{\mu \nu \rho}(\beta_1, \beta_2, k_2, k_3) &=&  \Delta^{\mu \nu \rho}(k_2, k_3) - \frac{i (\beta_1 + \beta_2)}{4 \pi^2} 
\varepsilon^{\mu \nu \rho \sigma} (k_{2}^{\sigma} - k_{3}^{\sigma})   \nonumber\\
\Delta^{\nu \mu \rho}(\beta_3, \beta_4, k_2, k_3) &=&  \Delta^{\nu \mu \rho}(k_1, k_3) - \frac{i (\beta_3 + \beta_4)}{4 \pi^2} 
\varepsilon^{\nu \mu \rho \sigma} (k_{1}^{\sigma} - k_{3}^{\sigma})   \nonumber\\
\Delta^{\rho \mu \nu}(\beta_5, \beta_6, k_1, k_2) &=&  \Delta^{\rho \mu \nu}(k_1, k_2) - \frac{i (\beta_5 + \beta_6)}{4 \pi^2} 
\varepsilon^{\rho \mu \nu \sigma} (k_{1}^{\sigma} - k_{2}^{\sigma}),   \nonumber\\
\eeqn
and redefining the shifts by setting 
\beqn
\beta_5 + \beta_6 = \overline{\beta}_1 \qquad \beta_1+ \beta_2 = \overline{\beta}_3 \qquad \beta_3+ \beta_4 = \overline{\beta}_2 
\eeqn
we obtain
\beqn
k^{\lambda} \Delta^{\lambda \mu \nu \rho}(k_1, k_2, k_3) &=& - \frac{i\overline{\beta}_1 }{ 4 \pi^2} 
\varepsilon^{\rho \mu \nu \sigma}(k_1^\sigma - k_2^\sigma) - \frac{i\overline{\beta}_2 }{ 4 \pi^2} 
\varepsilon^{\nu \mu \rho \sigma}(k_1^\sigma - k_3^\sigma)   \nonumber\\
&&-  \frac{i\overline{\beta}_3 }{ 4 \pi^2} 
\varepsilon^{\mu \nu \rho \sigma}(k_2^\sigma - k_3^\sigma).   
\eeqn
Finally, using the Bose symmetry on the r.h.s. (indices $\mu, \nu, \rho$) 
of the original diagram we obtain 
\beqn
\overline{\beta}_1 = \overline{\beta}_2 = \overline{\beta}_3, 
\eeqn
which is the correct Ward identity: $k^{\lambda} \Delta^{\lambda \mu \nu \rho } = 0$.
%
%  
%%%%%%%%%%%%%%%%%%%%%%%%%%%%%%%%%%%%%%%%
\begin{figure}[t]
{\centering \resizebox*{5cm}{!}{\rotatebox{0}
{\includegraphics{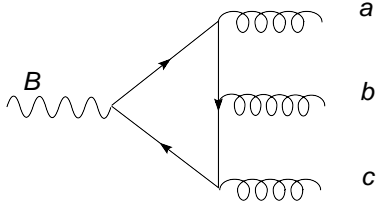}}}\par}
\caption{The tetragon diagram in non abelian case.}
\label{ward_non_ab}
\end{figure}
%%%%%%%%%%%%%%%%%%%%%%%%%%%%%%%%%%%%%%%%
%
We have shown that the correct choice of the CS shifts in tetragon 
diagrams, fixed by the requirements of Bose symmetries of the 
corresponding  amplitude and of the underlying 3-point functions, gives the correct Ward identities for these correlators. This is not unexpected, since 
the anomaly appears only at the level of 3-point functions, 
but shows how one can work in full generality with these amplitudes and 
determine their correct structure. It is also interesting to underline the modifications that take place once this study is extended to the non-abelian case.
In this case (shown in Fig.~\ref{ward_non_ab}) one obtains the same result already shown for the axial abelian Ward identity, but modified by color factors. We obtain
\beqn
&&Tr( \{ T^{a},T^{b}\} T^{c}) \, [ - \Delta^{\rho \mu \nu} +\Delta^{\rho \mu \nu}(\beta) ] 
+ Tr( \{ T^{c},T^{b}\} T^{a})  \, [- \Delta^{\mu \nu \rho} +\Delta^{\mu \nu \rho}(\beta) ]     \nonumber\\
&&+ Tr( \{ T^{a},T^{c}\} T^{b}) \, [- \Delta^{\nu \mu \rho} +\Delta^{\nu \mu \rho}(\beta)]    \nonumber\\
&=& d^{abc} \, [ - \Delta^{\rho \mu \nu} +\Delta^{\rho \mu \nu}(\beta) ] 
+ d^{cba} \, [- \Delta^{\mu \nu \rho} +\Delta^{\mu \nu \rho}(\beta) ]     \nonumber\\
&&+ d^{acb} \, [- \Delta^{\nu \mu \rho} +\Delta^{\nu \mu \rho}(\beta)],
\eeqn
and we have used the definition of the symmetric d-tensor
\beqn
d^{abc} = Tr ( \left\{ T^{a}, T^{b} \right\} T^{c}).
\eeqn
 Simple manipulations give a result which is proportional to the result of the 
abelian case
\beqn
d^{abc} (  [ - \Delta^{\rho \mu \nu} +\Delta^{\rho \mu \nu}(\beta) ] 
+  [- \Delta^{\mu \nu \rho} +\Delta^{\mu \nu \rho}(\beta) ]+  [- \Delta^{\nu \mu \rho} +\Delta^{\nu \mu \rho}(\beta)]  ).
\eeqn
The vanishing of the shift-dependence is related to the Bose symmetry under exchange of the indices 
$$\{(a, \mu, k_{1}), (b, \nu, k_{2}), (c, \rho, k_{3}) \}.$$
This result is clearly expected, since the gauge current of B is abelian and behaves as a 
gauge-singlet current under the gauge interaction of $A$, the latter having been promoted to a 
non-abelian current.

%%%%%%%%%%%%%%%%%%%%%%%%%%%%%%%%%%
\section{Appendix. DR-HVBM}
%%%%%%%%%%%%%%%%%%%%%%%%%%%%%%%%%%
%
In this appendix we fill out some of the details the computation of the direct plus exchanged diagrams in Dimensional Regularization using the HVBM scheme for a partially anticommuting $\gamma_5$ 
\cite{HVBM}. There are various results presented in the previous literature on the computation of these diagrams, most of them using a 
momentum shift without actually enforcing a regularization, shift that brings the anomaly contribution to 
the axial-vector vertex of the triangle diagram, keeping the vector Ward identities satisfied, which takes to 
Rosenberg's parameterization (\ref{Ros}). We fill this gap by showing how the regularization works using an arbitrary tensor structure $T^{\la\mu\nu}$ rather than scalar amplitudes. We also keep 
the mass of the fermion arbitrary, so to obtain a general result concerning the mass dependence of the corrections to the anomaly contributions. We remind that momentum shift are allowed in DR-HVBM, once the integration measure is extended from 4 to $n= 4 -\epsilon$ dimensions and the Feynman parametrization can be used to reduce 
the integrals into symmetric forms. Symmetric integration can then be used exactly as in the standard DR 
case, but with some attention on how to treat the Lorenz indices in the two subspaces of dimensions 
4 and $n-4$, introduced by the regularization. These points are illustrated below.  
 
In the following we will use the notation $I_{x y}$ to denote the parametric integration after performing the loop 
integral 

\beq
I_{x y}\left[...\right]\equiv 2 \int_0^1\int_0^{1-x} dy \left[...\right].
\eeq
 
 There are various ways to implement $\gamma_5$ in D-dimensions, but the prescription 
that works best and is not so difficult to implement is the t'Hooft-Veltman-Breitenlohner-Maison (simply denoted as HVBM) prescription. In the HVBM prescription $\gamma_5$ is 
only partially anticommuting. The gamma algebra in this case is split into $n= 4 + (n-4)$, and the indices of the matrices are split accordingly: $\mu=(\tilde{\mu},\hat{\mu})$. There are now two subspaces, and the indices carrying a $\tilde{} $ are the four dimensional ones. The 4-dimensional part of the algebra is the same as usual, but now 
\beq
\left[\gamma^{\hat{\mu}},\gamma_5\right]_+=0, 
\eeq
where the commutators have been replaced by anticommutators. 
It is important to clarify some points regarding the use of symmetric integration. 
We recall that in DR the use of symmetric integration gives 

\beq
\int d^n q  \frac{q^{\hat{\alpha}},q^{\tilde{\alpha}}}{(q^2 - \Delta)^{L}}=0,
\eeq
and 
\beqa
\int d^n q  \frac{q^{\hat{\mu}} q^{\hat{\nu}}}{({q^2 - \Delta})^{L}}&=& 
g^{\hat{\mu} \hat{\nu}}\int d^n q  \frac{q^2/n}{{(q^2 - \Delta)}^{L}},  \nonumber \\
\int d^n q  \frac{q^{\tilde{\mu}} q^{\tilde{\nu}}}{({q^2 - \Delta})^{L}}&=& 
g^{\tilde{\mu} \tilde{\nu}}\int d^n q  \frac{q^2/n}{{(q^2 - \Delta)}^{L}}.
\eeqa
Integrals involving mixed indices are set to vanish. 
We now summarize other properties of this regularization. We denote by
\beq
g_{\mu\nu}, \qquad  \tilde{g}_{\mu\nu}, \qquad \hat{g}_{\mu\nu}  
\eeq
the n, n-4 and 4 dimensional parts of the metric tensor. An equivalent notation is to set 
$\hat{g}_{\mu\nu}=g_{\hat{\mu} \hat{\nu}}$ and $\tilde{g}_{\mu\nu}=g_{\tilde{\mu} \tilde{\nu}}$, 
$\gamma^{\hat{\mu}}=\hat{\gamma}^\mu$, etc.
The contraction rules are 

\beq
g_\mu^\mu=n, \qquad g_{\mu\la}g^{\la}_{\nu}=g_{\mu\nu}, \qquad {\hat{g}}_{\mu}^{\mu}= n-4, \qquad 
\tilde{g}_\mu^\mu=4, \qquad \tilde{g}_{\mu \lambda}{\hat{g}}^{\lambda \nu}=0.
\label{rule1}
\eeq
Other properties of this regularization follow quite easily. For instance, from 
\beq
\tilde{\gamma}_\mu=\gamma^\sigma \tilde{g}_{\sigma \mu}, \qquad \hat{\gamma}^\mu=\gamma_\la \hat{g}^{\la\mu}, 
 \eeq
using (\ref{rule1}) it follows straightforwardly that  
\beqa
\tilde{\gamma}_\mu\gamma_{a_1}\gamma_{a_2}...\gamma_{a_D}\hat{\gamma}^\mu &=&0, \nonumber \\
{\gamma}_\mu\gamma_{a_1}\gamma_{a_2}...\gamma_{a_D}\hat{\gamma}^\mu &=& 
\hat{\gamma}_\mu\gamma_{a_1}\gamma_{a_2}...\gamma_{a_D}\hat{\gamma}^\mu.
\eeqa

The definition of $\gamma_5$ involves an antisimmetrization over the basic gamma matrices 

\beq
\gamma_5 \equiv \frac{i}{4!} \epsilon_{\mu\nu\rho\sigma}\gamma^{\mu}\gamma^{\nu}\gamma^{\rho}\gamma^\sigma. 
\eeq
The definition is equivalent to the standard one 
$\gamma_5= i \gamma_0 \gamma_1 \gamma_2\gamma_3$. The $\epsilon$ tensor is a 4 dimensional 
projector that selects only the ${\bf \tilde{}}√É¬ü$ indices of a contraction,

\beq
 \epsilon_{\mu \nu \ro \si}\gamma^{\mu}\gamma^{\nu}\gamma^{\rho}\gamma^{\si}= 
 \epsilon_{\tilde{\mu}\tilde{ \nu} \tilde{\ro}\tilde{\si}}\gamma^{\tilde{\mu}}\gamma^{\tilde{\nu}}\gamma^{\tilde{\rho}}
\gamma^{\tilde{\si}}. 
\eeq
It is then easy to show that with this definition 
\beq
\left\{\gamma_5,\tilde{\gamma}^\mu\right\}=0, \qquad \left[\gamma_5,\hat{\gamma}^\mu\right]=0.
\eeq
These two relations can be summarized in the statement
\beq
\left\{\gamma_5,{\gamma}_\mu\right\}=2 \hat{\gamma}_\mu \gamma_5.
\eeq
We compute the traces and remove the hat-momenta of the two external vector currents. We illustrate some steps of the computation. 
We denote by $I[...]$ a typical momentum integral that appears in the computation 
\beq
I[...]\equiv \int \frac{d^n q}{(2 \pi)^n} \frac{[...]}{\left(q^2 - \Delta\right)^3},
\eeq
setting $n=4 - \epsilon$, for instance we get
\beqa
I\left[\epsilon[k_1, k_2, {\mu},{\nu}]\right] \hat{q}_{\la} &=&0,    \nonumber \\
I\left[\epsilon[k_2,{\la},{\mu},{\nu}] \right] \hat{q}\cdot\hat{q} &=& \epsilon[k_2, {\la}, {\mu}, {\nu}] (n - 4) I_2, \nonumber \\
I\left[\epsilon[k_2,q,{\mu},{\nu}]\right] \hat{q}_{\la}&=& 0, \nonumber \\
I \left[\epsilon[k_1,q,{\mu},{\nu}]\right] \tilde{q}_{\la} &=& \epsilon[k_1, {\la}, {\mu},{\nu}] I_2, \nonumber \\
I\left[\epsilon[k_2,q,{\mu},{\nu}]\right] q_{\la}&=& \epsilon[k_2,{\la},{\mu},{\nu}] I_2.
\eeqa
Denoting by D and E the direct and the exchanged diagram (before the integration over the Feynman parameters x,y), 
we obtain
\beq
D + E= - i I_{x y} \left[a_1 c_1 + a_2 c_2 + a_3 c_3 + a_4 c_4 + a_5 c_5 \right],
\eeq
where
\beqa
c_1 &=&-4i {I_2} [ n(-2 +x +y ) + 2(2 + x + y) ] \nonumber \\
&& +4 i {I_1} [ m_f^2 ( -2 + x + y) + sx(1 - x + xy -y + y^2)  ],  \nonumber \\
c_2 &=& - c_1, \nonumber \\
c_3 &=& 8i I_1 x(x-y-1)( {k_1}_{\la} + {k_2}_{\la}),   \nonumber \\
c_4 &=& 8i I_1( x + y -1)( y {k_1}_{\mu} - x {k_2}_{\mu}),    \nonumber \\
c_5 &=& 8i I_1 (x + y -1)(x {k_1}_{\nu} - y {k_2}_{\nu}),      \nonumber \\
a_1 &=& \epsilon[k_1,{\la},\mu,\nu],     \nonumber \\
a_2 &=&  \epsilon[k_2,{\la},\mu,\nu], \nonumber \\
a_3 &=&  \epsilon[k_1,k_2,\mu,\nu],   \nonumber \\
a_4 &=&  \epsilon[k_1,k_2,{\la},\nu],   \nonumber \\
a_5 &=&  \epsilon[k_1,k_2,{\la},\mu],     \nonumber \\
\eeqa
and introducing the dimensionally regulated expressions of $I_1$ and $I_2$ and expanding in $\epsilon$ we obtain
\beqa
c_1&= &\frac{1}{\epsilon}\frac{3 x+3 y-2}{4 \pi ^2} \nonumber \\
&& + \frac{x (x+y-1) (2 y-1) s+(3 x+3 y-2) \left(s x y  - m_f^2\right) 
\log
   \left(\frac{m_f^2 - s x y }{\mu ^2} \right)}{ 8 \pi ^2 \left( m_f^2 - s x y  \right)}, \nonumber \\
c_3 &=& - \frac{x(x-y-1)}{4 \pi^2 (s x y - m_f^2)}\left( k_{1\la} + k_{2\la}\right),
\nonumber \\
c_4 &=& 
\frac{(x+y-1) (k_{2\mu} x-k_{1 \mu} y)}{4 \pi ^2 \left(s x y-m_f^2 \right)},  \nonumber \\
c_5 &=& -\frac{(x+y-1) (k_{1 \nu} x- k_{2 \nu} y)}{4 \pi ^2 \left( s x y - m_f^2 \right)},
\eeqa
where $\mu$ is the renormalization scale in the $\overline{MS}$ scheme  
with $\overline{\mu}^2 = \mu^2 e^{\bf \gamma}/(4 \pi)$  and ${\bf \gamma}$ is the Euler-Mascheroni constant.
The pole singularity is related to tensor structures which have a lower mass dependence on $k_1$ and $k_2$ 
($a_1$ and $a_2$) which involve loop integrations with an additional powers of $q$ and are, therefore, UV divergent. 
However, the pole contributions vanish after integration over the Feynman parameter, since
\beq
I_{x y}\left[ 3 x + 3 y -2 \right]=0.
\eeq
Performing the integration over the Feynman parameters we obtain the result 
reported below in Eq.~(\ref{firstres}).

\subsection{The vanishing of a massive {\bf AAV/VVV}} 
The vanishing of {\bf AAV } in DR in the general case 
(with non-vanishing fermion masses) can be established by a direct 
computation, beside using C-invariance (Furry's theorem). The vanishing of this diagram is due to 
the specific form of all the Feynman parameters which multiply every covariant structure in the corresponding tensor amplitude. Denoting by $X^{\lambda\mu\nu} $ any of these generic structures, the parametric integral is of the form 

\beq
\Delta_{AAV}^{\lambda\mu\nu}= X^{\lambda\mu\nu}\int_{0}^{1} dx \int_0^{1-x} dy \frac{f(x,y)}{\Delta(x,y)} + \dots
\eeq
with $f(x,y)$ antisymmetric in $x,y$ and $\Delta(x,y)$ symmetric, giving a vanishing result. 
For {\bf VVV} the result is analogous.
%
%
%%%%%%%%%%%%%%%%%%%%%%%%%%%%%%%%%%%
\subsection{{\bf AVV} and shifts}  
%%%%%%%%%%%%%%%%%%%%%%%%%%%%%%%%%%%
%
%
If we decide to use a shift parameterization of the diagrams then the two values of the amplitudes 
$\ul{a_1}$ and $\ul{a_2}$, are arbitrary. This point has been discussed in the previous sections, 
although here we need to discuss with further detail and include in our analysis fermion mass effects as well. However, the use of Dimensional regularization is such to determine an equal distribution of the anomaly among diagrams of the form {\bf AAA}, no matter which pamaterization of the momentum we choose in the graph. Therefore, in this case, if a current is conserved, there is no need to add CS interactions or, equivalently, perform a shift in order to remove the anomaly from vertices which are vector-like.

The first significant parameterization of the anomalus diagram can be found in Rosenberg's paper, later used by Adler in his work on the axial anomaly. The shift is fixed by requiring CVC, which is practical matter rather than a fundamental 
issue. We will show that this method can be mappend into the DR-HVBM result using the Schouten identity. 
We start from Rosenberg's parameterization
\beqa
T^{\la\mu\nu} &=& A_1 \epsilon[k_1,\mu,\nu,\la] + A_2 \epsilon[k_2,\mu,\nu,\la] +
A_3 \epsilon[k_1,k_2,\mu,\la]{k_1}^{\nu}
 + A_4\epsilon[k_1,k_2,\mu,\la]k_2^{\nu}\nonumber \\
&& + A_5 \epsilon[k_1,k_2,\nu,\la]k_1^\mu 
+ A_6 \epsilon[k_1,k_2,\nu,\la]k_2^\mu
\label{Ros}
\eeqa
given in \cite{Rosenberg}. By power-counting, 2 invariant amplitudes are divergent, $A_1$ and $A_2$, while the $A_i$ with $i\geq 3$ are finite\footnote{
We will be using the notation $\epsilon[a,b,\mu,\nu]\equiv \epsilon_{\alpha\beta\mu\nu}
a^\alpha b^\beta$ to denote the structures in the expansion of the anomalous triangle diagrams}. 
In general $A_1$ and $A_2$ are given by parametric integrals which are divergent and there are two free 
parameters in these integrals, amounting to momentum shifts, that can be chosen to render $A_1$ and $A_2$ finite. 
It is possible to redefine the momentum shifts so that the divergences are removed, and this can be 
obtained by imposing the defining Ward identities (conservation of the two vector currents) in the diagrams
\beq
k_{1 \mu}T^{\la\mu\nu}=k_{2\nu}T^{\la\mu\nu}=0.
\eeq
This gives $A_1=s/2\,\, A_3$ and $A_2=s/2\,\, A_6$. The expressions of the invariant amplitudes 
$A_i$ are given in Rosenberg as implicit parametric integrals. 
They can be arranged in the form
\beqa
A_1 &=& -\frac{i}{4 \pi ^2}+ i C_0(m_f^2,s) \nonumber \\
A_2 &=& \frac{i}{4 \pi ^2}- i C_0(m_f^2,s) \nonumber \\
A_3 &=& - \frac{i}{2 s \pi^{2} } + \frac{2i}{s} C_0(m_f^2,s)\nonumber \\
A_4 &=& 
\frac{i}{s \pi ^2}  -  i f(m_f^2,s)\nonumber \\
A_5 &=&-A_4 \nonumber \\
A_6 &=& -A_3
\label{As}
\eeqa
where we have isolated the mass-independent contributions, which will appear in the anomaly, 
from the mass corrections dependent on the fermion mass ($m_f$), and we have defined 
\beq
C_0(m_f^2,s)= \frac{ {Li}_2\left(\frac{2 }{1 - \sqrt{ 1 - 4 m_f^2
   /s}} \right) m_f^2}{2 s \pi ^2} + \frac{ {Li}_2\left(\frac{2 }{1 + \sqrt{1 - 4 m_f^2/ s}}\right) m_f^2}{2 s \pi
   ^2},
\label{C0}
\eeq
\beq
f(m_f^2,s)=\frac{ \sqrt{1- 4 m_f^2/s}  \,\tanh ^{-1}\left(\frac{{1}}{\sqrt{1- 4
   m_f^2/s}}\right)}{s \pi ^2}.
\label{fm}
\eeq
Eqs.~(\ref{C0}) and (\ref{fm}) have been obtained integrating the parametric expressions of Rosenberg.

The axial vector Ward identity is obtained from the contraction 
\beqa
&& \left( k_{1\la} + k_{2\la}\right) T^{\la\mu\nu}   \nonumber\\
&=&\Big(  - \frac{i}{2  \pi ^2}  
+ \frac{i {Li}_2 \left( \frac{2 }{ 1 - \sqrt{1 - 4 m_f^2 / s } } \right) m_f^2 }{ s \pi^2 } 
+ \frac{i {Li}_2 \left( \frac{2}{ 1 + \sqrt{1 - 4 m_f^2 /s } } \right) m_f^2}{ s \pi^2}  \Big)   
\epsilon[k_1,k_2,\mu,\nu]   \nonumber\\
&=& \left( - \frac{i}{2  \pi ^2} + 2 \,i\, C_0(m_f^2,s)
\right) \times \epsilon[k_1,k_2,\mu,\nu]
\label{finitemass}
\eeqa
where the first contribution is the correct value of the anomaly. The 
remaining term, expressed in terms of dilogarithmic functions, is related 
to the scalar 3-point function, as shown below. 

\subsection{DR-HVBM scheme}
\subsubsection{{\bf AVV} case }
In this case if we use DR we obtain
\beqa
T_{\la\mu\nu} &=& -i \tau_1\,\,\Big( \epsilon[k_1,{\la},\mu,\nu] - \,\,\epsilon[k_2,{\la},\mu,\nu]\Big) 
- i {\tau_2}\left( k_{1\la} + k_{2 \la}\right)\,\, 
\epsilon[k_1,k_2,\mu,\nu] \nonumber \\
&& - i {\tau_3}\left(k_{1\mu} - k_{2 \mu}\right) \,\, \epsilon[k_1,k_2,{\la},\nu] 
- i {\tau_3}\left(k_{1\nu} - k_{2 \nu}\right)\,\, \epsilon[k_1,k_2,{\la},\mu] 
\eeqa
\beqa
\tau_1 &=& -  \frac{ {Li}_2 \left( \frac{2}{ 1 - \sqrt{1 - 4 m_f^2 /s }} \right) m_f^2 }{ 4 s \pi^2} 
 - \frac{{Li}_2 \left( \frac{2 }{ 1 + \sqrt{1 - 4 m_f^2 / s} } \right) m_f^2 }{4 s \pi ^2}   \nonumber\\
&& \,+ \frac{3}{8 \pi ^2}  - \frac{ \sqrt{4 m_f^2 /s - 1 }  \,\,  \tan ^{-1}  \left(  \frac{ 1 }{ \sqrt{ 4 m_f^2 /s - 1 } } \right) }
{4  \pi^{2} }   \\
\tau_2 &=&
- \frac{{Li}_2 \left( \frac{2 }{ 1 - \sqrt{1 - 4 m_f^2 / s} } \right) m_f^2 }{ 2 s^{2} \pi^2} 
- \frac{ {Li}_2 \left( \frac{2}{ 1 + \sqrt{1 - 4 m_f^2 / s} } \right) m_f^2 }{ 2 s^{2} \pi^2 } \nonumber \\
&&
+ \frac{ \sqrt{ 4 m_f^2 /s - 1 } \,\,\tan^{-1} \left( \frac{ 1 }{ \sqrt{4 m_f^2 /s - 1 }} \right)}{ 2 s \pi^2} 
-\frac{1}{4 s \pi ^2}\nonumber \\
\tau_3 &=&
\frac{ {Li}_2 \left( \frac{2 }{ 1 - \sqrt{1 - 4 m_f^2 / s } } \right) m_f^2 }{2 s^{2} \pi^2 } 
+ \frac{ {Li}_2 \left( \frac{2}{ 1 + \sqrt{1 - 4 m_f^2 /s } } \right) m_f^2}{2 s^{2} \pi^2} \nonumber \\
&&
+ \frac{ \sqrt{ 4 m_f^2 /s - 1 } \,\, \tan ^{-1}\left( \frac{ 1 }{ \sqrt{4 m_f^2 /s - 1 }}  \right) }{ 2 s \pi^2 }
-\frac{3}{4 s \pi ^2}.
\label{firstres}
\eeqa
The expressions above require a suitable analytic continuation in order to cover all the kinematic 
range of the external invariant (virtuality) $s$. The position of the branch cut 
in the physical region is at $\sqrt{s}=2 m$, corresponding to an s-channel cut, where the virtual axial-vector line can produce two on-shell collinear massive fermions.

It is interesting to see how the vector and the axial-vector Ward identities are satisfied 
for a generic fermion mass $m$. For the vector Ward identity we get
\beqa
k_{1\mu}T^{\mu\nu\la} &=& \frac{i}{2} \left({\tau_3} s+2 {\tau_1}\right)
   {\epsilon[k_1,k_2,\la,\nu]} \nonumber \\
k_{2\nu}T^{\mu\nu\la} &=&- \frac{i}{2}  \left({\tau_3} s+2 {\tau_1}\right)
   {\epsilon[k_1,k_2,\la,\nu]}.
\eeqa
One can check directly that the combination $({\tau_3} s+2 {\tau_1})$ 
vanishes so that $k_{1\mu}T^{\mu\nu\la}=k_{2\nu}T^{\mu\nu\la}=0$.

The second and third term in (\ref{finitemass}) are related to the scalar 3-point function 
\beq
C_{00}(k^2,k_1^2,k_2^2,m_f^2,m_f^2,m_f^2)=\int d^4 q \frac{1}{\left(q^2 - m_f^2\right) 
\left( (q + k_1)^2 - m_f^2\right)\left( (q + k_1 + k_2)^2 - m_f^2\right)} \nonumber \\
\eeq

\beqa
C_{00}(k^2,0,0,m_f^2,m_f^2,m_f^2) &=& -\frac{1}{k^2}\left(Li_2\left(\frac{1}{r_1}
\right) + 
Li_2\left(\frac{1}{r_2}\right)\right) \nonumber \\
r_{1,2}&=& \frac{1}{2}\left[ 1 \pm \sqrt{1 - 4\frac{m_f^2}{k^2}}\right]
\eeqa
giving the equivalent relation 
\beq
\left( k_{1\la} + k_{2\la}\right) T^{\la\mu\nu}= \left(  - \frac{i}{2  \pi ^2}  
+ \frac{i {Li}_2 \left( \frac{2 }{ 1 - \sqrt{1 - 4 m_f^2 / s } } \right) m_f^2 }{ s \pi^2 } 
+ \frac{i {Li}_2 \left( \frac{2}{ 1 + \sqrt{1 - 4 m_f^2 /s } } \right) m_f^2}{ s \pi^2}  \right)   
\epsilon[k_1,k_2,\mu,\nu].
\eeq
Our result for $T^{\la\mu\nu}$ can be easily matched to other 
parameterizations obtained by a shift of the momentum in the loop integral 
performed in 4 dimensions. We recall that in this case one needs 
to impose the defining Ward identities on the amplitude, rather than obtaining them from a regularization, as in the case of the HVBM scheme. Before doing this, we present the analytically 
continued expressions of (\ref{firstres}) which 
are valid for $\sqrt{s} > 2 m^{}_{f}$ 
and are given by
\beqa
\tau_1 &=&  - \frac{1}{2} C_0(s,m_f^2) + \frac{3}{8 \pi^2} -
\frac{1}{4 \pi^2}\sqrt{1 - 4 m_f^2/s}\,\,\tanh^{-1}
\left(\frac{1}{\sqrt{1 - 4 m_f^2/s}}\right)    \nonumber \\
&=&  - \frac{1}{2} C_0(s,m_f^2) + \frac{3}{8 \pi^2} - \frac{s}{4} f(m_f^2,s),  \nonumber\\
\tau_2 &=&
- \frac{1}{s} C_0(s,m_f^2) - \frac{1}{4 s \pi^2 } +
\frac{1}{2 s \pi^2}\sqrt{1 - 4 m_f^2/s}\,\,\tanh^{-1}
\left(\frac{1}{\sqrt{1 - 4 m_f^2/s}}\right) \nonumber \\
&=& - \frac{1}{s} C_0(s,m_f^2) - \frac{1}{4 s \pi^2 } + \frac{1}{2}  f(m_f^2,s),    \nonumber\\
\tau_3 &=&
\frac{1}{s} C_0(s,m_f^2)  - \frac{3}{4 s\pi^2} + 
\frac{1}{2 s \pi^2}\sqrt{1 - 4 m_f^2/s}\,\,\tanh^{-1}
\left(\frac{1}{\sqrt{1 - 4 m_f^2/s}}\right) \nonumber \\
&=& \frac{1}{s} C_0(s,m_f^2)  - \frac{3}{4 s\pi^2} + \frac{1}{2}  f(m_f^2,s).
\label{ts}
\eeqa

\subsubsection{The {\bf AAA} diagram }
The second case that needs to be worked out in DR is that of a 
triangle diagram containing 3 axial vector currents. We use the HVBM scheme for $\gamma_5$. The 
analysis is pretty similar to the case of a single $\gamma_5$. 
In this case we obtain 
\beqa
T_3^{\la\mu\nu} &=& - i \left( I_{xy}[c_1] \epsilon[k_1,\la,\mu,\nu] + I_{xy}[c_2] \epsilon[k_2,\la,\mu,\nu]  
+ I_{xy}[c_3] \epsilon[k_1,k_2,\mu,\nu]
\left( k_1^{\la} + k_2^{\la}\right)    \right. \nonumber \\
&&  \left.  + I_{xy}[c_4^{\mu}] \epsilon[k_1,k_2,\la,\nu] + I_{xy}[c_5^{\nu}]\epsilon[k_1,k_2,\la,\mu] \right)
\eeqa
where $I_ {xy}$ is the integration over the Feynman parameters. Also in this case the coefficients $c_1$ and $c_2$ are divergent 
and are regulated in dimensional regularization. 
We obtain
\beqa
c_1 &=&
4 i \left({I_2} (n-6) (3 x+3 y-2)+{I_1} \left((-3 x-3
   y+2) m_f^2+ s x \left(y^2-y+x (y-1)+1\right)\right)\right)
\nonumber \\
c_2 &=&-c_1 \nonumber \\
c_3 &=& 8 i I_1 x( x-y-1) \nonumber \\
c_4 &=&-8 i {I_1} (x+y-1) (x {k_2}^{\mu}
   - y {k_1}^{\mu})
\nonumber \\
c_5 &=&8 i {I_1} (x+y-1) (x {k_1}^{\nu} - y {k_2}^{\nu}),
\eeqa
which in DR become 
\beqa
c_1 &=&
\frac{3 x+3 y-2}{4 \pi ^2 \epsilon} +
\frac{(3 x+3 y-2) \left(s x y-m_f^2\right) \log
   \left(\frac{m_f^2-s x y}{\mu ^2}\right)-s x
   (x+y-1) (2 y+1)}{8 \pi ^2 \left(m_f^2-s x y\right)} \nonumber \\
c_3 &=&\frac{x (x-y-1)}{4 \pi ^2 \left(m_f^2-s x y\right)} \nonumber \\
c_4 &=&-\frac{(x+y-1) ({{k_2}^{\mu}} x-{{k_1}^{\mu}} y)}{4 \pi
   ^2 \left(m_f^2-s x y\right)} \nonumber \\
c_5 &=&\frac{(x+y-1) ({{k_1}^{\nu}} x-{{k_2}^{\nu}} y)}{4 \pi
   ^2 \left(m_f^2-s x y\right)}.
\eeqa
After integration over x and y the pole contribution vanishes. We obtain 

\beqa
T^{(3)}_{\la\mu\nu} &=& -i \left(  \tau_1^{(3)}\,\,\left( \epsilon[k_1,{\la},\mu,\nu] - \,\,\epsilon[k_2,{\la},\mu,\nu]\right) 
+  {\tau_2^{(3)}}\left( k_{1\la} + k_{2 \la}\right)\,\, 
\epsilon[k_1,k_2,\mu,\nu]    \right. \nonumber \\
&&\left.   +  {\tau_3^{(3)}}\left(k_{1\mu} - k_{2 \mu}\right) \,\, \epsilon[k_1,k_2,{\la},\nu] 
+  {\tau_3^{(3)}}\left(k_{1\nu} - k_{2 \nu}\right)\,\, \epsilon[k_1,k_2,{\la},\mu]     \right)
\eeqa
\beqa
\tau_1^{(3)} &=& \frac{3 {Li}_2\left(\frac{2}{ 1 - \sqrt{1 - 4 m_f^2 /s}}
\right) m_f^2}{4 s \pi ^2}  + 
\frac{3 {Li}_2 \left(\frac{2 }{1 + \sqrt{1 - 4 m_f^2 / s}} \right) m_f^2}{4 s \pi ^2}    \nonumber\\
&& + \frac{\left( 64 \,m_f^4 / s^2 - 20\, m_f^2 /s  + 1  \right) \,\,\tan^{-1} \left( \frac{1}{ \sqrt{ 4 m_f^2 /s -1 } } 
\right) }{ 4  \pi^2 \sqrt{4 m_f^2 /s - 1 } } 
-\frac{4
   m_f^2}{s \pi ^2} +\frac{5}{24 \pi ^2},
\label{tau1}
\eeqa

\beqa
\tau_2^{(3)} &=& 
- \frac{{Li}_2 \left(\frac{2 }{1 - \sqrt{1 - 4 m_f^2 /s} } \right) m_f^2}{2 s^2 \pi^2} 
-\frac{ {Li}_2 \left( \frac{2}{1 + \sqrt{1  - 4 m_f^2 / s}} \right) m_f^2}{ 2 s^2 \pi ^2}  \nonumber \\
&& + \frac{ \sqrt{ 4 m_f^2 /s  - 1 } \tan ^{-1} \left( \frac{  1 }{\sqrt{4 m_f^2 /s - 1 }} 
\right)}{2 s \pi^2 }  -  \frac{1}{4 s \pi^2},
\label{tau2}
\eeqa

\beqa
\tau_3^{(3)} &=&
\frac{ {Li}_2 \left( \frac{2 }{1 - \sqrt{1 - 4 m_f^2 / s} } \right) m_f^2}{2 s^2 \pi^2 } 
+ \frac{ {Li}_2 \left( \frac{2}{ 1 + \sqrt{1 - 4 m_f^2 / s }}  \right) m_f^2}{ 2 s^2 \pi ^2} \nonumber \\
&& + \frac{ \sqrt{ 4 m_f^2 /s  - 1 } \tan^{-1} \left( \frac{  1  }{ \sqrt{4 m_f^2 / s - 1} } 
\right)}{2 s \pi ^2}-\frac{3}{4
   s \pi ^2}.
\label{tau3}
\eeqa
We present the analytically continued expressions of relations~(\ref{tau1}, \ref{tau2}, \ref{tau3}) valid for $\sqrt{s} > 2 m^{}_{f}$
\beqn
\tau_1^{(3)} &=& \frac{3 {Li}_2\left(\frac{2}{ 1 - \sqrt{1 - 4 m_f^2 /s}}
\right) m_f^2}{4 s \pi ^2}  + 
\frac{3 {Li}_2 \left(\frac{2 }{1 + \sqrt{1 - 4 m_f^2 / s}} \right) m_f^2}{4 s \pi ^2}    \nonumber\\
&& - \frac{\left( 64 \,m_f^4 / s^2 - 20\, m_f^2 /s  + 1  \right) \,\,\tanh^{-1} \left( \frac{1}{ \sqrt{1 - 4 m_f^2 /s} } 
\right) }{ 4  \pi^2 \sqrt{ 1 - 4 m_f^2 /s } } 
-\frac{4
   m_f^2}{s \pi ^2} +\frac{5}{24 \pi ^2},   
\eeqn
\beqn
\tau_2^{(3)} &=& 
- \frac{{Li}_2 \left(\frac{2 }{1 - \sqrt{1 - 4 m_f^2 /s} } \right) m_f^2}{2 s^2 \pi^2} 
-\frac{ {Li}_2 \left( \frac{2}{1 + \sqrt{1  - 4 m_f^2 / s}} \right) m_f^2}{ 2 s^2 \pi ^2}  \nonumber \\
&& + \frac{ \sqrt{1 - 4 m_f^2 /s} \tanh^{-1} \left( \frac{  1 }{\sqrt{1 - 4 m_f^2 /s }} 
\right)}{2 s \pi^2 }  -  \frac{1}{4 s \pi^2},  
\eeqn
\beqn
\tau_3^{(3)} &=&
\frac{ {Li}_2 \left( \frac{2 }{1 - \sqrt{1 - 4 m_f^2 / s} } \right) m_f^2}{2 s^2 \pi^2 } 
+ \frac{ {Li}_2 \left( \frac{2}{ 1 + \sqrt{1 - 4 m_f^2 / s }}  \right) m_f^2}{ 2 s^2 \pi ^2} \nonumber \\
&& + \frac{ \sqrt{  1 - 4 m_f^2 /s } \tanh^{-1} \left( \frac{  1  }{ \sqrt{ 1 - 4 m_f^2 / s} } 
\right)}{2 s \pi ^2}-\frac{3}{4 s \pi ^2}.
\eeqn
In the massless case, the contribution to the Ward identity is given by 
\beqa
k_3^\lambda T^{AAA}_{\la\mu\nu} &=& -\frac{i}{6 \pi^2}\epsilon[k_1,k_2,\mu,\nu]\nonumber \\ 
k_1^\mu T^{AAA}_{\la\mu\nu}&=& -\frac{i}{6 \pi^2}\epsilon[k_2,k_3,\nu,\la] \nonumber \\
k_2^\nu T^{AAA}_{\la\mu\nu}&=& -\frac{i}{6 \pi^2}\epsilon[k_3,k_1,\la,\mu] 
\eeqa
where we have chosen a symmetric distribution of (outgoing) momenta $(k_1,k_2,k_3)$ attached to vertices 
$(\mu,\nu,\la)$, with $k_3=-k=-k_1 - k_2$.

\subsection{Equivalence of the shift-based (CVC) and of DR-HVBM schemes}
The equivalence between the HVBM result and the one obtained 
using the defining Ward identities (\ref{As}) can be shown 
using the Schouten relation 
\beqa
&& k_i^{\mu_1} \epsilon[\mu_2,\mu_3,\mu_4,\mu_5] +k_i^{\mu_2} \epsilon[\mu_3,\mu_4,\mu_5,\mu_1] 
+ k_{i}^{\mu_3} \epsilon[\mu_4,\mu_5,\mu_1,\mu_2] \nonumber \\
&&+  k_i^{\mu_4} \epsilon[\mu_5,\mu_1,\mu_2,\mu_3] 
+ k_i^{\mu_5} \epsilon[\mu_1,\mu_2,\mu_3,\mu_4]=0, 
\eeqa
that allows to remove the $k_{1,2}^{\la}$ terms in terms of other contributions 
\beqa
k_1^\la \epsilon[k_1,k_2,\mu,\nu] &=& \frac{s}{2}\epsilon[k_1,\mu,\nu,\la] -
k_1^{\mu} \epsilon[k_1,k_2,\nu,\la] + k_1^\nu \epsilon[k_1,k_2,\mu,\la] \nonumber \\
k_2^\la \epsilon[k_1,k_2,\mu,\nu] &=& -\frac{s}{2}\epsilon[k_2,\mu,\nu,\la] -
k_2^{\mu} \epsilon[k_1,k_2,\nu,\la] + k_2^\nu \epsilon[k_1,k_2,\mu,\la].
\eeqa
The result in the HBVM scheme then becomes 
\beqa
T^{\la\mu\nu} &=& -i \left( \tau_1 + \frac{s}{2}\tau_2  \right) \epsilon[k_1,\mu,\nu,\la] -i 
\left( -\tau_1 - \frac{s}{2}\tau_2\right) \epsilon[k_2,\mu,\nu,\la] \nonumber \\
&& -i
\left(\tau_2 - \tau_3\right) \epsilon[k_1,k_2,\mu,\la]{k_1}^{\nu}
 -i \left(\tau_2 + \tau_3\right)\epsilon[k_1,k_2,\mu,\la]k_2^{\nu}\nonumber \\
&& -i \left(-\tau_2 -\tau_3\right) \epsilon[k_1,k_2,\nu,\la]k_1^\mu 
-i \left(\tau_3 -\tau_2\right) \epsilon[k_1,k_2,\nu,\la]k_2^\mu
\eeqa
and it is easy to check using (\ref{As}) and (\ref{ts}) that the invariant 
amplitudes given above coincide with those given  by Rosenberg. Therefore we have the correspondence 
\beqa
A_1 &=& -i (\tau_1 + \frac{s}{2}\tau_2 ) \nonumber \\
A_2 &=&-i(  - \tau_1 - \frac{s}{2}\tau_2 ) \nonumber \\
A_3 &=& -i ( \tau_2 -\tau_3 ) \nonumber \\
A_4 &=&-i ( \tau_2 + \tau_3  ) \nonumber \\
A_5 &=&-i(  - \tau_2 -\tau_3 ) \nonumber \\
A_6 &=& -i( \tau_3 - \tau_2 ). 
\label{idr}
\eeqa

A similar correspondece holds between the Rosenberg parameterization of 
{\bf AAA} and the corresponding DR-HVBM result
\beqa
A^{(3)}_1 &=&- i( \tau^{(3)}_1 + \frac{s}{2}\tau_2 ) \nonumber \\
A^{(3)}_2 &=&-i(  - \tau^{(3)}_1 - \frac{s}{2}\tau^{(3)}_2 ) \nonumber \\
A^{(3)}_3 &=&-i(  \tau^{(3)}_2 -\tau^{(3)}_3 ) \nonumber \\
A^{(3)}_4 &=&-i( \tau^{(3)}_2 + \tau^{(3)}_3 )\nonumber \\
A^{(3)}_5 &=& -i( - \tau^{(3)}_2 -\tau^{(3)}_3 )\nonumber \\
A^{(3)}_6 &=&-i( \tau^{(3)}_3 - \tau^{(3)}_2 ). 
\label{idr1}
\eeqa

%%%%%%%%%%%%%%%%%%%%%%%%%%%%%%%%%%%%%%%%%%%%%%%%%%%%%%%%%%%%%%%%%%%%%%%%%%%%%%%
\section{Appendix: The Chern-Simons and the Wess Zumino vertices}
%%%%%%%%%%%%%%%%%%%%%%%%%%%%%%%%%%%%%%%%%%%%%%%%%%%%%%%%%%%%%%%%%%%%%%%%%%%%%%%
The derivation of the vertex is CS straightforward and is given by
\beqn
&&\int dx \, dy \, dz\, T_{CS}^{\la\mu\nu}(z,x,y) \, B^{\lambda}(z) A^{\mu}(x) A^{\nu}(y)    \nonumber\\
&=&\int dx \, dy \, dz \, \int \frac{dk_1}{(2 \pi)^4} \frac{dk_2}{(2 \pi)^4} e^{- i k_1 (x - z) - i k_2 (y - z)} 
 \, \varepsilon^{\lambda \mu \nu \alpha} \,( k^{\alpha}_1 -k_2^\alpha)\, B^{\lambda}(z) A^{\mu}(x) A^{\nu}(y)   \nonumber\\
&=& \int dx \, dy \, dz \,  i \left( \frac{\partial}{\partial x^{\alpha}}- \frac{\partial}{\partial y^{\alpha}} \right) 
\left( \int  \frac{dk_1}{(2 \pi)^4} \frac{dk_2}{(2 \pi)^4} e^{- i k_1 (x - z) - i k_2 (y - z)} \right) B^{\lambda}(z) 
A^{\mu}(x) A^{\nu}(y) \varepsilon^{\lambda \mu \nu\alpha }     \nonumber\\
&=& (- i)  \int dx \, dy \, dz \,  \int \frac{dk_1 \, dk_2}{(2 \pi)^8} e^{- i k_1 (x - z) - i k_2 (y - z)} 
B^{\lambda}(z)\left(  \frac{\partial}{\partial x^{\alpha}} A^{\mu}(x)  A^{\nu}(y) - \frac{\partial}{\partial y^{\alpha}} 
A^{\nu}(y)  A^{\mu}(x) \right)  \varepsilon^{\lambda \mu \nu \alpha} \nonumber\\
&=& (-i) \int dx \, dy \, dz \, \delta (x-z) \delta (y-z) B^{\lambda}(z)
\left(  \frac{\partial}{\partial x^{\alpha}}A^{\mu}(x)  A^{\nu}(y) - 
\frac{\partial}{\partial y^{\alpha}}A^{\nu}(y)  A^{\mu}(x)\right) \varepsilon^{\lambda \mu \nu \alpha} \nonumber\\
&=& i \int dx A^{\lambda}(x) B^{\nu}(x) F^{A}_{\rho \sigma}(x) \varepsilon^{\lambda \nu \rho \sigma}.
\eeqn
Proceeding in a similar way we obtain the expression of the Wess-Zumino vertex
\beqn
&& \int d^{4}x \, d^{4}y  \, d^{4}z  \,  \int \frac{d^{4}k_1}{(2 \pi)^{4}}  \frac{d^{4}k_2}{(2 \pi)^4} \varepsilon^{\mu \nu \rho \sigma} 
k_{1}^{\rho} k_{2}^{\sigma} e^{- i k_{1} \cdot (x-z) - i k_{2} \cdot (y - z)} \,  b(z) B^{\mu}(x) B^{\nu}(y)   \nonumber\\
&=&  \int d^{4}x \, d^{4}y  \, d^{4}z  \,  \int \frac{d^{4}k_1}{(2 \pi)^{4}}  \frac{d^{4}k_2}{(2 \pi)^4} \varepsilon^{\mu \nu \rho \sigma} 
 \left( \frac{1}{- i} \right)  \frac{\partial}{\partial x^{\rho}}   e^{- i k_{1} \cdot (x-z)}
 \left( \frac{1}{- i} \right)  \frac{\partial}{\partial y^{\sigma}} e^{ - i k_{2} \cdot (y - z)} 
\,  b(z) B^{\mu}(x) B^{\nu}(y)  \nonumber\\
&=&  (- 1) \int d^{4}x \, d^{4}y  \, d^{4}z  \, \delta^{(4)} (x-z) \delta^{(4)}(y -z) \, b(z) \,  
\frac{ \partial B^{\mu}}{\partial x^{\rho}}(x) \,  \frac{\partial B^{\nu}}{\partial y^{\, \sigma}}(y)  \,
 \varepsilon^{\mu \nu \rho \sigma} \nonumber\\
&=&  - \frac{1}{4}  \int  d^{4}x \, b(x) \,  F^{B}_{\rho \mu}(x) \, F^{B}_{\sigma \nu}(x) \,  \varepsilon^{\mu \nu \rho \sigma} 
=  \frac{1}{4}  \int  d^{4}x \, b \,  F^{B}_{\rho \mu} \, F^{B}_{\sigma \nu} \,  \varepsilon^{\rho \mu \sigma \nu} 
\eeqn
so that we find the following correspondence between Minkowsky space and momentum space for the Green-Schwarz vertex
\beqn
4 \varepsilon^{\mu \nu \rho \sigma} k^{\rho}_{1} k^{\sigma}_{2} \;\; \; \; \;
\leftrightarrow \;\; \; \; \; b F^{B} \wedge F^{B}.
\eeqn

%
%
%%%%%%%%%%%%%%%%%%%%%%%%%%%%%%%%%%%%%%%%%%%%%%%%%%%%%%%%%%%%%%
\section{Appendix: Computation of the Effective Action}
%%%%%%%%%%%%%%%%%%%%%%%%%%%%%%%%%%%%%%%%%%%%%%%%%%%%%%%%%%%%%%

%
In this appendix we illustrate the derivation of the variation of the effective action for typical anomalous contributions involving 
{\bf AVV} and {\bf AAA} diagrams. We consider the case of the A-B model described in the first 
few sections. We recall that we have the relations 
\beq
\delta B^\mu\,=\, \partial_{\mu}\theta_B\nonumber\\
\qquad \delta A^\mu\,=\, \partial_{\mu}\theta_A.
\eeq

We obtain 
%%%%here%%%%%

\beqn
\delta_B \mathcal{S}_{BAA}&=&  \delta_B  \int d^4x\,d^4y\,d^4z \, T_{\bf AVV}^{\lambda\mu\nu}(z,x,y)\,
B^\lambda(z)\,A^\mu(x)\,A^\nu(y)\nonumber\\
&=&- \int d^4x\,d^4y\,d^4z \, \partial_{z^\lambda}T_{\bf AVV}^{\lambda\mu\nu}(z,x,y)\,A^\mu(x)\,A^\nu(y)\,
\theta^{}_B(z)\nonumber\\
&=&- i a_3(\beta) \varepsilon^{\mu\nu\alpha\beta}\int d^4x\,d^4y\,d^4z \, \partial_{x^\alpha}\partial_{y^\beta}[\delta(x-z)\,\delta(y-z)]\,
A^\mu(x)\,A^\nu(y)\theta_B(z)\nonumber\\
&=&- i a_3(\beta) \varepsilon^{\mu\nu\alpha\beta}\int d^4x  \, \partial_{x^\alpha}A^\mu(x)\,
\partial_{x^\beta}A^\nu(x) \, \theta^{}_B(x)\nonumber\\
&=& i \frac{ a^{}_{3}(\beta) }{4}\int dx \, \theta^{}_B \, F_{\alpha\mu}^A F_{\beta\nu}^A \varepsilon^{\alpha\mu\beta\nu},
\eeqn
\beqn
\delta_A \mathcal{S}_{BAA}&=&  \delta_A  \int d^4x\,d^4y\,d^4z \, T_{\bf AVV}^{\lambda\mu\nu}(z,x,y)\,
B^\lambda(z)\,A^\mu(x)\,A^\nu(y)\nonumber\\
&=&- \int d^4x\,d^4y\,d^4z \, \partial_{x^{\,\mu}} T_{\bf AVV}^{\lambda\mu\nu}(z,x,y)
\,B^\lambda(z)\,\theta^{}_A(x)\,A^{\nu}(y)\nonumber\\
&&- \int d^4x\,d^4y\,d^4z \, \partial_{y^{\,\nu}} T_{\bf AVV}^{\lambda\mu\nu}(z,x,y)
\,B^\lambda(z)\,A^{\mu}(x)\,\theta^{}_{A}(y)\nonumber\\
&=& i a_1(\beta) \varepsilon^{\lambda\nu\alpha\beta}\int d^4x\,d^4y\,d^4z \, \partial_{x^{\,\alpha}} \partial_{y^{\,\beta}}
[\delta(x-z)\,\delta(y-z)]\,
B^\lambda(z)\,\theta^{}_{A}(x)\,A^\nu(y)\nonumber\\
&=& - i a_1(\beta) \varepsilon^{\lambda\nu\alpha\beta}\int d^4x\, \, \partial_{x^\alpha}
B^\lambda(x)  \,\partial_{x^\beta} A^\nu(x) \,  \theta^{}_{A}(x)\nonumber\\
&&+  i a_1(\beta) \varepsilon^{\lambda\mu\alpha\beta} \int d^4x\, \, 
\partial_{x^\beta} \,B^\lambda(x) \,\partial_{x^\alpha} \, A^{\mu} \,  \theta^{}_{A}(x)\nonumber\\
&=&  i \frac{ a^{}_{1}(\beta) }{4} \,2 \,\int d^{4} x \, \theta^{}_A \, F_{\alpha\lambda}^B F_{\beta\nu}^A \,
\varepsilon^{\alpha\lambda\beta\nu},
\eeqn

%%%%%%%%%%%%%%%%%%%%%%%%%%%%%%%%%%%%%%%%%%%%%%%%%%%%%%%%%%%%%%%%%
\section{Decay of a pseudoscalar: the triangle $\chi B B$}
%%%%%%%%%%%%%%%%%%%%%%%%%%%%%%%%%%%%%%%%%%%%%%%%%%%%%%%%%%%%%%%%%
%
The computation is standard and the result is finite. There are no problems 
with the handling of $\gamma_5$ and so we can stay in 4 dimensions.

We first compute the triangle diagram with the position of zero mass fermion $m_f=0$
 \beq 
 \int \frac{ d^4 q }{ ( 2 \pi )^4 } \frac{ Tr \left[ \gamma^{5} ( \slash{q} - \slash{k}) \gamma^{\nu} ( \slash{q} - \slash{k_1})
 \gamma^{\mu} \slash{q} \right] }{q^2 ( q - k )^2  (q - k_1)^2 }  \mbox{+ exch.}
\eeq
which trivially vanishes because of the $\gamma$-algebra. 
Then the relevant contribution to the diagram comes to be proportional to the mass $m_f \neq 0$, 
as we are now going to show.

We set $k=k_1 + k_2$ and set on-shell the B-bosons: $k_1^2 = k_2^2 = M_{B}^2$, so that 
$k^2 = 2 M_{B}^2 + 2 k_1 \cdot k_2 = m_{\chi}^2$

The diagram now becomes

\beq
\int \frac{ d^4 q }{ ( 2 \pi )^4 } \frac{ Tr \left[ \gamma^{5} ( \slash{q} - \slash{k} + m_f ) \gamma^{\nu} ( \slash{q} - 
\slash{k_1} + m_f)  \gamma^{\mu} ( \slash{q} + m_f ) \right] }{ \left[ q^2 - m^2_f \right] \left[ ( q - k )^2 - m^{2}_{f} \right]
\left[ (q - k_1)^2 - m^{2}_{f} \right] }
\label{diagr}
\eeq
Using a Feynman parameterization we obtain

\beqn
&=& 2 \int_{0}^{1}dx \int_{0}^{1-x} \frac{1}{ \left[ q^2 - 2 q [ k_2 y + k_1 (1-x) ] 
+ [\, y m_{\chi}^2 -m_{f}^2 + m_{B}^2(1-x-y)\,]   \right]^3}  \nonumber\\
&=&  2 \int_{0}^{1}dx \int_{0}^{1-x} dy \frac{1}{ \left[ q^2 - 2 q \Sigma +  D \right]^3} = 
2 \int_{0}^{1}dx \int_{0}^{1-x} dy \frac{1}{ \left[ (q - \Sigma)^2 -( \Sigma^2 - D) \,\right]^3}  \nonumber\\
&=&  2 \int_{0}^{1}dx \int_{0}^{1-x} dy \, \frac{1}{ \left[ (q - \Sigma)^2 - \Delta  \,\right]^3}.
\eeqn

We define
\beq 
\Sigma = y k_2 + k_1 ( 1 - x )   
\eeq
and 
\beq 
D =  y m_{\chi}^2 -m_{f}^2 + M_{B}^2(1-x-y),
\eeq
for the direct diagram and the function

\beq
\Delta = \Sigma^2 - D =  m_f^2 - x\,y \,m_{\chi}^2 + m_B^2 (x+y)^2 - x M_B^2 - y m_B^2 \equiv \Delta(x,y,m_f,m_\chi,M_B)
\eeq
and perform a shift of the loop momentum 
\beqn
q^\prime = q - \Sigma
\eeqn
obtaining

\beq
 2\int_{0}^{1}dx \int_{0}^{1-x}dy \int \frac{ d^D q }{ (2 \pi)^D  } \frac{ Tr [ \gamma^{5} ( \slash{q} + \slash{\Sigma}
- \slash{k} + m_{f}) \gamma^{\nu} ( \slash{q} + \slash{\Sigma} - \slash{k_1} + m_{f} ) \gamma^\mu ( \slash{q} + \slash{\Sigma} 
 + m_{f} ) ] }{ [ q^2 - \Delta ]^3 } 
\label{A}
\eeq
Using symmetric integration we can drop linear terms in q, together with
$q^\mu q^\nu = \frac{1}{D} q^2 g^{\mu\nu}$. Adding the exchanged diagram and after a routine calculation we obtain the amplitude for the decay
\beqn 
\Delta^{\mu\nu}=  \epsilon^{\alpha \beta \mu \nu }  k_1^{\alpha} k_2^{\beta}
m_f \left( \frac{1}{ 2 \pi^2}  \right) I(m_f,m_\chi,m_B)
\eeqn
with

\beq
I=\int^{1}_{0} dx \int^{1-x}_{0} dy  \frac{1 - 2 x - 2 y}{\Delta(x,y,m_f,m_\chi,m_B)}. 
\eeq

%%%%%%%%%%%%%%%%%%%%%%%%%%%%%%%%%%%%%%%%%%%%%%%%%%%%%%%%%%%%%%%%%%%%%%%%%%%%%%%%%%%%%%%%%%%%%%%%%%%%%%%%%%%%%%%%
%%%%%%%%%%%%%%%%%%%%%%%%%%%%%%%%%%%%%%%%%%%%%%%%%%%%%%%%%%%%%%%%%%%%%%%%%%%%%%%%%%%%%%%%%%%%%%%%%%%%%%%%%%%%%%%%

\end{document}